\documentclass[aps,onecolumn,nofootinbib,preprintnumbers]{revtex4}
\usepackage{amsmath,amssymb,epsfig}
\usepackage{times}
\usepackage{latexsym}
\usepackage{multirow}
\usepackage{url}
\usepackage{color}

\usepackage{graphicx}
\usepackage{epsfig}
\usepackage{amssymb}
\usepackage{color}
\usepackage{slashed}
\usepackage{soul}

\DeclareMathAlphabet   {\mathsc}{OT1}{cmr}{m}{sc}

\def\[{\left [}
\def\]{\right ]}
\def\({\left (}
\def\){\right )}
\newcommand{\lang}{\left\langle}
\newcommand{\rang}{\right\rangle}
\newcommand{\lbr}{\left\{}
\newcommand{\rbr}{\right\}}
\newcommand{\oline}[1]{\overline{#1}}

\newcommand{\Lag}{\mathcal{L}}

\newcommand{\GeV}      {~\mathrm{GeV}}
\newcommand{\TeV}      {~\mathrm{TeV}}

\newcommand{\PL}       {\mathsc{pl}}

\newcommand{\STR}      {\mathsc{str}}

\newcommand{\diff}{\mbox{d}}
\newcommand{\order}{\mathcal{O}}
\newcommand{\re}{{\rm Re}}

\newcommand{\gappeq}{\mathrel{\rlap {\raise.5ex\hbox{$>$}}
{\lower.5ex\hbox{$\sim$}}}}
\newcommand{\lappeq}{\mathrel{\rlap{\raise.5ex\hbox{$<$}}
{\lower.5ex\hbox{$\sim$}}}}

\newcommand{\comment}[1]{}

\begin{document}

\thispagestyle{empty}

\title{Mirage Models Confront the LHC:\\ II. Flux-Stabilized Type~IIB String Theory}
\author{Bryan L. Kaufman and Brent D. Nelson}
\affiliation{Department of Physics, Northeastern University, Boston, MA 02115, USA}

\begin{abstract}
We continue the study of a class of string-motivated effective supergravity theories in light of current data from the CERN Large Hadron Collider (LHC). In this installment we consider Type~IIB string theory compactified on a Calabi-Yau orientifold in the presence of fluxes, in the manner originally formulated by Kachru, et al. We allow for a variety of potential uplift mechanisms and embeddings of the Standard Model field content into $D3$ and $D7$ brane configurations. We find that an uplift sector independent of the K\"ahler moduli, as is the case with anti-$D3$ branes, is inconsistent with data unless the matter and Higgs sectors are localized on $D7$ branes exclusively, or are confined to twisted sectors between $D3$ and $D7$ branes. We identify regions of parameter space for all possible $D$-brane configurations that remain consistent with PLANCK observations on the dark matter relic density and measurements of the CP-even Higgs mass at the~LHC. Constraints arising from LHC searches at $\sqrt{s} = 8\,{\rm TeV}$ and the LUX~dark matter detection experiment are discussed. The discovery prospects for the remaining parameter space at dark matter direct detection experiments are described, and signatures for detection of superpartners at the~LHC with $\sqrt{s} = 14\,{\rm TeV}$ are analyzed.
\end{abstract}

\maketitle

\renewcommand{\thepage}{\arabic{page}}
\setcounter{page}{1}
\def\thefootnote{\arabic{footnote}}
\setcounter{footnote}{0}

\section{Introduction}

The recent pause in data-taking at the Large Hadron Collider (LHC) has allowed the two general-purpose detector collaborations to update their various supersymmetry search results to the full $\sqrt{s}=8\TeV$ data set of roughly 20~fb$^{-1}$. Thus far these searches have failed to provide any evidence of a signal beyond Standard Model backgrounds. As a result, limits can be placed on the parameter space of models of supersymmetry breaking within the context of the field content of the Minimal Supersymmetric Standard Model (MSSM). Absent any theoretical guidance, this parameter space is vast, though the number of parameters relevant for LHC~observables has been estimated to be roughly $\order(20)$~\cite{Djouadi:1998di,Brhlik:1998gu,Binetruy:2003cy,ArkaniHamed:2005px,Berger:2008cq,AbdusSalam:2009qd}. While this represents some improvement, it remains difficult to interpret the LHC data without recourse to certain simplified models, such as minimal supergravity~\cite{msugra}, in which the parameter space is greatly reduced.

Many indirect arguments suggest that supersymmetry will eventually be discovered at an energy scale within reach of the LHC. Most of these arguments stem from low-energy considerations. But for those inclined to accept that string theory is likely to provide a unified description of particle physics and gravitation, an additional motivation is the generic presence of $N=1$ supersymmetry in realistic models of particle physics from compactification of superstring theory to four dimensions. Thus, string theory provides a top-down motivation for studying supersymmetry at the LHC. Conversely, it can be argued that constraints on supersymmetry from null results at the LHC have consequences for which compactifications can be deemed `realistic' in the landscape of string vacua. This paper represents a continuation of a sequence of studies which take this viewpoint.

We have chosen to focus on models which exhibit some form of the so-called `mirage' pattern of superpartner masses~\cite{Choi:2007ka}. Roughly speaking, such models of supersymmetry breaking involve a hierarchy between the size of the gravitino mass (the order parameter of supersymmetry breaking in supergravity) and the size of supersymmetry breaking in the observable sector, parameterized by the vacuum expectation value (VEV) of the auxiliary field of some chiral super-multiplet. Usually, this chiral superfield represents one of the geometrical moduli of the compactified manifold whose stabilization via non-perturbative effects ultimately breaks supersymmetry in the vacuum.

We began our inquiry in Reference~\cite{Kaufman:2013pya}, with the case of heterotic string theory, in which the dilaton is stabilized via gaugino condensation in a hidden sector. The mirage pattern emerges in this case when one uses non-perturbative corrections to the dilaton action to modify the K\"ahler metric and thereby tune the resulting vacuum energy to vanish at the minimum of the effective potential~\cite{Gaillard:2007jr}. This construction was the first manifestation of the mirage pattern~\cite{Binetruy:1996xja,Binetruy:1996nx}, and its relatively simple parameter space made it a natural first point of investigation. Indeed, the model can be viewed as a concrete realization of the `generalized dilaton domination' scenario~\cite{Casas:1996zi} from the early days of string model building.

Some years later, the issue of geometrical moduli stabilization was studied in certain constructions of Type~IIB string theory compactified on Calabi-Yau orientifolds by Kachru, Kallosh, Linde and Trivedi (KKLT)~\cite{Kachru:2003aw}. In this class of theories NS~and RR~three-form fluxes are introduced to stabilize many of the moduli directly upon compactification. The presence of this flux warps the bulk geometry of the Calabi-Yau, resulting in a ``throat'' of the Klebanov-Strassler type~\cite{Klebanov:2000hb}. In the simplest implementation of the model, a hidden sector gaugino condensate is presumed to exist on a set of $D7$ branes at the infrared end of this throat, and is thus ``sequestered'' from the observable sector, in the language of Randall and Sundrum~\cite{Randall:1998uk}. At this point, all geometrical moduli are stabilized, but the minimum is, in fact, supersymmetric with a negative vacuum energy. Supersymmetry is broken explicitly by postulating the presence of anti-$D3$ branes at the tip of the Klebanov-Strassler throat, which can produce the desired minimum while only slightly perturbing the engineered moduli stabilization.

Not long after the theoretical basis for the KKLT framework was established, the phenomenology of the supersymmetry breaking was studied by Choi et al.~\cite{Choi:2004sx,Choi:2005ge} and the name `mirage mediation' was given to the pattern of soft supersymmetry breaking~\cite{Choi:2005uz}.\footnote{In the earliest work on the phenomenology of the KKLT scenario, the pattern of supersymmetry breaking was given the more cumbersome, if more precise, name ``mixed modulus--anomaly mediated supersymmetry breaking'' by Choi et al.} For our purposes, we continue to use the phrase `mirage model' to designate any theory in which  the ratios of the soft supersymmetry-breaking gaugino masses at the electroweak scale are governed by the approximate formula
\begin{equation} M_1\,:\,M_2\,:\,M_3 \, = \, (1.0+0.66\alpha)\, : \,
(1.93+0.19\alpha) \, : \, (5.87-1.76\alpha)\, ,
\label{mirage_ratios} 
\end{equation}
where the parameter $\alpha$ is determined by the model-dependent dynamics which stabilize the relevant moduli in the theory. In the heterotic case studied in~\cite{Kaufman:2013pya}, the remaining modulus requiring stabilization in the four-dimensional effective field theory was the dilaton, which is absent from the non-canonical kinetic terms of the observable sector matter fields, at leading order. In the Type~IIB context this remaining modulus is instead one of the K\"ahler moduli which, depending on how the MSSM is embedded in the compactified space, generally do appear in the observable sector K\"ahler metric. As a result, though the gaugino sector will show remarkable similarities between the two constructions, the phenomenology of the scalar sector will generally be very different. In some sense, the KKLT~scenario might be described as the ``generalized (K\"ahler) modulus-dominated'' scenario, in that the phenomenology will be heavily influenced by the effective modular weights of the chiral supermultiplets.

It should be emphasized, that in the decade since the original idea first appeared, many variants on the KKLT-based flux compactification model now exist, which remedy various theoretical or phenomenological shortcomings of the original paradigm. For example, inclusion of perturbative $\alpha'$ corrections to the K\"ahler potential leads to stabilization in a wholly different region of parameter space~\cite{Balasubramanian:2005zx}. Such Large (or LARGE) Volume Scenarios~\cite{Cicoli:2008va} have many compelling features, but are not (strictly-speaking) `mirage models', and thus we will not consider them further here~\cite{Conlon:2007xv}. Alternatively, one can retain the tree-level K\"ahler potential but include an O'Raifertaigh sector~\cite{Kallosh:2006dv} or Polonyi sector~\cite{Lebedev:2006qq} to address supersymmetry breaking and the vacuum energy density. The latter case has been used to engineer the mirage pattern in the heterotic context without recourse to K\"ahler stabilization~\cite{Lowen:2008fm,Krippendorf:2012ir}. Finally, one can imagine including a messenger sector which provides gauge mediation, and thus a `deflected' mirage mediation~\cite{Everett:2008qy,Everett:2008ey}. Many of these cases offer a rich model space with better phenomenological prospects than the original paradigm considered here. But we nevertheless choose to study the original KKLT~model in large part because of the central role played by the effective modular weights of the matter fields -- a role that can be obscured when additional structure is added to the construction.\\

This paper continues with a review of the physics of moduli stabilization in Type~IIB models in which compactification occurs on a Calabi-Yau orientifold with non-vanishing background flux. For those who work often in this field, the content of Section~\ref{kklt} will be very familiar, though the section does help to introduce notation and set the conventions which we will use throughout the remainder of the work. The actual soft supersymmetry-breaking terms which will define the model we consider appear in Section~\ref{sec:soft}. These terms suggest a space of free parameters, some of whom are governed by the manner in which the Standard Model field content is embedded into a system of $D$-branes in the compact space. We will not advocate for any particular construction, but instead perform a scan over all possibilities in Section~\ref{scan}, requiring (among other things) that the lightest supersymmetric particle (LSP) is a neutralino, that the thermal abundance of this lightest neutralino is no larger than the upper bound set by recent data from the PLANCK satellite, and that the lightest CP-even Higgs mass be within the range $124.1\,{\rm GeV} \leq m_h \leq 127.2\,{\rm GeV}$. The last condition will prove to greatly restrict the parameter space, particularly the value of the parameter $\alpha$ in~(\ref{mirage_ratios}). Little of the surviving parameter space would have been otherwise accessible at the LHC with center-of-mass energy $\sqrt{s} = 8\,{\rm TeV}$. We therefore devote the remainder of the study to the prospects for direct detection of the relic neutralino dark matter in future large scale liquid xenon detectors in Section~\ref{sec:dark}, and of superpartners generally at the~LHC with $\sqrt{s} = 14\,{\rm TeV}$ in Section~\ref{LHC}.

\section{The KKLT Model}
\label{kklt}

\subsection{K\"ahler Modulus Stabilization}

The original model of Kachru et al.~is an example of Type~IIB string theory compactified on a Calabi-Yau (CY) manifold in the presence of background fluxes. It is presumed that these fluxes fix the value of the dilaton and the complex structure moduli, leaving only the K\"ahler moduli in the low-energy four-dimensional effective theory~\cite{Giddings:2001yu}. In what follows we will take the simple case considered in~\cite{Kachru:2003aw}, in which there is a single, overall K\"ahler modulus $T$ parameterizing the overall size of the compact space. The existence of such a limit in the moduli space of any given Calabi-Yau is sufficiently generic to warrant the simplifying assumption, and (as we will see shortly) is roughly consistent with gauge coupling unification at or near the string scale.

The Standard Model can be realized on stacks of $D3$ branes, stacks of $D7$ branes, or strings stretched between some combination thereof. For gauge fields living on $D3$ branes, the gauge coupling is determined by the vacuum expectation of the dilaton $S$, while for gauge theories living on $D7$ branes, the gauge coupling is determined by the K\"ahler modulus $T$. We will consider only the latter case for the gauge degrees of freedom in this paper.
The K\"ahler potential for the modulus $T$ is taken to be $K(T,\oline{T}) = -3\ln (T+\oline{T})$. For gauge theories with group $\mathcal{G}_a$, living on $D7$ branes which wrap four-cycles in the CY manifold, the gauge coupling is determined by the K\"ahler modulus $T$ via the (universal) gauge kinetic function $f_a = T$. Note that, with these assumptions, 
\begin{equation}
 <\re \; t>\; = 1/g_{\STR}^{2} , \label{f2}
\end{equation}
where $t=T|_{\theta=0}$ is the lowest component of the superfield $T$, and $g_{\STR}$ is the universal gauge coupling at the string scale.

In $N=1$ supergravity theories the scalar potential is determined by the auxiliary fields $F^N$, associated with the chiral supermultiplet $Z^N$, and the auxiliary field $M$ of the supergravity multiplet. It is easy to solve the equations of motion for these auxiliary fields, thereby relating these quantities to the K\"ahler potential and superpotential via
\begin{equation}
F^M = - e^{K/2} K^{M\oline{N}} \left(\oline{W}_{\oline{N}} +
K_{\oline{N}} \oline{W} \right), \; \; \oline{M} = -3e^{K/2}
\oline{W} \label{EQM}
\end{equation}
with $W_{\oline{N}} = \partial W / \partial \oline{Z}^{\oline{N}}$,
$K_{\oline{N}} = \partial K / \partial \oline{Z}^{\oline{N}}$ and
$K^{M \oline{N}}$ being the inverse of the K\"ahler metric $K_{M
\oline{N}}=
\partial^2 K / \partial Z^M \partial \oline{Z}^{\oline{N}}$.  The
scalar potential is given by
\begin{equation}
V= K_{M\oline{N}} F^M  \oline{F}^{\oline{N}} - \frac{1}{3} M \oline{M}
\, \label{pot}
\end{equation}
where repeated indices are summed. Note that the final term in~(\ref{pot}) involves the gravitino mass explicitly, via the vacuum relation
\begin{equation} \lang M \rang = -3\lang e^{K/2} W \rang = -3m_{3/2} \, . \label{Mvev} \end{equation}

In the effective supergravity theory just below the string compactification scale, the presence of the three-form fluxes is represented by a constant $W_0$ in the effective superpotential. Combined with the effect of gaugino condensation in the hidden sector the total effective superpotential is then
\begin{equation}
W = W_0 + \sum_i A_i e^{-a_i T}\, , \label{WKKLT} \end{equation}
where the label $i$ runs over the various condensing gauge groups. When the non-perturbative corrections arise from gaugino condensation, we expect the $a_i$ to be related to the beta-function coefficient of the hidden sector gauge group, with a normalization such that $a = 8\pi^2/N$ for the group $SU(N)$. For simplicity, let us assume a single condensate from the gauge group $\mathcal{G}_+$ with coefficients $A_+=1$ and $a=a_+$.\footnote{To make contact with the notation from the heterotic model of Reference~\cite{Kaufman:2013pya}, one need only make the identification $ a_+ \to \frac{3}{2b_+}$.}
Minimizing the resulting scalar potential $V(t,\bar{t})$ 
generates a non-vanishing value for $\lang t + \bar{t} \rang$ at which the auxiliary field $F^T$ vanishes~\cite{Choi:2004sx}.
Restoring the Planck units to the second term in~(\ref{pot}) we see that the vacuum must therefore have an energy density given by $\lang V \rang = -3 m_{3/2}^2 M_{\PL}^2$. The size of the VEV for ${\rm Re}\,t$, as well as the size of the gravitino mass $m_{3/2}$, are determined by the size of the constant term $W_0$ in~(\ref{WKKLT}). In particular we have~\cite{Choi:2004sx}
\begin{eqnarray} \lang a_+ {\rm Re}\,t \rang &\simeq& \ln(A_+/W_0) \nonumber \\
m_{3/2} & \simeq & M_{\PL} \frac{W_0}{(2\lang{\rm
Re}\,t\rang)^{3/2}}\, . \label{relations} \end{eqnarray}
%
An acceptable phenomenology requires that the constant $W_0$ be finely-tuned to a value $W_0 \sim \order(10^{-13})$ in Planck units. That such a fine-tuning is possible at all is a particular feature of Type~IIB compactifications with three-form fluxes, as was noted by the original KKLT collaboration. Combining the two relations in~(\ref{relations}) we see that the model will assume an appropriate value of $W_0$ such that
\begin{equation} \lang a_+ {\rm Re}\,t \rang \simeq
\ln(M_{\PL}/m_{3/2}) \, . \end{equation}

\subsection{The Uplift Sector and Parameter $\alpha$}
\label{uplift}

The remaining component to the model is the inclusion of some additional `uplift' sector which generates supersymmetry breaking in the observable sector while producing a Minkowski (or slightly de Sitter) vacuum. Here a number of theoretical tools are at hand, but it is illustrative to begin with the canonical method employed in the original KKLT paper: the inclusion of anti-$D3$ branes which break supersymmetry {\em explicitly}. By sourcing the supersymmetry breaking at the end of the warped throat, it is reasonable to expect that the vacuum stabilization for the K\"ahler modulus $t = T|_{\theta=0}$ is thus largely unaffected. Being an explicit breaking of supersymmetry it is not possible to perfectly capture the effects of the anti-$D3$ branes in the form of corrections to the supergravity effective Lagrangian in superspace. However, it can be approximated~\cite{Choi:2005ge,Choi:2005uz} by assuming a correction to the pure-supergravity part of the action
\begin{equation} \Lag \ni -2\int \diff^4 \theta E  \to
-2 \int \diff^4 \theta \[ E + P(T,\oline{T})\] \label{Lag}
\end{equation}
which gives rise to a new contribution to the scalar potential for the modulus $T$.

When the modulus-dependence of $P(T,\oline{T})$ is trivial, and $P(T,\oline{T})=C$, then the resulting scalar potential contribution is simply
\begin{equation} V_{\rm lift} = \frac{C}{(t+\bar{t})^2} \, .
\label{Vlift1} \end{equation}
Such is the case with the anti-$D3$ brane scenario, since the positions of the  anti-$D3$ branes are fixed at a tip of the Klebanov-Strassler throat which is insensitive to the overall size modulus $T$. The coefficient $C$ can be calculated in this case, and is related to the warping factor which defines the throat. The dependence on ${\rm Re}\,T$ in~(\ref{Vlift1}) then arises exclusively from the consistency of $N=1$ supergravity under K\"ahler $U(1)$ transformations~\cite{Binetruy:2000zx,Choi:2005uz}.

One can generalize away from explicit supersymmetry breaking via anti-$D3$ branes to traditional D-term~\cite{Achucarro:2006zf,Choi:2006bh,Dudas:2006vc} or F-term~\cite{Dudas:2006gr,Abe:2006xp} soft supersymmetry breaking mechanisms, for which the superspace form in~(\ref{Lag}) is applicable. In general, we expect these mechanisms to involve explicit dependence on the K\"ahler modulus $T$ which we may parameterize via
\begin{equation} P(T,\oline{T})=C(T+\oline{T})^{n}\, . \label{Pgen}
\end{equation}
Under these circumstances the addition to the scalar potential is generalized from~(\ref{Vlift1}) to
\begin{equation} V_{\rm lift} = \frac{C}{(t+\bar{t})^{(2-n)}}\, .
\label{Vlift2} \end{equation}
The auxiliary field for the K\"ahler modulus no longer vanishes in the `lifted' vaccum, but instead satisfies the approximate solution
\begin{equation} M_0 \equiv \lang \frac{F^T}{t+\bar{t}} \rang \simeq
m_{3/2} \frac{2-n}{a_+ \lang t + \bar{t}\rang}\, . \label{MKKLT}
\end{equation}
Note that we will assume that the original solution for the value of the lowest component $t=T|_{\theta=0}$ is changed by only a negligible amount by the addition of the uplift sector.

The quantity $M_0$ in~(\ref{MKKLT}) serves as an order parameter of supersymmetry breaking in the observable sector. Soft supersymmetry breaking masses will generally be of this size, as we will describe in the next section. We note that provided the VEVs in~(\ref{relations}) can be arranged, we can identify a hierarchy defined by the ratio
\begin{equation} r = \frac{m_{3/2}}{M_0} \simeq a_+ \lang t +
\bar{t}\rang \simeq \ln(M_{\PL}/m_{3/2}) \gg 1 \, .
\label{rvalB} \end{equation}
It is this hierarchy that will ultimately generate the mirage pattern of gaugino masses. The outcome is not unlike the case of K\"ahler-stabilized heterotic string theory considered in~\cite{Kaufman:2013pya}.\footnote{Indeed, the connection between the ratio $r$ in~(\ref{rvalB}) and the parameter $a_{\rm np}$ of~\cite{Kaufman:2013pya} can be made explicit in the case where $n=0$
\begin{equation} r = a_+ \lang {\rm Re}\, t\rang
 =  a_+ \frac{1}{g_{\STR}^2} \quad \to \quad \frac{3}{2b_+ g_{\STR}^2}
 =  \frac{\sqrt{3}}{2a_{\rm np}}\, . \label{connection}
\end{equation}
Further details can be found in the appendix to Ref.~\cite{Altunkaynak:2009tg}.} Following Choi et al.~\cite{Choi:2005uz} we define the parameter $\alpha$ via
\begin{equation} \alpha \equiv
\frac{m_{3/2}}{M_0\ln\(M_{\PL}/m_{3/2}\)} \, , \label{alpha}
\end{equation}
and the implied value of $\alpha$ for an uplift sector~(\ref{Vlift2}) follows from the definition in~(\ref{alpha})
\begin{equation} \alpha = \frac{1}{1-n/2} +
\order\( 1/\ln(M_{\PL}/m_{3/2})\) \label{alphaKKLT}\, . \end{equation}
In the canonical case of anti-$D3$ branes, with $n=0$, we therefore have the prediction that $\alpha \simeq 1$ for this class of theories. While our discussion throughout the current subsection has anticipated that the parameter $n$ in~(\ref{Pgen}) is an integer (and, hence, that $\alpha$ is a rational number to leading order), we should note that significant departures from~(\ref{alphaKKLT}) can be obtained in cases where, for example, multiple condensates conspire to stabilize the K\"ahler modulus, with coefficients $a_i$ in~(\ref{WKKLT}) tuned accordingly~\cite{Choi:2005uz}. We will return to this issue when we discuss the parameter space of the model in Section~\ref{scan}.

\subsection{Observable Sector Soft Terms}
\label{sec:soft}

We can now directly write down the soft-supersymmetry breaking mass terms which we will consider in this paper, parameterized in terms of the two scales $M_0$ and $m_{3/2}$, or (equivalently) in terms of a single overall scale and the parameter $\alpha$.  For the sake of explicitness, we will follow~\cite{Choi:2005uz} and use the former convention. Given the assumption, stated above, that all gauge fields will arise from $D7$ branes, we can use the leading-order gauge kinetic function $f_a = T$ to obtain the gaugino masses at the boundary condition scale (taken to be the grand unification scale)
\begin{equation} M_a = M_0 + b_a g_{\rm STR}^2 M_g \, , \label{gaugmass}
\end{equation}
where we have defined~\cite{Falkowski:2005ck}
\begin{equation} M_g \equiv \frac{m_{3/2}}{16\pi^2} \, , \label{Mg} 
\end{equation}
and $b_a$ represents the beta-function coefficient for the Standard Model gauge group $\mathcal{G}_a$ with the normalization $b = \lbr 33/5,\, 1,\, -3 \rbr$. In~(\ref{gaugmass}) we include the so-called `anomaly mediated' contribution~\cite{Gaillard:1999yb,Bagger:1999rd}, as the hierarchy in~(\ref{rvalB}) will compensate for the loop factor, making the two terms competitive in size. This is the origin of the mirage pattern, and subsequent renormalization group (RG) evolution to the electroweak scale will produce the ratios in~(\ref{mirage_ratios}).

The soft-terms associated with the scalar sector of the theory will show a similar combination of tree-level and loop-level supergravity terms, with the latter arising through the super-conformal anomaly. To compute these, we will assume a leading order K\"ahler metric for matter field $Q_i$ given by
\begin{equation} K_{i\bar{j}} = \frac{\delta_{i\bar{j}}}{(T+\overline{T})^{n_i}} \, , \label{Kij}
\end{equation}
where $n_i$ is the modular weight of the field under $SL(2,Z)$ modular transformations. These weights can be inferred from the computation of string scattering amplitudes involving matter fields and geometrical moduli. These calculations have been performed in Type~IIB models, and in dual Type~IIA models with intersecting $D6$ branes~\cite{Lust:2004cx,Lust:2004fi,Lust:2004dn}. In brief, matter localized on stacks of $D3$ branes will have modular weights $n_i = 1$. Untwisted sectors localized on single stacks of $D7$ branes will have $n_i=0$, while twisted sectors stretched between $D3$ and $D7$ branes, or between different stacks of $D7$ branes, will exhibit a dependence on the overall K\"ahler modulus in the low-energy supergravity theory which can be represented in the form of~(\ref{Kij}) with $n_i=1/2$.
Details of the calculation of supersymmetry breaking soft terms in a general supergravity theory at one-loop can be found in Ref.~\cite{Binetruy:2000md}. Here we simply present the result in this particular effective theory~\cite{Choi:2005uz}
\begin{eqnarray} A_{ijk} &=& -(3-n_i - n_j - n_k)M_0 + \(\gamma_i + \gamma_j + \gamma_k \)M_g \label{Aterm} \\ 
m_i^2 &=& (1-n_i)M_0^2  - \theta_i M_0 M_g - \dot{\gamma}_i M_g^2 \, , \label{scalarmass}
\end{eqnarray}
where we have assumed that $m_{3/2}$ and $M_0$ are real. This can always be arranged in cases with a single condensate appearing in the superpotential~(\ref{WKKLT})~\cite{Choi:2005ge}. The various constants $\gamma_i$, $\dot{\gamma}_i$ and $\theta_i$ are collected in the Appendix.

\section{KKLT Parameter Scan}
\label{scan}

The KKLT model framework discussed in Section~\ref{kklt} involves two independent mass scales, given by the (normalized) gravitino mass $M_g$ in~(\ref{Mg}) and the modulus contribution $M_0$ in~(\ref{MKKLT}). Alternatively, one can work with either of the mass scales and the derived parameter $\alpha$ in~(\ref{alpha}). In exploring the parameter space of this model we will choose the latter, and use $M_0$ as the independent mass scale. The value of $m_{3/2}$ will then be computing by fitting to the expression in~(\ref{alpha}), and the calculated value will then be input into the high scale soft term expressions in~(\ref{gaugmass}), (\ref{Aterm}) and~(\ref{scalarmass}). By scanning on $M_0$ we will be better able to restrict our attention to the region that is of most interest to the LHC, and most motivated by fine-tuning considerations. 

In addition, one must specify the modular weights for the chiral supermultiplets that make up the MSSM field content. In this work we will allow only a limited amount of non-universality in assigning these weights. In particular, we will always assume that all matter multiplets arise from the same sector of the theory, so that they carry a universal modular weight $n_M$, while the two Higgs doublets may carry an independent modular weight which we will denote $n_H$. This assumption is consistent with the observed rates of flavor-changing neutral current processes, and with possible theoretical prejudices such as $SO(10)$ grand unification. 

Under these assumptions there are then nine possible combinations of modular weights to consider, which we can represent by the pair of weights $(n_M, n_H)$. Previous investigations into the phenomenology of the KKLT model have treated these discrete choices somewhat democratically~\cite{Baer:2006id,Baer:2006tb}, and we will do the same initially. However, we note that semi-realistic embeddings of the MSSM into Type~IIB orientifold compactifications tend to involve systems of open strings stretched between $D3$ and $D7$ branes, or among $D7$ branes at intersections~\cite{Lust:2004dn,Ibanez:2004iv,Marchesano:2004yq,Marchesano:2004xz}. Thus we will pay special attention in what follows to the four combinations of modular weights that do not involve $n_i = 1$ for either sector.

Finally, we will not address the origin of the supersymmetric Higgs mass parameter $\mu$, nor its accompanying soft-breaking parameter $B\mu$. Instead, we will perform the usual substitution of the known value of $M_Z$ and the continuous parameter $\tan\beta$ for these two quantities when addressing electroweak symmetry breaking (EWSB). This implies that our final parameter space involves a discrete choice of modular weights and three continuous parameters: $M_0$, $\alpha$ and $\tan\beta$. As mentioned in Section~\ref{uplift}, the value of $\alpha$ can be determined in explicit models of an uplift sector, but we will here prefer to allow the parameter to vary continuously. Nevertheless, we will be most interested in the original KKLT prediction $\alpha = 1$ and other special cases implied by the relation in~(\ref{alphaKKLT}).

\subsection{Global Scan}
\label{sec:global}

We therefore begin our survey of the LHC phenomenology of Type~IIB flux compactifications with nine scans on the three dimensional parameter space defined by $1000\,{\rm GeV} \leq M_0 \leq 5000 \, {\rm GeV}$, $0 \leq \alpha \leq 2$ and $2 \leq \tan\beta \leq 56$. 
The range of values in $\tan\beta$ reflect the range in which all three third-generation Yukawa couplings remain perturbative up to the boundary condition scale, for which we will follow standard practice and assume to be the grand unified scale $m_{\rm GUT} = 2 \times 10^{16}\, {\rm GeV}$. The lower bound on $M_0$ will ultimately reflect the need to achieve a mass for the lightest CP-even Higgs state of $m_h \gappeq 125~{\rm GeV}$. The upper bound on $M_0$ is arbitrary, but covers most of the region relevant for current and future searches for superpartners at the~LHC. The lower bound on the parameter $\alpha$ is the case of minimal supergravity, while the upper bound is near the value at which all three soft supersymmetry breaking gaugino masses in~(\ref{mirage_ratios}) are equal at the electroweak scale. Much larger values of $\alpha$ are in principle possible, but are phenomenologically challenged -- not least by the possibility that the lightest supersymmetric particle may become the gluino. Note that we will not consider negative values of $\alpha$. Arguments that give rise to the expression~(\ref{alphaKKLT}) would tend to disfavor uplift mechanisms that could generate negative values for this parameter. Nevertheless, if one takes a more phenomenological point of view, somewhat divorcing the soft terms in~(\ref{gaugmass}), (\ref{Aterm}) and~(\ref{scalarmass}) from the original string theory context, then such negative values may prove interesting. For a treatment of this extended parameter space, see~\cite{Falkowski:2005ck,Baer:2007eh}.

The need to perform nine distinct scans necessitates a two-stage approach: in this subsection we perform a global scan with a coarse subdivision of the parameter ranges studied. We will then determine phenomenologically relevant areas for a detailed, targeted scan, to be described in Section~\ref{sec:fine}. For the preliminary scan, we therefore allow $\alpha$ to vary between~0 and~2 in steps of size~0.1; $\tan\beta$ will range from~2 to~56 in unit steps, and $M_0$ is allowed to range from~1 to~5~TeV in 100~GeV steps. This results in 55 planes of constant $\tan\beta$, with 861 points per constant $\tan\beta$ plane, and approximately 425,000 points overall.

For each choice of the modular weights $(n_M, n_H)$, and value of the parameters $(M_0, \alpha, \tan\beta)$, the soft terms are computed from~(\ref{gaugmass}), (\ref{Aterm}) and~(\ref{scalarmass}). We note that when $n_M = 1$ it is not impossible for the squared scalar masses of the matter fields to be negative at the boundary condition scale. This is because the leading term in~(\ref{scalarmass}) then vanishes identically, and subleading terms generally give negative contributions to the scalar masses. In absolute value, the scalar masses will be comparable to, or slightly smaller than, the gaugino masses. Subsequent renormalization group evolution for the squared scalar masses of the matter fields generally drives all such terms to positive values by the electroweak scale. As a consequence, we will not consider this a fatal flaw for such a point in the parameter space, provided the squared soft masses at the low-energy scale are positive for the matter fields of the MSSM. 

The renormalization group equations are solved from the boundary condition scale to the electroweak scale using the package SOFTSUSY~3.3.9~\cite{Allanach:2001kg}. We will immediately exclude a combination of input parameters if the soft supersymmetry breaking scalar mass-squared parameter is negative for one or more of the matter fields at the electroweak scale. At this stage the radiatively-corrected Higgs potential is minimized and physical masses are calculated. We again eliminate a combination of input parameters if no solution to the conditions for electroweak symmetry-breaking can be found, or if the solution fails to converge adequately. Finally, we then ask that each model point have a neutralino LSP and sufficiently heavy superpartners to escape detection at LEP (namely a chargino heaver than $103.5\,{\rm GeV}$).

Having passed these minimal requirements, the electroweak scale spectrum is then passed to MicrOmegas~2.4.5~\cite{Belanger:2001fz,Belanger:2004yn} where the thermal relic abundance $\Omega_{\chi} h^2$ is computed for the stable neutralino. In addition, the rate for several rare decays are also computed, which can be directly compared to experimental results. For this model, and for the parameter range we investigate, the most important of these is the rate for the decay $B^0_s\rightarrow\mu^+\mu^-$. The first results from the LHCb~collaboration, using $1.0\,\rm{fb}^{-1}$ of data at $\sqrt{s}=7\rm{~TeV}$ and $1.1\,\rm{fb}^{-1}$ of data at $\sqrt{s}=8\rm{~TeV}$ report a branching ratio of $\mathcal{BR}\left( B^0_s \rightarrow\mu^+\mu^- \right)=\left(3.2^{+1.5}_{-1.2}\right)\times 10^{-9}$~\cite{Aaij:2012nna}. We take a generous 3$\sigma$~bound on this range, to avoid prematurely excluding any parameter space that may prove viable as more precise measurements are taken. As we will see below, this measurement tends to eliminate parameter space with light gauginos and high values of $\tan\beta$.

In addressing the issue of cold dark matter, we take a conservative approach and allow for the possibility of multi-component dark matter, of which the stable neutralino is but one component, and impose only an upper bound on the neutralino relic density. The final data release from the  WMAP~collaboration~\cite{Bennett:2012fp} gave a best-fit for the density of cold dark matter of $\Omega_{\rm CDM} h^2 = 0.1153\pm 0.0019$, when including data from `extended' CMB measurements, baryon acoustic oscillations, and direct measurements of the Hubble constant. Since that time, the PLANCK satellite has produced a slightly higher measurement~\cite{Ade:2013zuv} of $\Omega_{\rm CDM}h^2=0.1199\pm0.0027$. We will utilize this more recent measurement and choose to enforce a three-sigma upper bound on the calculation from MicrOmegas of $\Omega_{\chi}h^2\leq0.128$.

Finally, the initial discovery of the Higgs boson~\cite{:2012gu,:2012gk} has since been followed by more refined measurements of the mass of the Higgs field, using the complete data sets from both $\sqrt{s}= 7\,{\rm TeV}$ and 8~TeV center-of-mass energies. ATLAS reports a combined result of $m_h=125.5\pm0.2^{+0.5}_{-0.6}$ GeV~\cite{ATLAS:2013mma}, while CMS reports $m_h=125.3\pm0.4\pm0.5$ GeV~\cite{Chatrchyan:2013lba}. Adding the ATLAS uncertainties in quadrature as a back-of-the-envelope combination of errors, we find $m_h=125.5\pm0.54$~GeV, while a similar exercise for CMS gives $m_h=125.7\pm0.424$~GeV. Combining these two to arrive at an acceptable Higgs mass range, we allow $m_h=125.6^{+0.8}_{-0.7}$~GeV, leaving us with the range $124.1\,{\rm GeV}\leq m_h \leq 127.2\,{\rm GeV} $.

\begin{table}[t]
\begin{center}
\begin{tabular}{|c||c|c|c||c||c|c|c|}
\multicolumn{4}{c}{Bino-like LSP} & \multicolumn{4}{c}{Higgsino-like LSP} \\
\hline
 &  $n_H=0$ & $n_H=1/2$ & $n_H=1$ &  & $n_H=0$ & $n_H=1/2$ & $n_H=1$ \\ \hline\hline
$n_M=0$ & $\alpha=1.0-1.1$  & $\alpha=1.0-1.3$  & $\alpha=0-0.2$ 
 & $n_M=0$& $\alpha=2.0$ & $\alpha=1.9-2.0$ & --\\
 & $M_0=1.2-2.5$ & $M_0=1.4-2.0$ & $M_0=1.7-2.8$  &  
 & $M_0=2.5-3.4$   & $M_0=2.0-2.7$ &     \\
 & $\tan\beta=24-32$ & $\tan\beta=10-30$ &  $\tan\beta=51-52$  & 
 & $\tan\beta=48-51$ & $\tan\beta=42-48$  &    \\ \hline
$n_M=1/2$& $\alpha=1.0-1.8$ & $\alpha=0.5-0.8$ & $\alpha=0$ 
 & $n_M=1/2$ & $\alpha=1.0-1.8$ & $\alpha=1.5-1.8$ & $\alpha=2.0$ \\
 & $M_0=1.6-4.0$ & $M_0=1.8-2.5$  &  $M_0=2.3-3.0$ &  
 & $M_0=1.6-4.0$ & $M_0=2.4-5.0$  &  $M_0=4.6-5.0$ \\
 & $\tan\beta=6-50$ & $\tan\beta=12-35$ & $\tan\beta=53-54$  &   
 & $\tan\beta=6-50$ & $\tan\beta=7-52$  &  $\tan\beta=34-45$ \\ \hline
$n_M=1$ & -- & -- & -- 
 & $n_M=1$  & $\alpha=0.7$ & $\alpha=0.8$  & $\alpha=1.1$  \\
 &    &     &      & 
 & $M_0=2.2-4.6$ &  $M_0=3.3-5.0$  & $M_0=4.8-5.0$     \\
  &    &     &      & 
  & $\tan\beta=6-29$ & $\tan\beta=8-46$ & $\tan\beta=18-29$ \\ \hline
\end{tabular}
\caption{Allowed combinations of $\alpha$, $M_0$ (in TeV) and $\tan\beta$, for each combination of modular weights $n_M$ and $n_H$, separated into regions in which the LSP is bino-like versus regions in which it is predominantly Higgsino-like. For the case of modular weight combination  $(n_M, n_H) = (1/2,0)$ the parameter space interpolates between these two cases, allowing for a mixed-wavefunction LSP. For this reason we have listed the allowed parameter space in both panels. An empty cell implies that the indicated type of neutralino wavefunction does not occur for any combination of parameters with that set of modular weight assumptions.}
\label{coanntable}
\end{center}
\end{table}

A combination of input parameters must meet all of the above requirements to be considered phenomenologically viable. Ultimately, the most restrictive conditions on the parameter space prove to be the upper bound on the thermal relic abundance of neutralinos and the measured value of the lightest CP-even Higgs mass. The latter will mostly require an overall increase in the supersymmetry-breaking mass scale parameterized by $M_0$, but the relic abundance constraint will have implications that vary from one set of modular weights to another, and will often single out particular values of the parameter $\alpha$. We therefore find it convenient to summarize our results in the form of Table~\ref{coanntable}, where we have grouped the allowed parameter space regions first in terms of the identity of the lightest neutralino, and secondarily in terms of the modular weights $n_M$ and $n_H$. 

The left panel in Table~\ref{coanntable} represents the parameter combinations in which the lightest neutralino is overwhelmingly bino-like throughout the parameter space. In these cases the correct thermal relic abundance for the neutralino is obtained primarily through co-annihilation between the LSP and the lightest stau. The right panel represents the parameter combinations in which the lightest neutralino is overwhelmingly Higgsino-like. Here the correct thermal relic abundance is obtained primarily through co-annihilation between the LSP and the lightest chargino and/or second-lightest neutralino. 
Each cell in the table represents a particular combination of modular weights $(n_M,n_H)$, and we give the rough range in the continuous parameters $\lbrace \alpha, M_0, \tan\beta \rbrace$ consistent with the conditions outlined above. Note that the total allowed parameter space for a given pair $(n_M,n_H)$ is the union of the regions in both panels of Table~\ref{coanntable}. 
For the particular combination $(n_M, n_H) = (\frac{1}{2},0)$ the parameter space is listed in both panels of the table. Here the wavefunction of the neutralino varies across the parameter space, from overwhelmingly bino-like to completely Higgsino-like, with some regions of mixed-wavefunction LSPs. In this case there is no significant co-annihilation at all, but the thermal relic density is nevertheless consistent with the WMAP~bound. 
Even with the relatively coarse step size at this stage in the analysis, it is clear that the allowed parameter space is given by disjoint sets of points. This will be of great utility when we investigate these spaces in greater resolution in the next section. 

Before we do so, however, it is convenient to identify some broad properties of the sorts of parameter combinations that remain viable in the fluxed Type~IIB model of~KKLT. First we note how tightly constrained are the cases in which either $n_M$ or $n_H =1$. Such cases represent constructions in which either the matter sector or the Higgs sector is confined exclusively to $D3$~branes. In general these cases tend to cluster around a single acceptable value of $\alpha$. For all modular weight combinations, the relatively large mass scales are necessitated by the requirement that the ultimate value of the Higgs mass be bounded by $m_h \geq 124.1 \, {\rm GeV}$. To understand this behavior, we recall from~(\ref{scalarmass}) that cases with $n_M, n_H=1$ will have scalar masses that are highly suppressed relative to gaugino masses, making it difficult to achieve large radiative corrections to the Higgs mass at the electroweak scale. The problem is exacerbated by the fact that when $n_M=1$ we expect the trilinear A-terms to be no larger than the gaugino mass at the boundary condition scale, making it difficult to achieve the `maximial mixing scenario' to boost the mass of the lightest CP-even Higgs~\cite{Carena:2002qg}. 

The basic texture of the panels in Table~\ref{coanntable} is also readily understood from the nature of the soft terms in~(\ref{gaugmass}) and~(\ref{scalarmass}). In general, for co-annihilation to be effective at reducing the abundance of a relic species, the co-annihilator should be within a few percent of the mass of the of the relic particle. For the case of stau/neutralino co-annihilation this requires a careful conspiracy between the values of the gaugino masses (governed by $\alpha$ and $M_0$) and the stau mass (governed by $M_0$, $n_M$ and $\tan\beta$). The two masses will be roughly equivalent when $\tan\beta$ is moderately large and $n_M = 0$, though $n_M = 1/2$ is also possible if the value of $\alpha$ and $\tan\beta$ compensate appropriately. For all such stau co-annihilation regions in the left panel of Table~\ref{coanntable}, the LSP is overwhelmingly bino-like in composition.

Meanwhile, processes involving co-annihilation among a system of degenerate gauginos are largely independent of the size of the scalar masses relative to the gauginos. Thus we find chargino co-annihilation processes in nearly all the allowed combinations of modular weights. For cases with $n_M=1$ we find that {\em all} the allowed parameter space involves neutralino/chargino co-annihilation. The mass degeneracy in the gaugino sector increases as $\alpha \to 2$ and in all of these cases the LSP is predominantly Higgsino-like in nature, with a high degree of mass degeneracy with other neutralinos and charginos. The low value of the $\mu$-parameter in these cases is being driven by large radiative corrections to the electroweak minimization conditions~\cite{Allanach:2012qd}, themselves the result of the very large scalar masses in this sector of the modular weight space.

\begin{figure}[t]
\begin{center}
\includegraphics[width=0.80\textwidth]{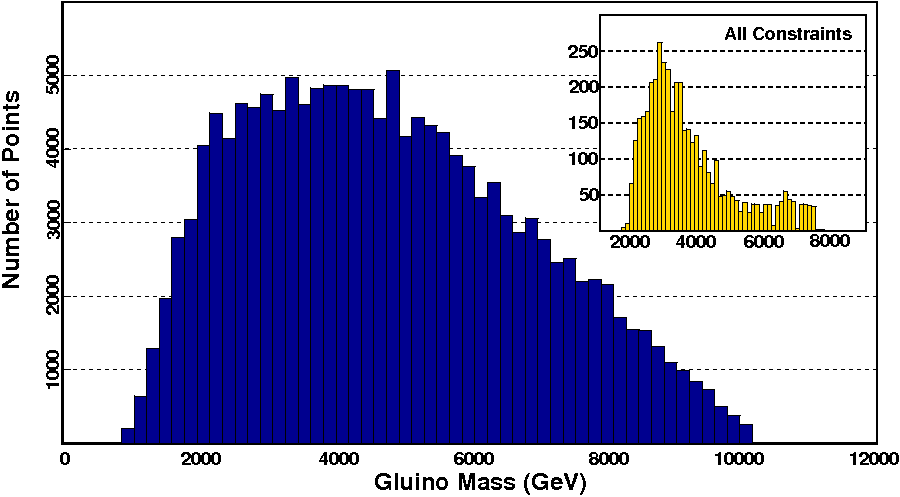}
\caption{\textbf{Histogram of Gluino Masses for All Modular Weight Combinations.} Distribution of gluino masses, in~GeV, aggregated across all modular weights in the global scan. Blue bars represent the distribution when only proper EWSB and neutralino LSP is imposed. The inset with the yellow bars shows the distribution after imposing the Higgs mass constraint and an upper bound on thermal neutralino relic density.}
\label{plot:gluinohisto}
\end{center}
\end{figure}

The large mass scales imposed by the Higgs mass constraint has a direct impact on the size of the gluino mass for these Type~IIB flux compactification models. This is illustrated in Figure~\ref{plot:gluinohisto}, where we show the range of resulting gluino masses, aggregated across all modular weight combinations in our global scan. The main body of the plot (blue bars) shows the distribution of gluino masses when only the most minimal phenomenological requirements are imposed: proper electroweak symmetry breaking and the demand that the LSP be the lightest neutralino. Even without requiring $m_h \geq 124.1 \, {\rm GeV}$, the distribution is highly skewed toward gluino mass values which are inaccessible at the LHC. After imposing the Higgs mass requirement and the upper bound on the thermal relic neutralino density, we arrive at the inset distribution (yellow bars). In this case, the lowest gluino masses have been eliminated, but the distribution peaks more sharply at values that, while challenging, are within reach for the LHC at $\sqrt{s} = 14\,{\rm TeV}$. Clearly, we do not expect any of the allowed parameter space to be eliminated with current data at $\sqrt{s} = 8\,{\rm TeV}$. We will return to both of these statements in Section~\ref{LHC}.

\begin{figure}[t]
\begin{center}
\includegraphics[width=0.80\textwidth]{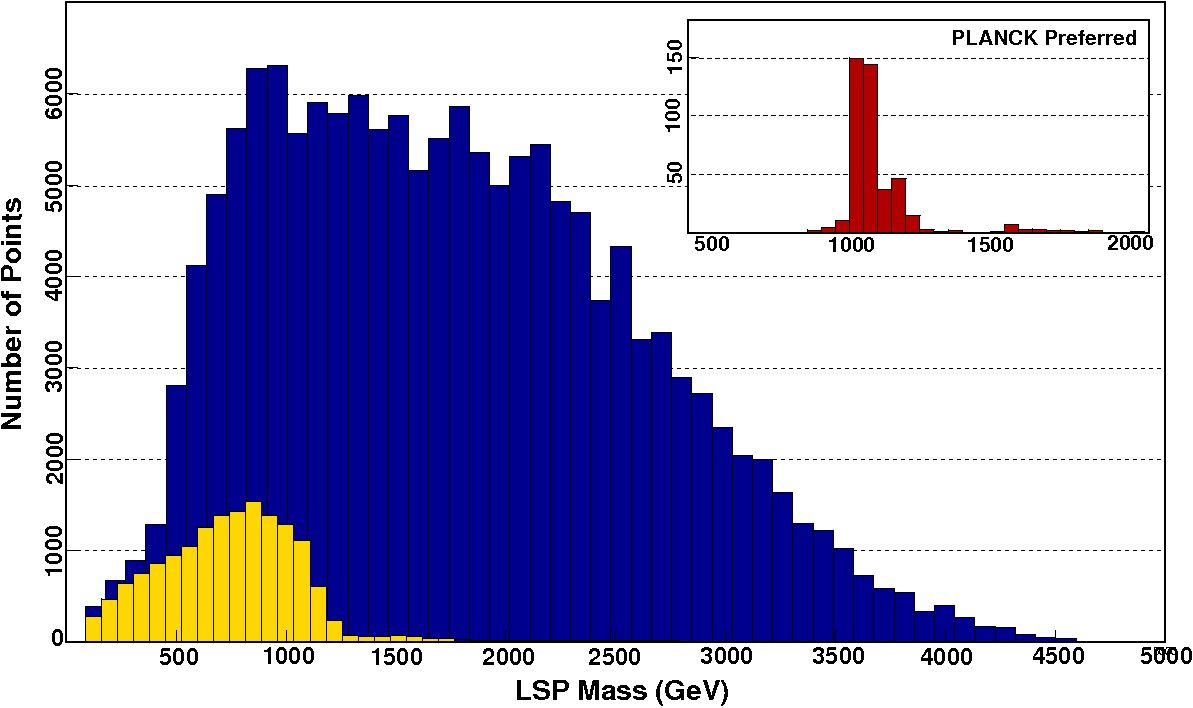}
\caption{\textbf{Histogram of LSP Masses for All Modular Weight Combinations.} Distribution of the mass of the lightest neutralino, in~GeV, aggregated across all modular weights in the global scan. Blue bars represent the distribution when only proper EWSB and neutralino LSP is imposed. The yellow bars show the distribution after imposing the Higgs mass constraint and an upper bound on the thermal neutralino relic density. The inset with the red bars is the result of requiring the relic density to be within 3$\sigma$ of the PLANCK measurement $\Omega_{\rm CDM}h^2=0.1199\pm0.0027$. }
\label{plot:LSPhisto}
\end{center}
\end{figure}

Before looking more closely at each of the regions designated in Table~\ref{coanntable}, we make a final observation regarding the distribution of LSP masses across all the modular weight combinations in our global scan. In Figure~\ref{plot:LSPhisto} the large distribution (blue bars) represents the mass of the lightest neutralino, aggregated over all modular weight combinations, in which we only require proper EWSB and that the LSP is, in fact, a neutralino. Like the case of the gluino mass in Figure~\ref{plot:gluinohisto}, the overall distribution is forced to exceptionally high values even before any further phenomenological considerations are made. The smaller distribution in the main plot (yellow bars) is the result of requiring $124.1\,{\rm GeV}\leq m_h \leq 127.2\,{\rm GeV} $ and the {\em upper bound} on the thermal relic density of this LSP, $\Omega_{\chi}h^2\leq0.128$. Relatively large values of the LSP mass are favored, though we expect to find many cases in which $m_{\chi} \sim 100\,{\rm GeV}$. However, if we instead require that this LSP represent {\em all} of the dark matter, such that the thermal relic density is within a 3$\sigma$~range about the PLANCK measurement of $\Omega_{\rm CDM}h^2=0.1199\pm0.0027$, we find that the LSP~mass is forced to be at or above 1~TeV. This is illustrated by the inset distribution (red bars), labeled `PLANCK Preferred' in Figure~\ref{plot:LSPhisto}. We will comment on the implication of these facts in Section~\ref{sec:dark}.

\subsection{Individual Targeted Scans}
\label{sec:fine}

Table~\ref{coanntable} makes it clear that there are distinct regions where either $\tilde\tau$ or $\tilde\chi^\pm$ co-annihilation depletes the dark matter content sufficiently so that both the dark matter relic density and Higgs mass are consistent with experimental observations. In the majority of cases these regions are non-intersecting, which allows us to narrow our search and examine smaller regions with a finer resolution. The spectrum of neutralino and chargino masses tends to be, at leading order, independent of the value of $\tan\beta$, and thus we find the value of $\tan\beta$ to be largely uncorrelated with that of $\alpha$ and $M_0$ for the Higgsino LSP cases of Table~\ref{coanntable}. For these combinations of modular weights, therefore, we can choose a fixed value of $\tan\beta$, then perform a scan over $M_0$ and $\alpha$ with ranges that correspond to the phenomenologically viable regions discovered in Section~\ref{sec:global}. For the bino LSP cases the value of $\tan\beta$ must be correlated with that of both $M_0$ and $\alpha$ so as to obtain a sufficient mass degeneracy between the lightest neutralino and the lightest stau. For these cases, then, we will continue to perform a three-dimensional scan, but restrict the ranges to the phenomenologically viable regions discovered in Section~\ref{sec:global}. 
Whereas in the global scan $M_0$ was scanned with $100\GeV$ intervals, these scans will proceed with $10\GeV$ intervals. Similarly, $\alpha$ is now scanned in steps of 0.01.

In what follows we will discuss each of the nine sets of modular weights $(n_M,n_H)$, occasionally grouping cases where the basic features are similar. Throughout we will wish to bear in mind those combinations of parameters that find the greatest motivation from underlying theories of moduli stabilization in Type~IIB string theory and attempts to realize the Standard Model field content in such models. As such, we will be particularly focused on systems in which both matter and Higgs representations involve systems of $D7$~branes, so that both $n_M$ and $n_H$ take values of zero or one-half. In addition, we recall that simple models of uplift sectors indicate a prediction for $\alpha$ given by~(\ref{alphaKKLT}), which suggests that certain rational numbers ($2$, $1$, $2/3$, $1/2$ , $2/5$ etc.) should be considered more reasonable values for $\alpha$ than other general values, modulo the higher order corrections that amount to a few percent for most of parameter space.

\subsubsection{The case $\(n_M,\, n_H\) = \(0,\,0\)$}

We begin, therefore with the case in which all matter arises from sectors confined to a single system of $D7$~branes, such that both modular weights vanish. Table~\ref{coanntable} indicates that this scenario involves a tightly-confined region with a Higgsino-like LSP and degenerate gauginos, and a second region near $\alpha=1$ with more moderate mass scales in which stau co-annihilation is the dominant mechanism for achieving the correct relic density of neutralinos.
It is instructive to consider the gaugino co-annihilation region first and in detail, since many of the properties that constrain this region will be repeated in the other cases we address.

When all $n_i =0$ the gauginos and scalars begin at the high scale with roughly equal masses. Thus, depending on the values of $\tan\beta$ and $\alpha$ it is possible for a squark or a slepton to emerge at the electroweak scale lighter than the lightest neutralino eigenstate. The gaps between the parameter space represented by the two panels in Table~\ref{coanntable} arise from precisely this phenomenon. For example, for $35 \lappeq \tan\beta \lappeq 45$ the stau is {\em always} the LSP, regardless of the value of $\alpha$. In addition, for $\alpha \gappeq 1.2$ the stop becomes the LSP for $\tan\beta \gappeq 5$. The origin of this behavior is evident from the final term in~(\ref{scalarmass}). The quantity $M_g$ carries an implicit factor of $\alpha$ relative to $M_0$, as can be seen from~(\ref{alpha}), so the final term in~(\ref{scalarmass}) becomes increasingly dominant at large $\alpha$. For first and second generation particles, this tends to reduce the boundary scale mass, since $\dot{\gamma} \sim g^4$, but for third generation particles we have $\dot{\gamma} \sim g^4 +g^2\lambda^2-\lambda^4$, which increases the boundary condition mass for fields with large third-generation Yukawa couplings. Thus, for the stop field, which naturally has a mass very near the LSP in this region of parameter space, the extra contribution from the $\lambda_t^4$ term when $\tan\beta$ is small can make the stop just slightly more massive than the neutralino. These surviving points at very low $\tan\beta$ and $\alpha \gappeq 1.2$ ultimately fail to deliver a Higgs mass that exceeds even the previous LEP bound of $m_h \geq 114.4\,{\rm GeV}$, and are thus eliminated from further study.

The gaugino co-annihilation region that opens up for very large values of $M_0$, $\alpha$ and $\tan\beta$ emerges for a very different reason. Here we are firmly in the region where radiative corrections to the EWSB potential are growing rapidly, with the radiatively corrected $\mu$ parameter diminishing rapidly as both $\tan\beta$ and $M_0$ increases. Thus, while the stop mass is dropping, the mass of the lightest neutralino -- dominated as it is by the value of $\mu$ in this area of parameter space -- is falling even faster. Eventually the system of highly-degenerate Higgsino-like neutralinos and charginos emerge as lighter in mass than the stop, and the system becomes viable. In our targeted scan we fixed $\tan\beta=48$ and find that the allowed region in the parameter $\alpha$ is highly constrained with $1.96 \leq \alpha \leq 2$. The lower bound corresponds to the requirement $M_0 \geq 2700\,{\rm GeV}$, while at $\alpha = 2$ we must require $M_0 \geq 2570\,{\rm GeV}$. The lower bounds on $M_0$ arise from the constraint on the process $B_s \to \mu^+ \mu^-$ and the lower bound on the Higgs mass $m_h$. In fact, over this entire allowed region the Higgs mass satisfies $m_h \leq 125.2\,{\rm GeV}$, despite the very large value of $\tan\beta$. The LSP neutralino is quite massive ($1201\,{\rm GeV}\leq m_{\chi_1^0} \leq 1585\,{\rm GeV}$) though the very large value of $\alpha$ results in a relatively light gluino ($1737\,{\rm GeV}\leq m_{\tilde{g}} \leq 3506\,{\rm GeV}$).

This leaves only the region in the left panel in Table~\ref{coanntable}, where the combination of $\alpha$ and $\tan\beta$ conspire to make the stau ever-so-slightly larger in mass than the neutralino LSP, thereby producing an acceptable relic abundance of cold dark matter. For our targeted scan we restrict the range in $\tan\beta$ to $24 \leq \tan\beta \leq 32$. For most of the allowed combinations of $\lbrace \alpha,M_0 \rbrace$, however, a very small range of $\tan\beta$ was allowed. For example, for $0.93 \leq \alpha \leq 1.07$ we find that we must require $\tan\beta = 31 \pm 1$ in order for the stau mass to be sufficiently close to the bino-like LSP mass to allow for an acceptable value of the thermal relic density for the neutralino.

\begin{figure}[t]
\begin{center}
\includegraphics[width=0.55\textwidth]{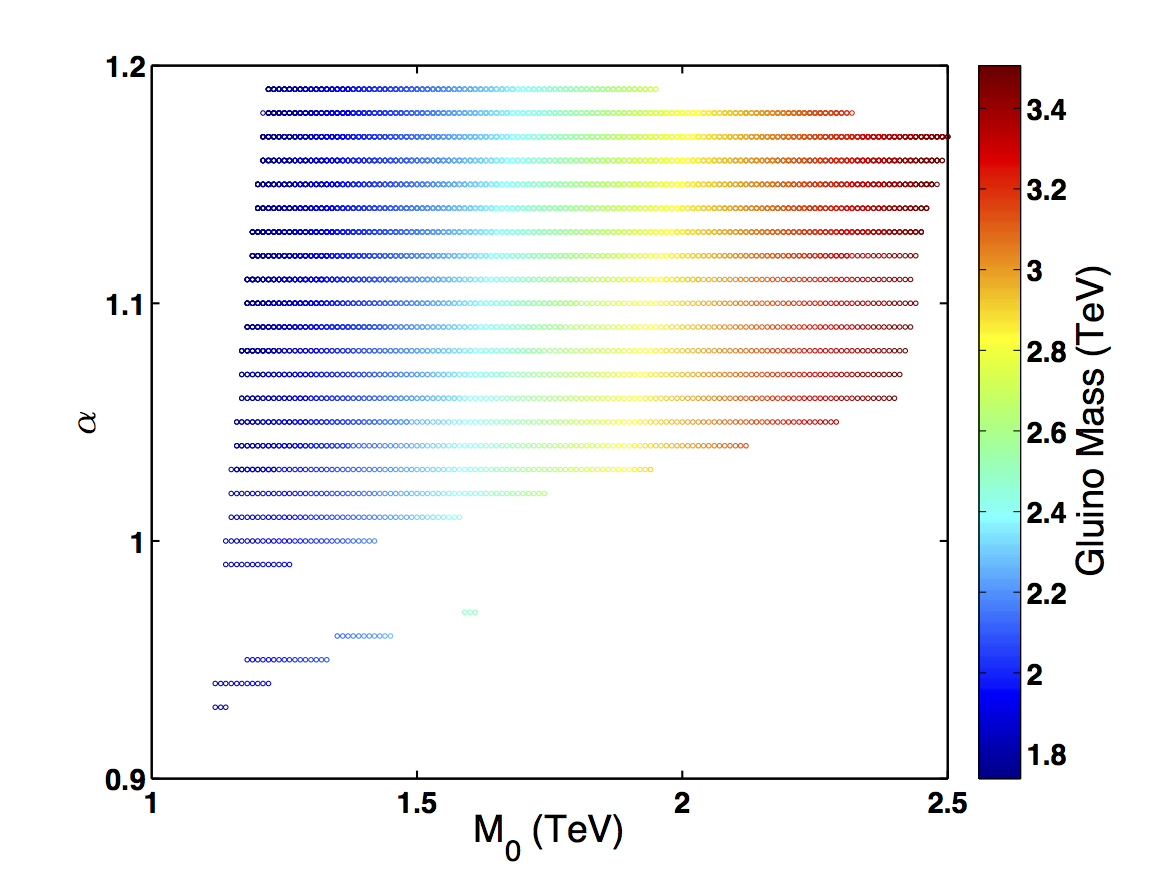}
\caption{\textbf{Allowed Parameter Space for $(n_M,n_H)$ =  $(0,0)$ for Bino-like LSP.} Parameter combinations of $\lbrace \alpha,M_0 \rbrace$ consistent with proper EWSB, Higgs mass measurements and upper bound on thermal relic abundance. The gluino mass, in units of~TeV, is given by the color as indicated by the scale to the right.}
\label{plot:n00}
\end{center}
\end{figure}

It is thus possible to show the allowed parameter space as a two-dimensional projection onto the $\lbrace \alpha,M_0 \rbrace$ plane, as in Figure~\ref{plot:n00}. In general, the lower bound on $\alpha$ for a fixed value of $M_0$ arises from the relic density requirement, which is only satisfied when the stau mass is sufficiently close to that of the lightest neutralino. The small disconnected region in Figure~\ref{plot:n00}, below $\alpha \simeq 1$, exists only for $\tan\beta=32$ and represents those points for which the lightest stau is slightly more massive than the lightest neutralino and the relic density is just slightly below our imposed upper bound.  The upper limit on $\alpha$, for a fixed value of $M_0$, occurs when there is no value of $\tan\beta$ for which the scalar top is {\em not} the lightest superpartner. 

Meanwhile, the lower bound on the Higgs mass, $m_h \geq 124.1\,{\rm GeV}$ produces the edge at lower $M_0$ values, while the upper bound $m_h \leq 127.2\,{\rm GeV}$ provides an upper bound on $M_0$ for a fixed value of $\alpha$. The gluino mass that arises for each parameter combination is largely insensitive to the chosen value of $\tan\beta$, so it is possible to display this quantity in the projected parameter space. This is indicated by the color in Figure~\ref{plot:n00}. Gluino masses for this part of the $(n_M,n_H) = (0,0)$ parameter space range from a low of 1737~GeV to a high of 3506~GeV, while the predominantly bino-like LSP takes a mass in the range $788\,{\rm GeV} \leq m_{\chi_1^0} \leq 1952\,{\rm GeV}$.

\subsubsection{The case $\(n_M,\, n_H\) = \(0,\,\frac{1}{2}\)$}

Incrementing the modular weight for the Higgs sector to the case $(n_M,n_H)$ =  $(0,1/2)$, we consider next the case where the matter fields of the MSSM remain confined exclusively to single stacks of $D7$~branes, but the Higgs sector is realized on stretched strings connecting $D7$~branes with either $D3$~branes or another set of $D7$~branes. The overall structure of the parameter space is similar to the previous case, so we allow ourselves the opportunity to be more succinct in the description. 

\begin{figure}[t]
\begin{center}
\includegraphics[width=0.45\textwidth]{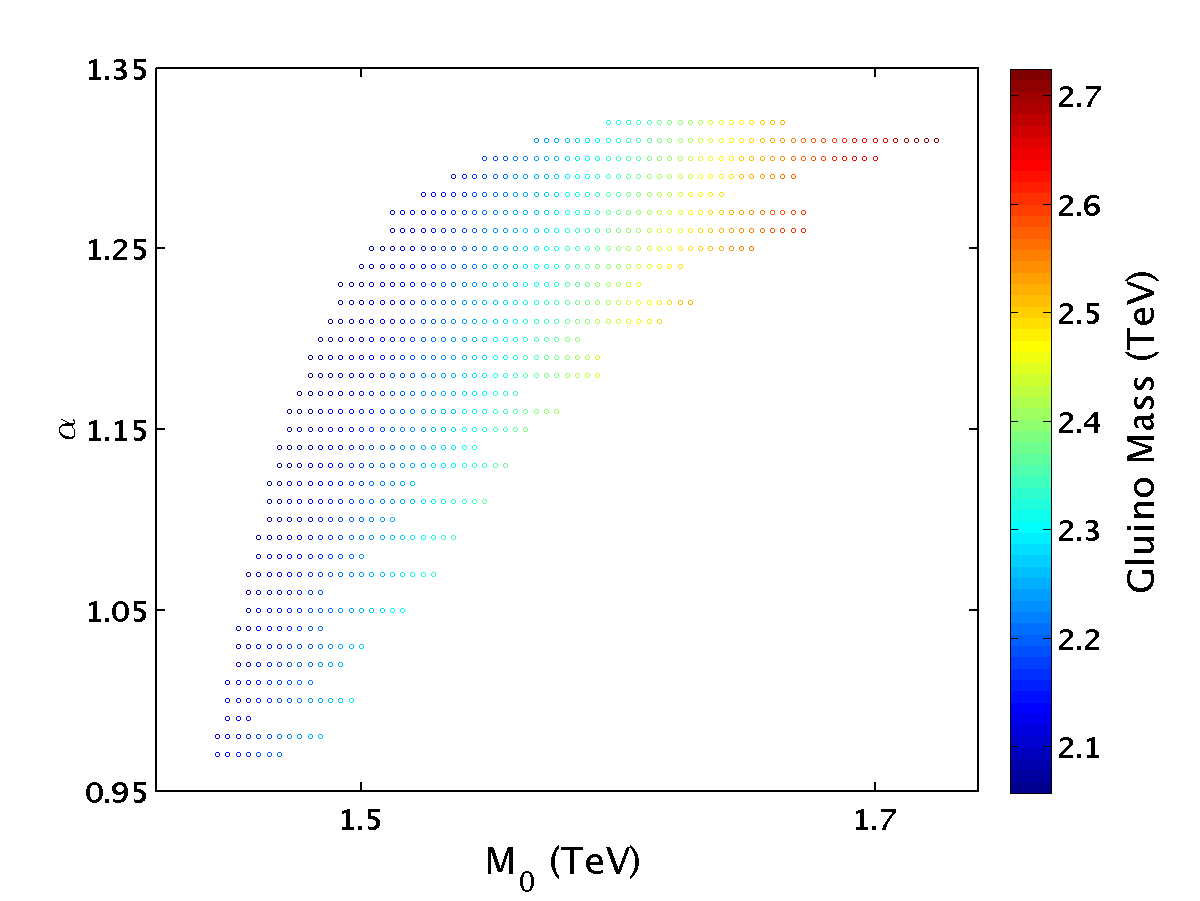}
\includegraphics[width=0.45\textwidth]{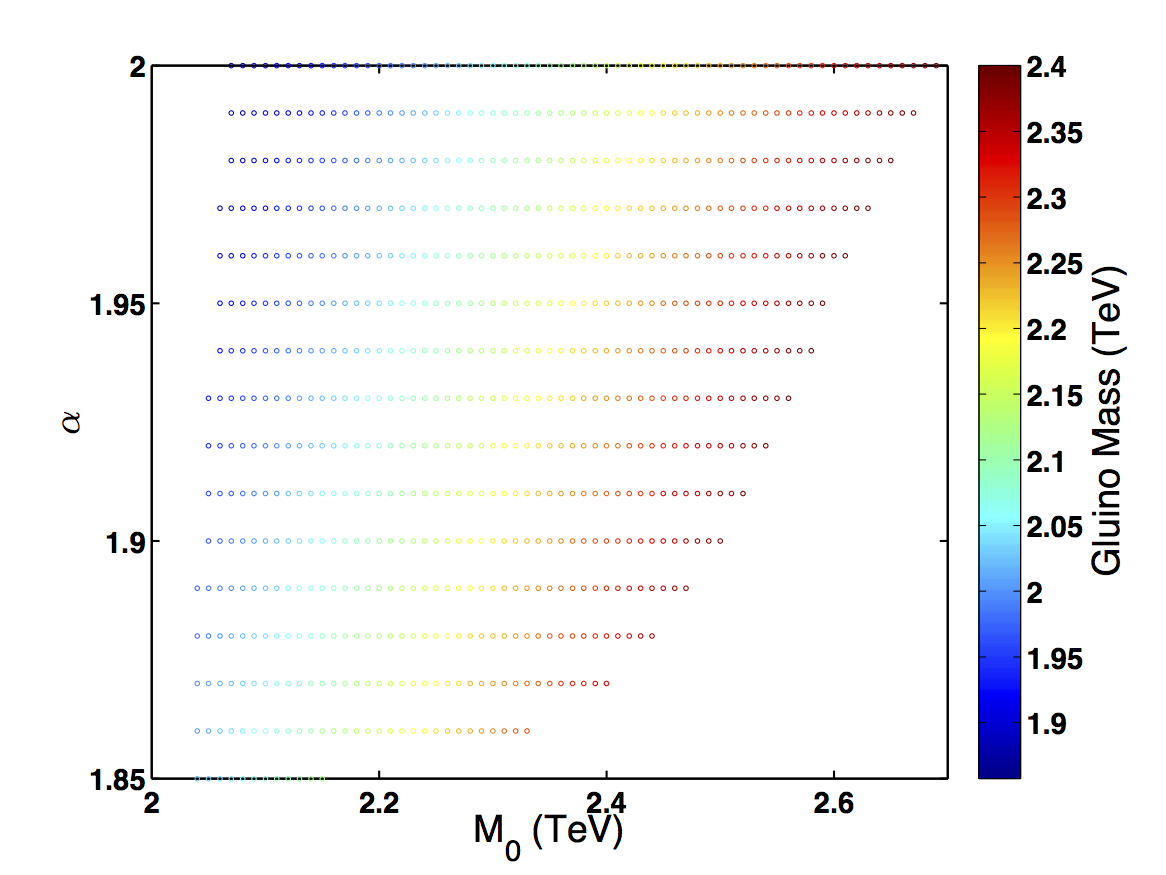}
\caption{\textbf{Allowed Parameter Space for $(n_M,n_H)$ =  $(0,\frac{1}{2})$.} Left panel is the bino-like LSP case with stau co-annihilation, summed over all values of $10\leq \tan\beta \leq 30$. The right panel is the Higgsino-like LSP case with chargino/neutralino co-annihilation and $\tan\beta=48$. The color gives the gluino mass in units of~TeV, as indicated by the scale to the right.}
\label{plot:n05}
\end{center}
\end{figure}

As with the $(n_M,n_H)$ =  $(0,0)$ case, we can identify two distinct regions in the allowed parameter space, described in the two panels of Table~\ref{coanntable}. The first region, near $\alpha \simeq 1$, consists of a bino-like LSP and covers a wide range of moderate values in the parameter $\tan\beta$. The relic density condition is satisfied in this region through stau-neutralino co-annihilation. The second region exists at $\alpha \gappeq 1.85$ and larger $\tan\beta$. Here the LSP is predominantly Higgsino-like with a degenerate sector of co-annihilating charginos and neutralinos in the early universe. These two regions are shown in the $\lbrace \alpha,M_0 \rbrace$ plane in Figure~\ref{plot:n05}. In both plots the resulting gluino mass is shown, in units of~GeV, by the color scale to the right of the plot.

The left panel represents the bino-like LSP region. Here the allowed parameter space is tightly bound in all directions by the requirement that the relic density be consistent with PLANCK observations, while avoiding the case in which the $\tilde\tau$ is light enough to become the LSP.  This requires a correlation between the value of $\alpha$ and that of $\tan\beta$, and thus the left panel in Figure~\ref{plot:n05} represents the projection onto the $\lbrace \alpha,M_0 \rbrace$ for all values of $10\leq \tan\beta \leq 30$. Generally speaking, the larger $\tan\beta$ values correspond to the smaller values of $\alpha$ in the figure. The right-most edge of the plot corresponds to the constraint $\Omega_{\chi}h^2\leq0.128$. The precise value of $M_0$ where this inequality is saturated depends somewhat on $\tan\beta$ -- hence the apparently uneven boundary for large $M_0$ values in the left panel of Figure~\ref{plot:n05}. The left edge (low $M_0$ values for a fixed value of $\alpha$) is the result of the imposition $m_h \geq 124.1 \, {\rm GeV}$. Not surprisingly, this constraint forces a larger lower-bound on $M_0$ for lower values of $\tan\beta$ and larger values of $\alpha$. The cutoff for $\alpha \simeq 1.3$ is the location in which the stau becomes the LSP. Throughout the entire region the LSP is bino-like with a mass near 1~TeV, and the mass difference $\Delta m = m_{\tilde{\tau}} - m_{\chi_1^0}$ nowhere exceeds 25~GeV.

The right panel represents the Higgsino-like region, where we have performed a two-dimensional scan fixing $\tan\beta = 48$. Here the parameter space is slightly larger than in the $(n_M,n_H)$ =  $(0,0)$ case, though still restricted to very large values of the parameter $\alpha$. The high value of $\tan\beta$ and relatively low value of $M_0$ in this case imply that the rate for $B_s \to \mu^+ \mu^-$ is generally large in this region of parameter space. In fact, throughout the region depicted in the right panel of Figure~\ref{plot:n05}, the branching fraction for this process is always above the central value reported by~LHCb by at least one standard deviation. The lower limit on $M_0$ for a fixed $\alpha$ value is set by the three-sigma upper bound on this process. The limit $\alpha\geq 1.85$, as well as the upper limit on $M_0$ for a fixed value of $\alpha$, is set by the upper limit on the thermal relic density of the neutralino.

\begin{figure}[t]
\begin{center}
\includegraphics[width=0.45\textwidth]{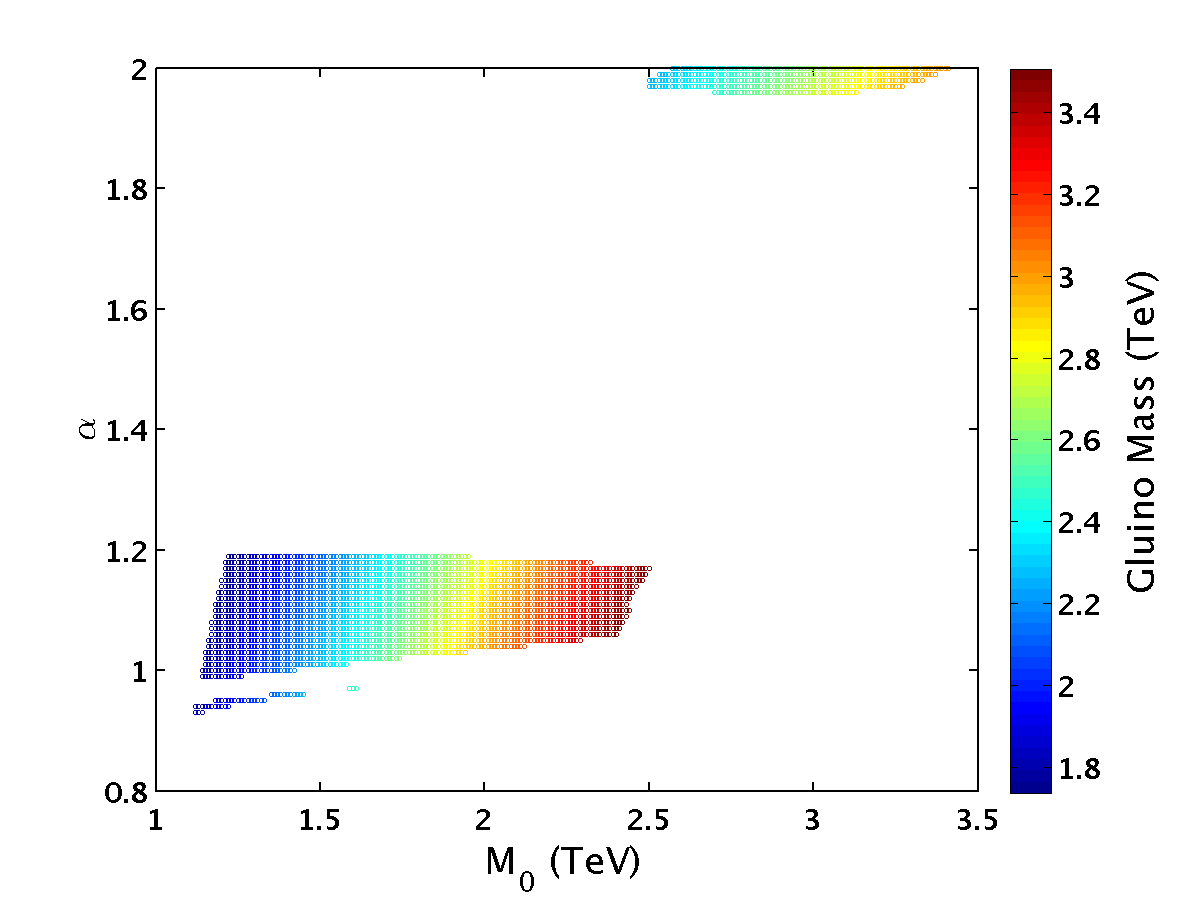}
\includegraphics[width=0.45\textwidth]{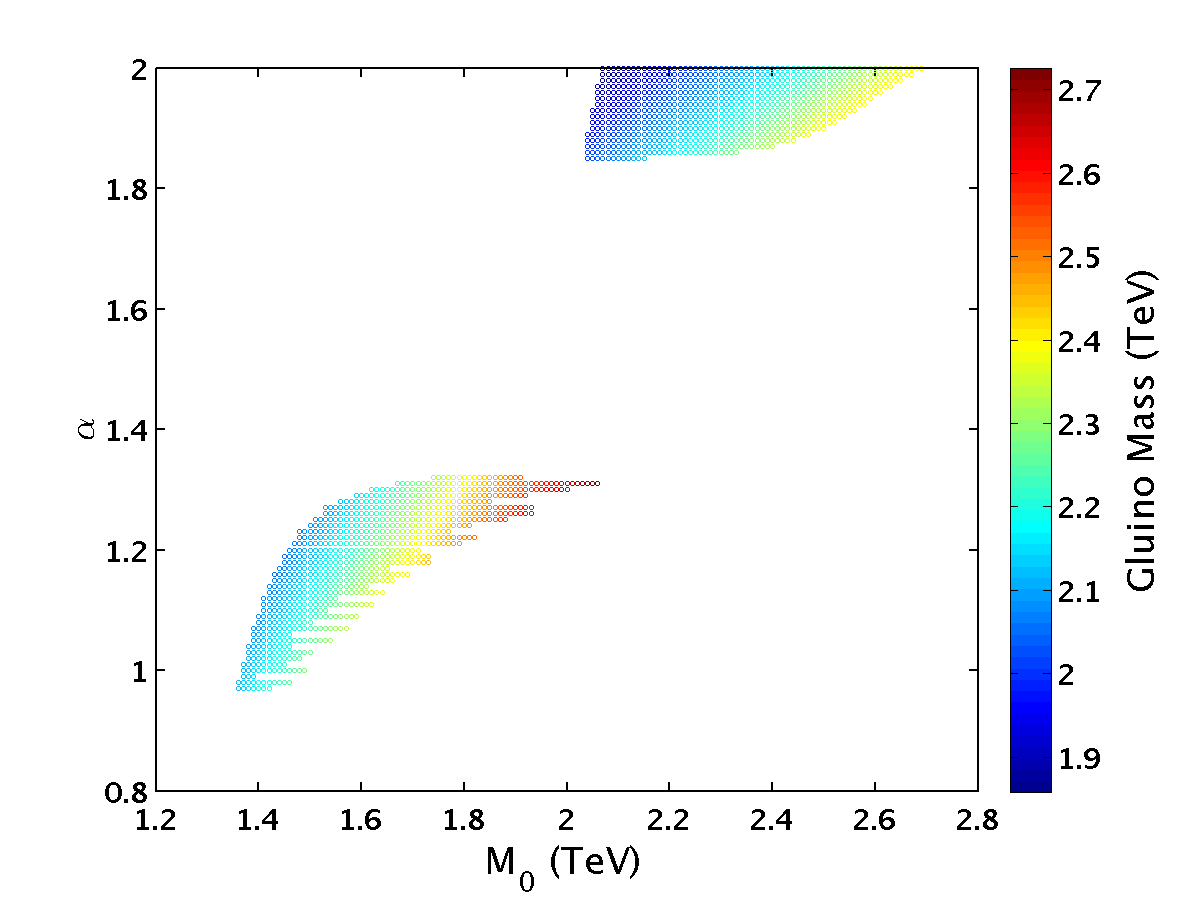}
\caption{\textbf{Total Allowed Parameter Space for $(n_M,n_H)$ =  $(0,0)$ (Left) and $(0,\frac{1}{2})$ (Right).} In both cases the parameter space is aggregated over all values of $\tan\beta$ in the targeted scans. The allowed region near $\alpha =1$ in both cases has a bino-like LSP, while the regions near $\alpha=2$ on both cases has a Higgsino-like LSP. The color indicates the gluino mass, in units of~TeV, as given by the color scale to the right.}
\label{plot:n00n05}
\end{center}
\end{figure}

In summary, the overall phenomenology of the two cases, $(n_M,n_H)$ =  $(0,0)$  and $(n_M,n_H)$ =  $(0,\frac{1}{2})$, is strikingly similar. In Figure~\ref{plot:n00n05} we have superimposed the bino-like and Higgsino-like spaces for each modular weight combination into a single plane. The general location of the allowed parameter space in the $\lbrace \alpha,M_0 \rbrace$ plane is nearly identical. As a consequence, the predicted masses for the LSP neutralino are the same: $1200\,{\rm GeV} \lappeq m_{\chi_1^0} \lappeq 1600\,{\rm GeV}$ for the Higgsino-like region at large $\alpha$, and $m_{\chi_1^0} \simeq 1000\,{\rm GeV}$ for the bino-like region near the KKLT~prediction of $\alpha = 1$. 

\subsubsection{The case $\(n_M,\, n_H\) = \(\frac{1}{2},\,  0\)$}

\begin{figure}[t]
\begin{center}
\includegraphics[width=0.45\textwidth]{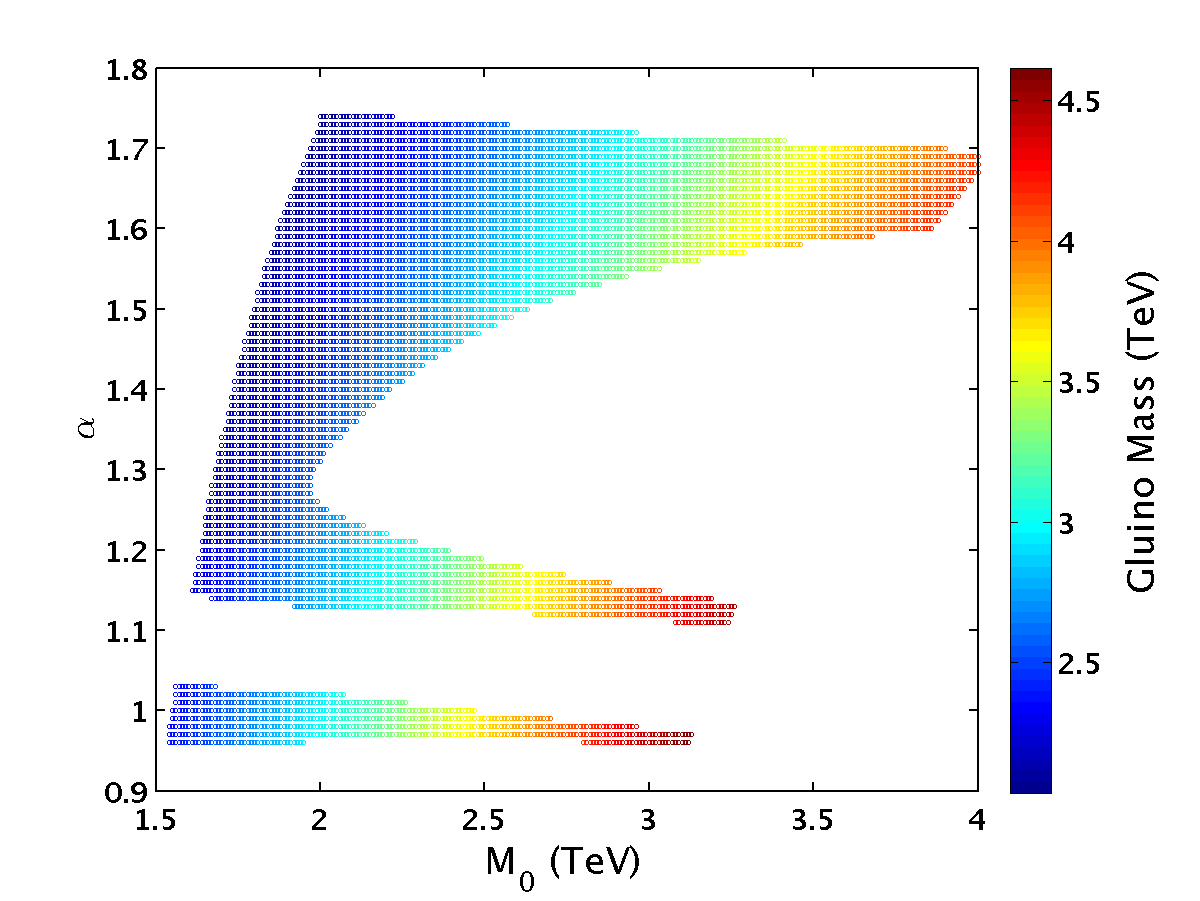}
\includegraphics[width=0.45\textwidth]{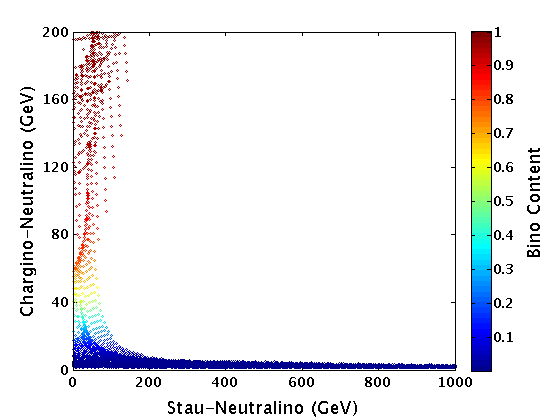}
\caption{\textbf{Allowed Parameter Space (Left) and LSP Phenomenology (Right) for $(n_M,n_H)$ =  $(\frac{1}{2},0)$ and $\tan\beta=16$.} The left panel gives allowed parameter combinations of $\lbrace \alpha,M_0 \rbrace$ consistent with proper EWSB, Higgs mass measurements and upper bound on thermal relic abundance. The gluino mass, in units of~TeV, is given by the color as indicated by the scale to the right. The right panel plots the mass difference between the lightest chargino and lightest neutralino versus the mass difference between the lightest stau and the lightest neutralino. The color in this panel indicates the fraction of the LSP wavefunction that is bino-like.}
\label{plot:n50}
\end{center}
\end{figure}

The reciprocal case $(n_M,n_H)$ =  $(\frac{1}{2},0)$ is distinctly different from the previous two in that the wavefunction of the LSP varies across the allowed parameter space, from bino-like to Higgsino-like. It was for this reason that the allowed parameter space for this combination of modular weights was listed in both panels in Table~\ref{coanntable}. 
For our targeted scan we have chosen to fix $\tan\beta=16$ and scan over the region $0.9\leq \alpha \leq 1.8$ and $1500\,{\rm GeV} \leq M_0 \leq 4000\,{\rm GeV}$.
The allowed parameter space, after imposing the Higgs mass and dark matter constraints, is given in the left panel of Figure~\ref{plot:n50}. The gap at $\alpha \simeq 1.1$ is a region where the stau is the LSP. In this region of $\alpha$ the eigenvectors of the neutralino mass matrix undergo a level-crossing: at $\alpha = 1.0$ we have $M_2 > \mu > M_1$ at the electroweak scale, while at $\alpha = 1.2$ we have $M_2 > M_1 > \mu$. During the transition the stau briefly becomes the LSP, before once again becoming the next-to-lightest neutralino. On the edges of this region stau co-annihilation is important, but elsewhere the mass gap between the stau and the lightest neutralino increases rapidly. 
As with Figure~\ref{plot:n00}, the left-most edge and right-most edges are the Higgs mass contours of $m_h = 124.1\,{\rm GeV}$ and $m_h = 127.2\,{\rm GeV}$, respectively. The curved exclusion region from $1.2 \lappeq \alpha \lappeq 1.6$ is eliminated by an over-abundance of dark matter, as are values of $\alpha \lappeq 0.9$. The upper bound on $\alpha$ represents the point at which electroweak symmetry breaking fails to occur. As with previous figures, the gluino mass is indicated by the color key to the right of the plot. 

This particular combination of modular weights is unique in that the wavefunction of the LSP interpolates between fully bino-like and fully Higgsino-like throughout the allowed parameter space. Below the gap at $\alpha \simeq 1.1$, the LSP is~93\% to~98\% bino-like, with the remainder of the wavefunction being Higgsino-like. After the level-crossing occurs, however, the wavefunction for $\alpha \simeq 1.15$ becomes `well-tempered'~\cite{ArkaniHamed:2006mb,Baer:2006te} with a composition roughly 94\% Higgsino, 4\% bino and 2\% wino. The Higgsino content then steadily increases as $\alpha$ increases, until at $\alpha \simeq 1.4$ it becomes more than~99\% Higgsino-like. This progression is illustrated in the right panel of Figure~\ref{plot:n50}, in which we show the distribution (in units of~GeV) for the two key mass differences for co-annihilation. The mass difference between the lightest stau and lightest neutralino is plotted on the horizontal axis, while that between the lightest chargino and the lightest neutralino is plotted on the vertical axis. The imposition of the relic density constraint $\Omega_{\chi}h^2\leq0.128$ forces a tight correlation between the spectrum and the wavefunction of the LSP, as indicated by the bino fraction given by the color in the figure. Clearly, one or both of the co-annihilation mechanisms (stau and neutralino/chargino) is operative throughout the parameter space.

\subsubsection{The case $\(n_M,\, n_H\) = \(\frac{1}{2},\,  \frac{1}{2}\)$ }

\begin{figure}[t]
\begin{center}
\includegraphics[width=0.45\textwidth]{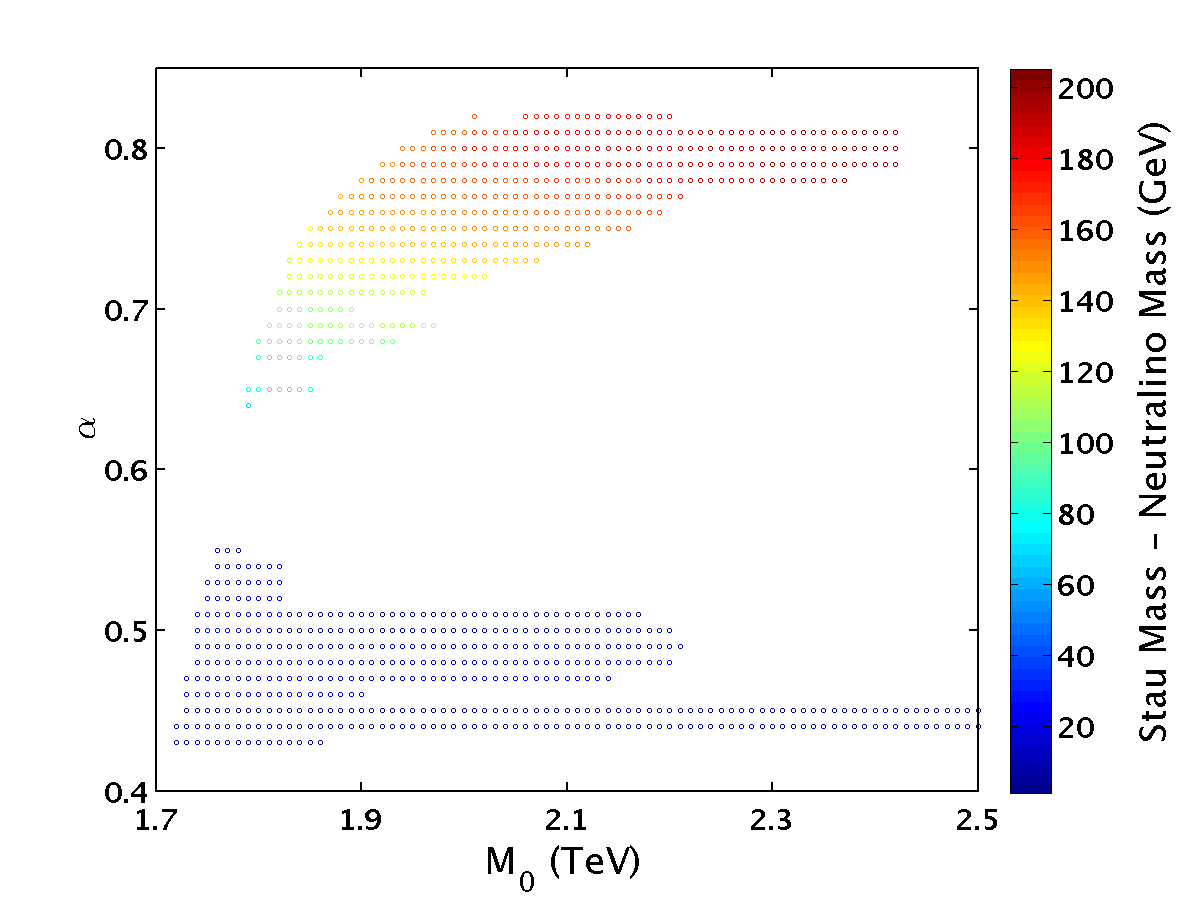}
\includegraphics[width=0.45\textwidth]{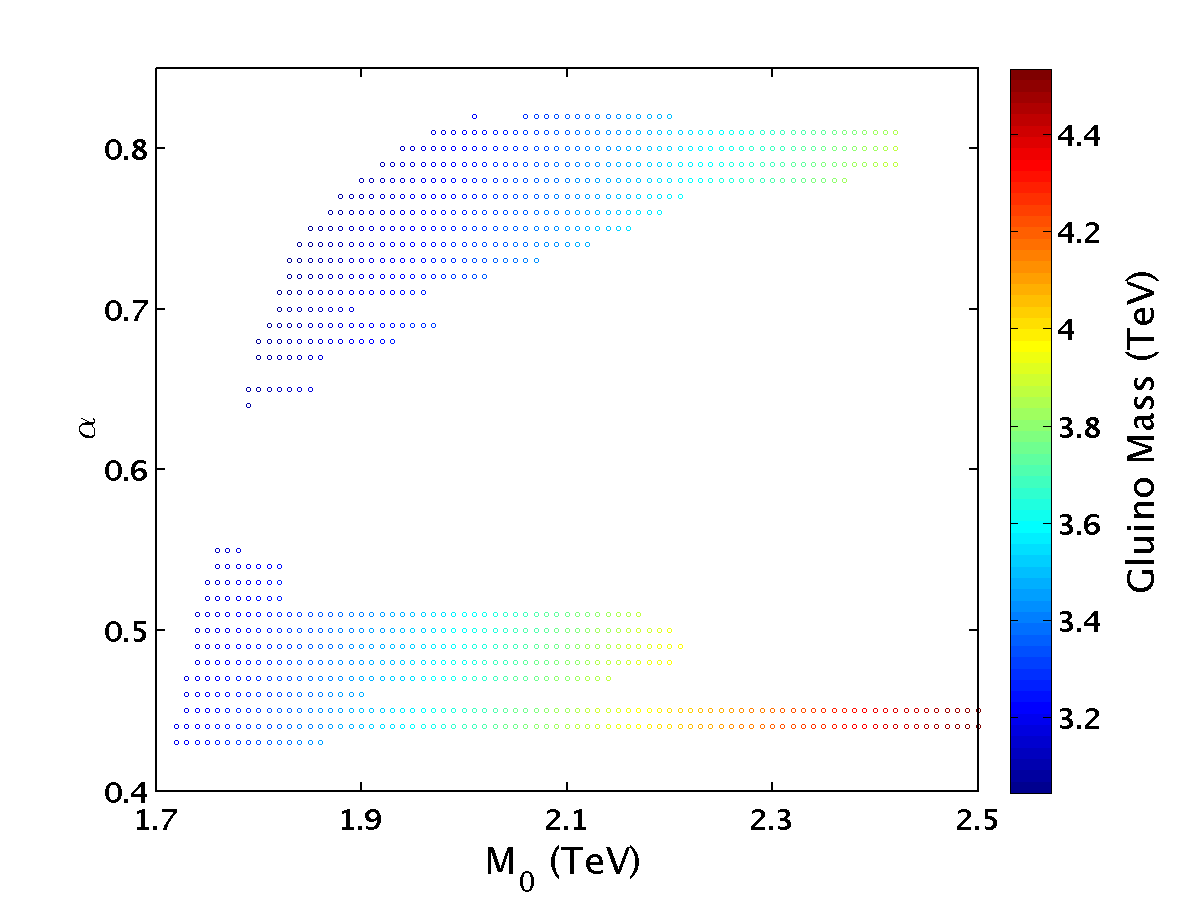}
\caption{\textbf{Allowed Parameter Space for $(n_M,n_H)$ =  $(\frac{1}{2},\frac{1}{2})$ for the Bino-like LSP Case.} Left panel gives the mass degeneracy between the lightest stau and the lightest neutralino, in units of~GeV, as indicated by the color scale to the right of the plot. The right panel gives the gluino mass, in units of~TeV, as indicated by the color scale to the right of the plot.}
\label{plot:n55bino}
\end{center}
\end{figure}

We next consider the case where $n_M = n_H = 1/2$. We expect some similarities with the case where $n_M = n_H =0$, given the universal treatment of scalar masses. Indeed, we find precisely two, well defined, and distinct regions separated clearly in the parameter $\alpha$, as indicated in Table~\ref{coanntable}. As before, the two regions are separated by points in parameter space where either the stau or the stop is the lightest supersymmetric particle, and are thus eliminated. As with the previous modular weight combinations, the bino-like LSP region exists over a range of $\tan\beta$ values which are correlated with the allowed values of the parameter $\alpha$. We thus performed a targeted scan over the range $11\leq \tan\beta \leq 35$, $0.4 \leq \alpha \leq 0.9$, and $1700\, {\rm GeV} \leq M_0 \leq 2500\,{\rm GeV}$. The allowed parameter space, after imposing the Higgs mass and dark matter constraints, is given as a projection onto the $\lbrace \alpha,M_0 \rbrace$ plane in the two panels of Figure~\ref{plot:n55bino}.

As with the $(n_M,n_H)$ =  $(0,\frac{1}{2})$ case, we have an inverse relationship between the value of $\alpha$ and the value of $\tan\beta$ necessary to achieve sufficient mass degeneracy between the lightest neutralino and the lightest stau. This mass gap is indicated by the color scale to the right of the plot in the left panel of Figure~\ref{plot:n55bino}. Though the LSP is over 99\% bino-like throughout this region, the area of significant stau co-annihilation is confined to $\alpha \leq 0.6$. This lower disconnected region exists only for $33 \leq \tan\beta \leq 35$, while the upper region spans $12 \leq \tan\beta \leq 28$. For the points in the gap there is no value of $\tan\beta$ for which the stau is not the LSP. The remainder of the features are similar to that of Figure~\ref{plot:n50}. The left-most edge is the contour where $m_h = 124.1\,{\rm GeV}$. In fact, the Higgs mass never exceeds $125.7\,{\rm GeV}$ throughout the allowed parameter space. The right most edge is the boundary where $\Omega_{\chi}h^2 = 0.128$. As this quantity depends on $\tan\beta$ indirectly via the stau mass, the right edge varies considerably when we project all values of $\tan\beta$ onto the  $\lbrace \alpha,M_0 \rbrace$ plane. Overabundance of thermal relic neutralinos also eliminates all $\tan\beta$ and $M_0$ values for $\alpha \leq 0.42$.
The right panel of Figure~\ref{plot:n55bino} gives the gluino mass value in units of~GeV, according to the color scale to the right of the plot.

\begin{figure}[t]
\begin{center}
\includegraphics[width=0.45\textwidth]{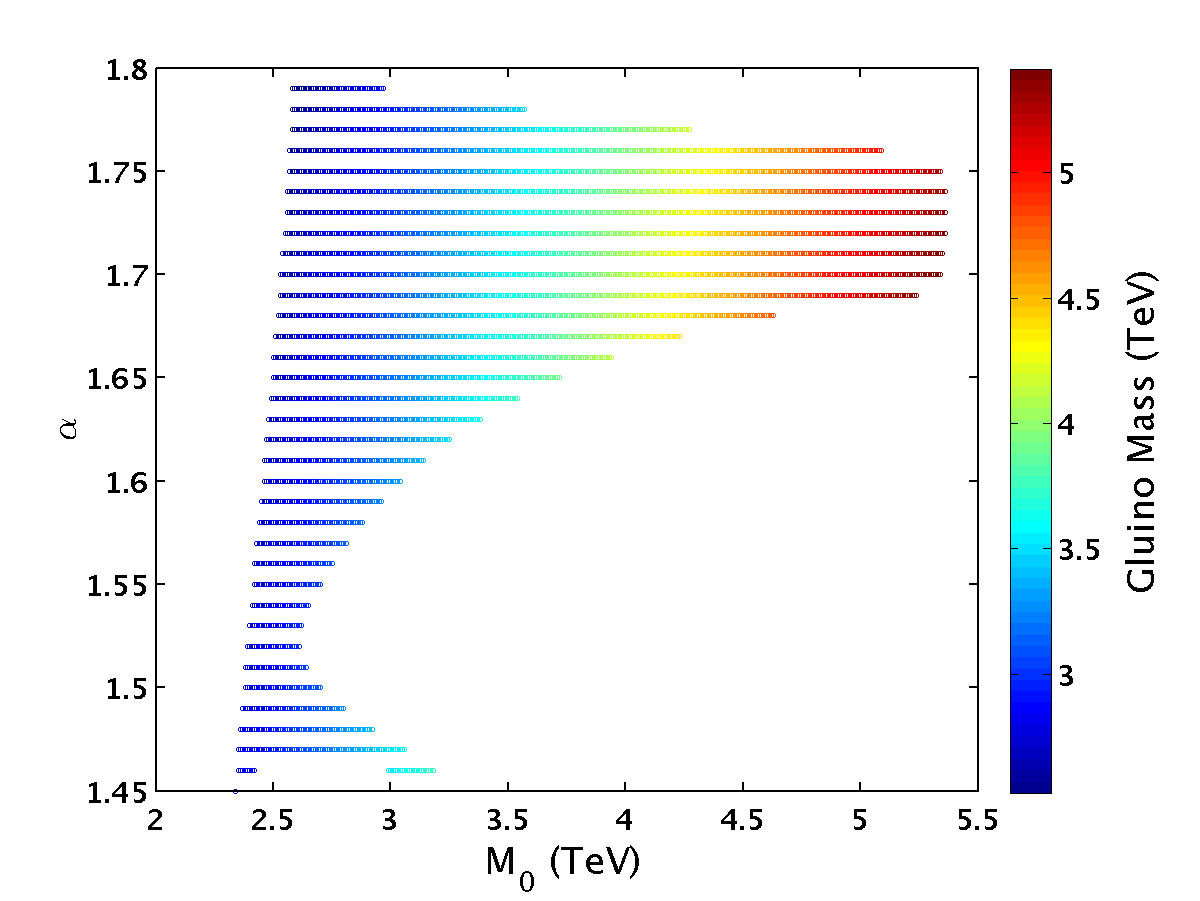}
\includegraphics[width=0.45\textwidth]{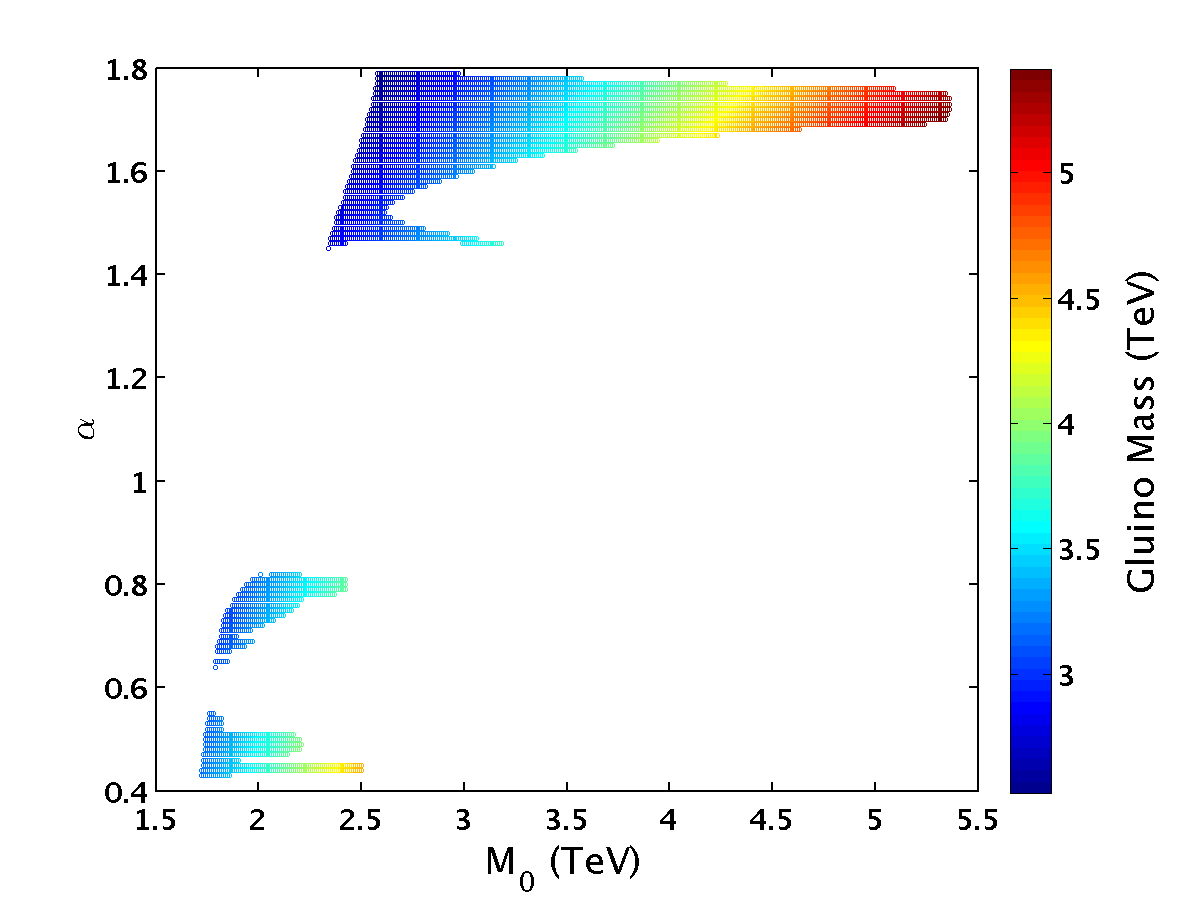}
\caption{\textbf{Allowed Parameter Space for $(n_M,n_H)$ =  $(\frac{1}{2},\frac{1}{2})$ for the Higgsino-like LSP Case (Left) and Combined Cases (Right).} Both panels give the allowed parameter combinations of $\lbrace \alpha,M_0 \rbrace$ consistent with proper EWSB, Higgs mass measurements and upper bound on thermal relic abundance. The gluino mass, in units of~TeV, is given by the color as indicated by the scale to the right of each panel. The left panel is solely the part of the parameter space with a Higgsino-like LSP, while the right panel combines this space with the region shown in Figure~\ref{plot:n55bino}.}
\label{plot:n55hino}
\end{center}
\end{figure}

For the gaugino co-annihilation region we adopt $\tan\beta=30$ and perform a targeted scan with $1.2 \leq \alpha \leq 1.9$ and $1000\,{\rm GeV} \leq M_0 \leq 6000\,{\rm GeV}$. The allowed parameter space, after imposing the Higgs mass and dark matter constraints, is given in  the left panel of Figure~\ref{plot:n55hino}. The gluino mass is indicated by the color, in units of~TeV. For $\alpha$ values less than about $\alpha = 1.4$, the stau is generally the LSP for $\tan\beta = 30$, while for $\alpha \gappeq 1.8$ there fails to be an adequate solution to the EWSB conditions.  For these large values of $\alpha$ the value of $\mu$ rapidly approaches zero and the LSP mass tracks this value. The dark matter constraint favors larger $\alpha$ values and smaller values of $M_0$, which then tends to conflict with the lower bound on the Higgs mass. The left edge of the parameter space represents the locus of points where $m_h =124.1\,{\rm GeV}$, while the cut-off at $M_0 \sim 5000\,{\rm GeV}$ arises from the upper bound we impose on the Higgs mass of $m_h \leq 127.2\,{\rm GeV}$. The curved edge is the upper bound on the neutralino relic density. Overall the characteristics of the parameter space are similar to those of the $(n_M,n_H)$ =  $(\frac{1}{2},0)$ in Figure~\ref{plot:n50}. The right panel of Figure~\ref{plot:n55hino} combines the bino-like and Higgsino-like regions into a single plot, to give a sense of proportion to the allowed parameter space.

\subsubsection{The cases $\(n_M,\, n_H\) = \(0,\, 1\)$  and $\(n_M,\, n_H\) = \(\frac{1}{2},\, 1\)$}

\begin{figure}[t]
\begin{center}
\includegraphics[width=0.45\textwidth]{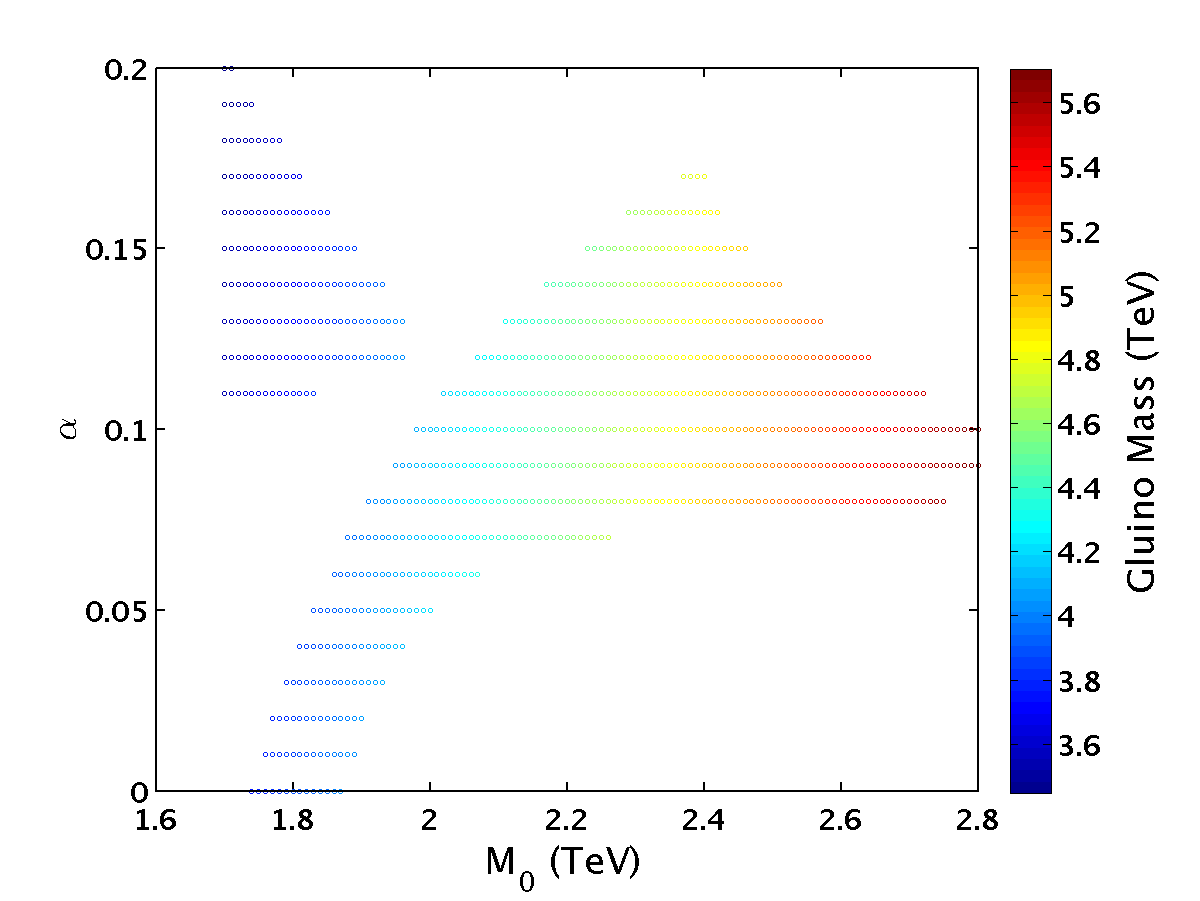}
\includegraphics[width=0.45\textwidth]{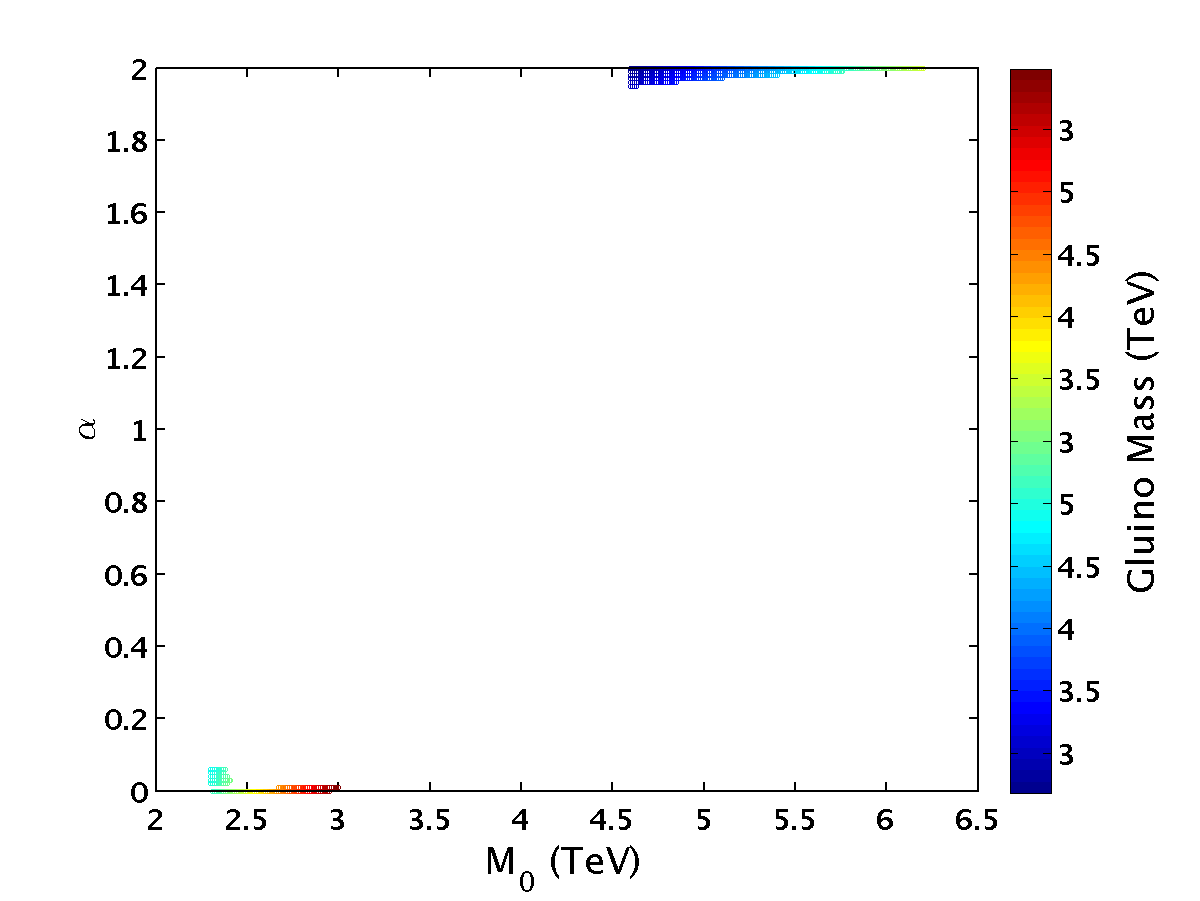}
\caption{\textbf{Allowed Parameter Space for $(n_M,n_H)$ =  $(0,1)$ (Left) and for $(n_M,n_H)$ =  $(\frac{1}{2},1)$ (Right).}  In the left panel the two regions correspond to two different choices of the parameter $\tan\beta$. The region in the upper left has $\tan\beta = 51$, while the lower, disconnected, region has $\tan\beta = 52$. In the right panel, the region at $\alpha \simeq 0$ is the bino-like case, while the region at $\alpha \simeq 2$ is the Higgsino-like case. As before, the gluino mass, in units of~TeV, is indicated by the color scale to the right of each plot.}
\label{plot:nx1}
\end{center}
\end{figure}

The remaining two subsets we consider arise when one or both of the MSSM states -- the matter fields or the Higgs fields -- are confined to stacks of $D3$~branes where the effective modular weight is unity. We begin with the case $(n_M,n_H)$ =  $(0,1)$, where Table~\ref{coanntable} indicates the existence of a small region of allowed parameter space with $\alpha \simeq 0$ and very large values of $\tan\beta$. In this region the lightest stau and lightest neutralino are highly degenerate and the LSP is purely bino-like. Because this region is so tightly confined we have chosen to perform two small targeted scans, one each at $\tan\beta=51$ and $\tan\beta=52$, over the range $0 \leq \alpha \leq 0.2$ and $1700\,{\rm GeV} \leq M_0 \leq 2800\,{\rm GeV}$. The result of the scan is shown in the left panel of Figure~\ref{plot:nx1}.

The smaller region at lower $M_0$ and $\alpha\geq 0.1$ corresponds to $\tan\beta = 51$. Here the Higgs mass satisfies  $124.1\,{\rm GeV} \leq M_0 \leq 124.8\,{\rm GeV}$. For the other region, with $\tan\beta = 52$ the Higgs mass is between~124.1 and~126.1~GeV. As with previous plots, the right-most boundaries on $M_0$ for both $\tan\beta$ values arise from the relic density constraint $\Omega_{\chi}h^2 \leq 0.128$. The left most constraint for $\tan\beta=51$ represents the contour where $m_h = 124.1\,{\rm GeV}$, while for $\tan\beta=52$ it is the contour where the stau becomes the LSP. The upper bound on $\alpha$ arises where the contour of $\Omega_{\chi}h^2 = 0.128$ intercepts the Higgs mass constraint ($\tan\beta=51$) or the stau LSP contour ($\tan\beta = 52$).

The modular weight combination $(n_M,n_H)$ =  $(\frac{1}{2},1)$ is even more tightly constrained. Though nominally there are two allowed regions in the $\lbrace \alpha,M_0 \rbrace$ plane, they are both concentrated at extreme values of the parameter $\alpha$, as anticipated in Table~\ref{coanntable}. The bino-like and Higgsino-like regions are shown simultaneously in the right panel of Figure~\ref{plot:nx1}. Just as in the case $(n_M,n_H)$ =  $(0,1)$, the bino-like case with stau co-annihilation is concentrated at $0 \leq \alpha \leq 0.01$ for $\tan\beta=54$ and $0.02 \leq \alpha \leq 0.06$ for $\tan\beta = 53$. This is therefore essentially the case of minimal supergravity, with a large hierarchy between gauginos and scalars, and a mass scale $2500\,{\rm GeV} \lappeq M_0 \lappeq 3000\,{\rm GeV}$. The Higgs mass in this region is confined to $124.1\,{\rm GeV} \leq m_h \leq 125.1\,{\rm GeV}$. The Higgsino-like region exists only for $\alpha \geq 1.95$ and $4600\,{\rm GeV} \leq M_0 \leq 6200\,{\rm GeV}$ for our choice of $\tan\beta=41$. The Higgs mass ranges over the same values as the bino-like case, and the gluino mass is roughly 4-5~TeV in both regions of the parameter space.

\subsubsection{The cases $\(n_M,\, n_H\) = \(1,\, 0\)$, $\(1,\,\frac{1}{2}\)$ and $\(n_M,\, n_H\) = \(1,\, 1\)$}

\begin{figure}[t]
\begin{center}
\includegraphics[width=0.5\textwidth]{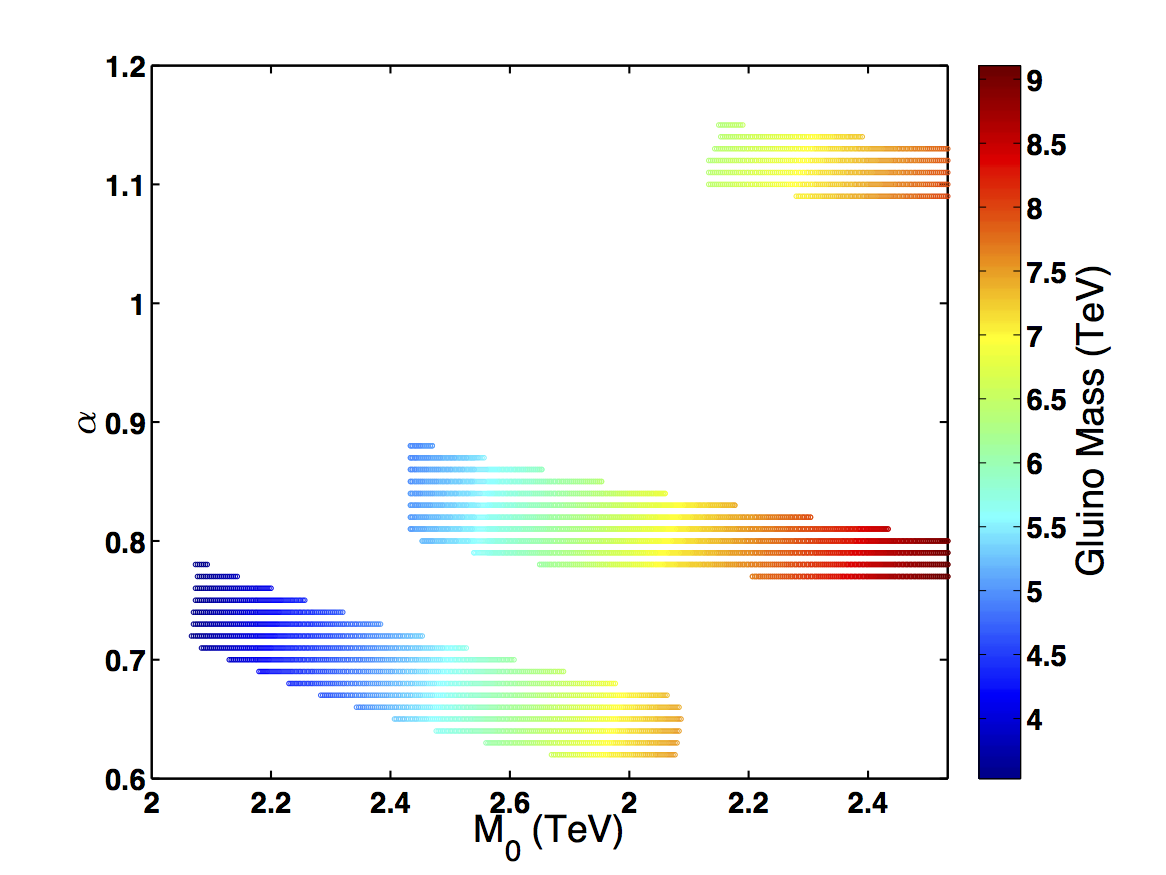}
\caption{\textbf{Allowed Parameter Space for the Cases  $(n_M,n_H)$ =  $(1,0)$, $(1,\frac{1}{2})$ and $(1,1)$.} All three of these cases involve a Higgsino-like LSP. The lowest region corresponds to $(n_M,n_H)$ =  $(1,0)$, while the region at the highest values of $\alpha$ corresponds to $(n_M,n_H)$ =  $(1,1)$. As with the other figures, the gluino mass, in units of TeV, is indicated by the color scale to the right of each plot.}
\label{plot:n1x}
\end{center}
\end{figure}

The last three cases with unit modular weight for the matter sector share many of the same overall features, allowing us to group their treatment into a single, brief discussion. All three cases give rise to a single region of parameter space with a Higgsino-like LSP and degenerate neutralinos and charginos. A wide range of $\tan\beta$ values are available, but the parameter $\alpha$ is tightly constrained in all three modular weight combinations. For our targeted scan we have chosen $\tan\beta = 15$ for $(n_M,n_H)$ =  $(1,0)$ and  $(n_M,n_H)$ =  $(1,\frac{1}{2})$, and $\tan\beta = 24$ for the  $(n_M,n_H)$ =  $(1,1)$ case. The results of the targeted scans are given in a single plot in Figure~\ref{plot:n1x}, with the gluino mass again given in~TeV by the color scale to the right. The allowed regions are defined by $0.62 \leq \alpha \leq 0.78$, $0.77 \leq \alpha \leq 0.88$ and $1.09 \leq \alpha \leq 1.15$ for $n_H = 0, \frac{1}{2}$ and~1, respectively.

In all three cases the boundary for low $M_0$ values continues to be the Higgs mass constraint $m_h \geq 124.1\,{\rm GeV}$. The maximum value for $M_0$ in the low-$\alpha$ case of  $(n_M,n_H)$ =  $(1,0)$ is given by the upper bound on the Higgs mass we have chosen of $m_h \leq 127.2\,{\rm GeV}$. For the other two cases it is given by the value for $M_0$ at which $\mu^2 \rightarrow 0$ and EWSB fails to occur. Failure to achieve proper EWSB is also the origin of the upper bound on the parameter $\alpha$ for fixed $M_0$ values in all three cases. Finally, the lower bound on $\alpha$ for a fixed value of $M_0$ is always dictated by the constraint $\Omega_{\chi}h^2 \leq 0.128$.

Considering these three cases together, the shifting of the parameter space to higher values of $M_0$ is relatively easy to understand. The maximum value for the squark masses, at the boundary condition scale, occurs when $n_M=0$, while the masses are greatly suppressed in the other extreme when $n_M =1$. As a consequence, the overall mass scale $M_0$ must be increased significantly as $n_M$ increases to achieve sufficiently large radiative corrections to the lightest CP-even Higgs mass to satisfy the LHC~measurements. As a result, the gluino mass gets progressively larger in each parameter space, as indicated by the color scheme in Figure~\ref{plot:n1x}. Though the minimum gluino mass is increasing from 3500~GeV for  $(n_M,n_H)$ =  $(1,0)$ to 6300~GeV for  $(n_M,n_H)$ =  $(1,1)$, the range of allowed values in $\alpha$ imply that the LSP neutralino masses should fall into similar mass ranges: roughly $100\,{\rm GeV} \leq m_{\chi_1^0} \leq 1100\,{\rm GeV}$ in all three cases. 
That such small LSP masses are possible is interesting, and ultimately reflects the fact that the $\mu$ parameter is smaller in these cases than equivalent parameter points when $n_H \neq 1$. 
%

\subsection{Summary of Targeted Scan Results}

\begin{figure}[t]
\begin{center}
\includegraphics[width=0.45\textwidth]{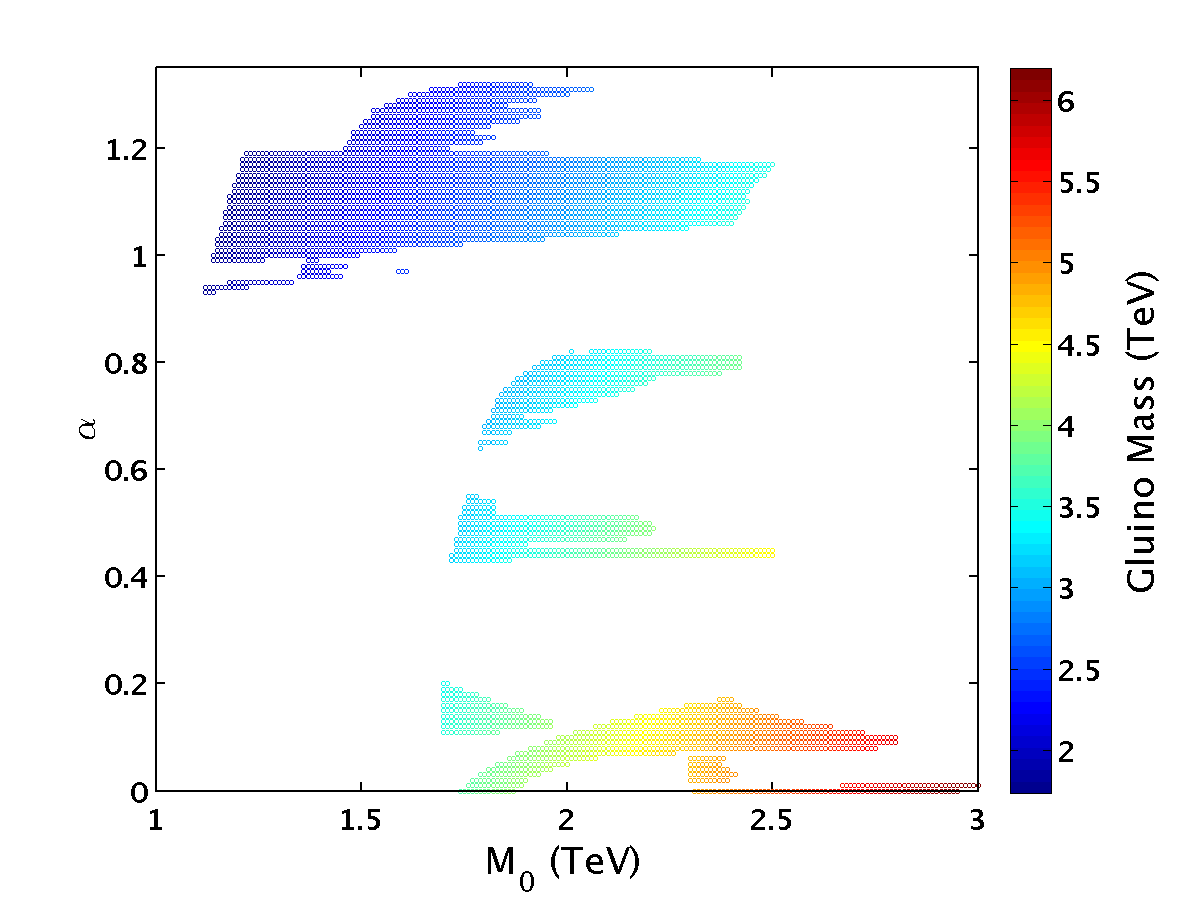}
\includegraphics[width=0.45\textwidth]{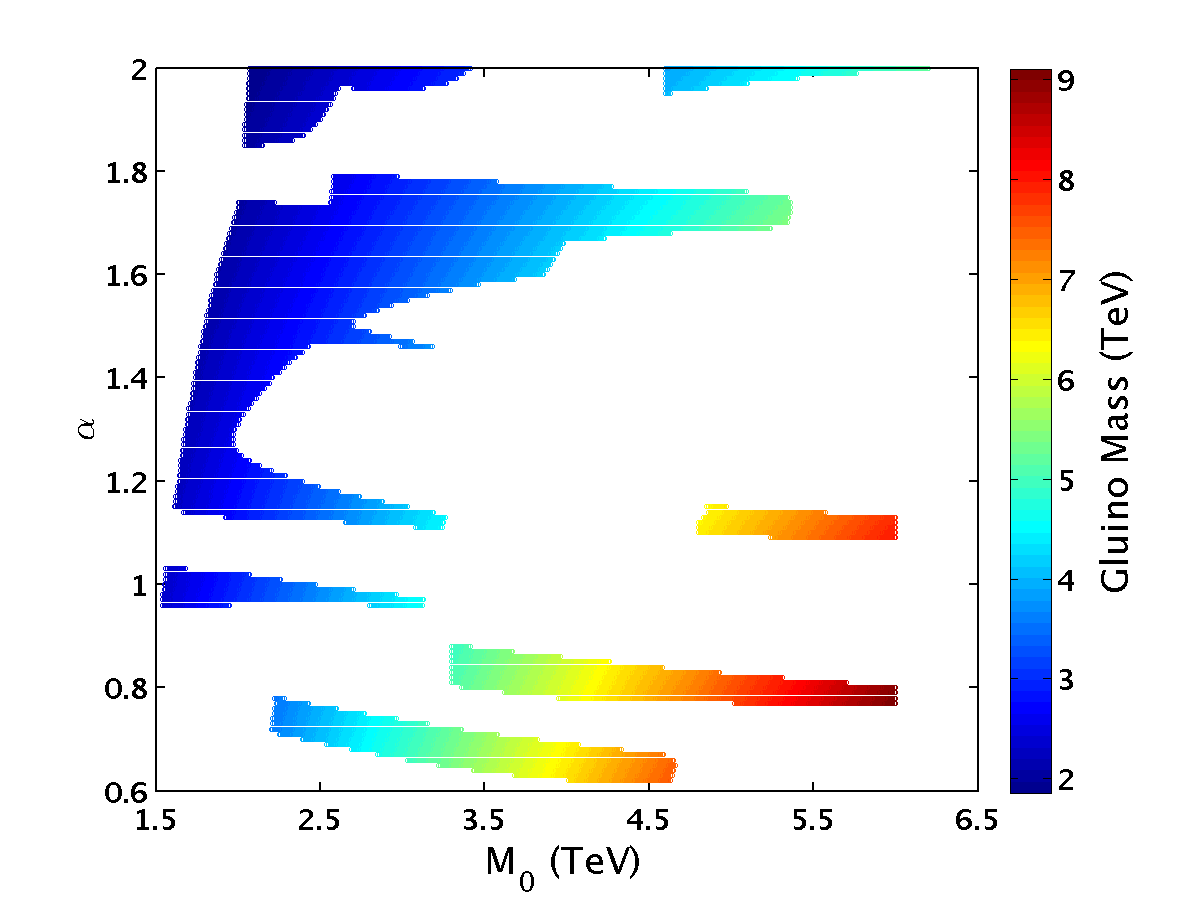}
\caption{\textbf{Allowed Parameter Space for All Modular Weight Combinations in the Flux Compactified Type~IIB Model.} Left panel aggregates all the cases with a bino-like LSP. Right panel aggregates all cases where the LSP is Higgsino-like. All points shown are reproduced form early plots in this section. The gluino mass, in units of~TeV, is indicated for each point by the color scheme to the right of the plot.}
\label{plot:nxx}
\end{center}
\end{figure}

Having completed our survey of the nine individual combinations of modular weights, it is useful to consider some of the broad features that emerge in the overall parameter space. We have identified thirteen distinct regions in the space of parameters $\lbrace \alpha, M_0 , \tan\beta \rbrace$ over the nine combinations. We then conducted targeted fine-resolution scans for each of these thirteen regions, the projection of which onto the  $\lbrace \alpha,M_0 \rbrace$ plane is shown in Figure~\ref{plot:nxx}. The left panel is the sum of all cases in which the LSP neutralino is predominantly bino-like. In these cases each region represents a range of $\tan\beta$ values, roughly given by the ranges in Table~\ref{coanntable}. The right panel is the sum of all cases in which the LSP neutralino is predominantly Higgsino-like. Here we have chosen specific values of $\tan\beta$ in each scan. Thus, we can expect the figure to represent a reasonable approximation to the total parameter space, but only an approximation. We expect that some points on the boundaries of these regions would be viable for different choices of $\tan\beta$ within the ranges set in Table~\ref{coanntable}.

\begin{table}[t]
\begin{center}
\begin{tabular}{|lr||c|c|c||} \hline
 &   &  $n_H = 0$ & $n_H = \frac{1}{2}$ & $n_H =1$ \\ \hline \hline
 &  & $0.93 \leq \alpha \leq 1.19$ & $0.97 \leq \alpha \leq 1.32$ & $0 \leq \alpha \leq 0.2$ \\ 
 & Bino & $1737 \leq m_{\tilde{g}} \leq 3506$ & $2057 \leq m_{\tilde{g}} \leq 2725$ & $3448 \leq m_{\tilde{g}} \leq 5706$ \\ 
 &  & $788 \leq m_{\chi_1^0} \leq 1952$ & $976 \leq m_{\chi_1^0} \leq 1684$ & $756 \leq m_{\chi_1^0} \leq 1323$ \\
$n_M=0$ &  &  &  &  \\  
 &  & $1.96 \leq \alpha \leq 2$ & $1.85 \leq \alpha \leq 2$ &  \\ 
 & Higgsino & $2256 \leq m_{\tilde{g}} \leq 3017$ & $1857 \leq m_{\tilde{g}} \leq 2400$ &  \\ 
 &  & $1201 \leq m_{\chi_1^0} \leq 1585$ & $1203 \leq m_{\chi_1^0} \leq 1605$ &  \\
 \hline
 &  &  & $0.48 \leq \alpha \leq 0.82$ & $0 \leq \alpha \leq 0.06$ \\ 
 & Bino &  & $3045 \leq m_{\tilde{g}} \leq 4535$ & $4747 \leq m_{\tilde{g}} \leq 6193$ \\ 
 &  & $0.96 \leq \alpha \leq 1.74$ & $965 \leq m_{\chi_1^0} \leq 1629$ & $1006 \leq m_{\chi_1^0} \leq 1327$ \\
$n_M=\frac{1}{2}$ &  & $2038 \leq m_{\tilde{g}} \leq 4612$ &  &  \\  
 &  & $965 \leq m_{\chi_1^0} \leq 1629$ & $1.45 \leq \alpha \leq 1.79$ & $1.95 \leq \alpha \leq 2$ \\ 
 & Higgsino &  & $2528 \leq m_{\tilde{g}} \leq 5413$ & $3873 \leq m_{\tilde{g}} \leq 5199$ \\ 
 &  &  & $107 \leq m_{\chi_1^0} \leq 1505$ & $826 \leq m_{\chi_1^0} \leq 1089$ \\
 \hline
 &  &  &  &  \\ 
 & Bino &  &  &  \\ 
 &  &  &  &  \\
$n_M=1$ &  &  &  &  \\  
 &  & $0.62 \leq \alpha \leq 0.78$ & $0.77 \leq \alpha \leq 0.88$ & $1.09 \leq \alpha \leq 1.15$ \\ 
 & Higgsino & $3529 \leq m_{\tilde{g}} \leq 7527$ & $4938 \leq m_{\tilde{g}} \leq 9107$ & $6353 \leq m_{\tilde{g}} \leq 7998$ \\ 
 &  & $106 \leq m_{\chi_1^0} \leq 1070$ & $105 \leq m_{\chi_1^0} \leq 1109$ & $122 \leq m_{\chi_1^0} \leq 1012$ \\
 \hline
\hline
\end{tabular}
\caption{\footnotesize \textbf{Summary Table for All Modular Weight Combinations in the Flux Compactified Type~IIB Model}. For each combination of modular weights we give the allowed range in the model parameter $\alpha$ consistent with all phenomenological constraints. In combinations where two such ranges exist, we separate them into bino-like LSP and Higgsino-like LSP cases. For each range in the parameter $\alpha$ we provide the resulting range in gluino mass values and LSP mass values, in units of~GeV.}
\label{targetscan}
\end{center}
\end{table}

A summary of our targeted scan results is given in Table~\ref{targetscan}. The upper left quartet of $n_M,\,n_H$ = $0,\,\frac{1}{2}$, in which all fields of the MSSM are realized through strings ending on $D7$~branes, gives the richest phenomenology and largest allowed parameter space. Bino-like and Higgsino-like dark matter candidates are possible, as is the case of a well-tempered neutralino in the specific $(n_M,n_H)$ =  $(\frac{1}{2},0)$ case. The bino-like cases, with stau co-annihilation in the early universe, are also the ones most consistent with the original KKLT hypothesis of anti-$D3$~brane uplift mechanisms, for which we expect $\alpha =1$. We find that the K\"ahler modulus dependence of the uplift mechanism, in the form of the exponent $n$ in~(\ref{Pgen}), can also take values of $n= \pm 1$ and $n=-2$ and remain consistent for certain values of the modular weights. 

With the lightest gluino in Table~\ref{targetscan} being over 1700~GeV, it is unlikely that any of these points have superpartners which should have generated detectable excess over backgrounds at the LHC in data collected thus far. We will confirm this statement in Section~\ref{LHC} below. We note, however, that models with an $\order(100\,{\rm GeV})$ Higgsino-like LSP, whose relic density is comparable to the PLANCK observation on $\Omega_{\rm CDM} h^2$, would likely have produced a detectable signal at current-generation liquid xenon dark matter detectors. We therefore begin our analysis of future discovery channels with this class of experiment.

\section{Prospects for Dark Matter Direct Detection Experiments}
\label{sec:dark}

Even with the discovery of the Higgs, and increasingly stringent measurements of the dark matter relic density, model points with bino-like and/or Higgsino-like LSPs remain from every combination of modular weights we considered. One may now ask if any of these points, though not yet excluded by direct searches for superpartners, could nevertheless be detected in the near future. To answer this question, we focus on two types of experiments: LHC collider searches and dark matter direct detection experiments. We will consider each combination of modular weights independently, or group by LSP type when appropriate.

To date, discovery prospects for heavy neutralino dark matter ($100\,{\rm GeV} \lappeq m_{\chi} \lappeq 1000\,{\rm GeV}$) have been dominated by the liquid xenon direct detection experiments: the Xenon100 Dark Matter Project in Gran Sasso, Italy~\cite{Aprile:2011dd}, and the South Dakota-based LUX experiment~\cite{Akerib:2012ys}. The former recently released data for 224.6~live days of exposure on a 34~kg target~\cite{Aprile:2012nq}. On October~30, the LUX experiment released a preliminary result from 85.3~live days of exposure on a 118~kg target~\cite{Akerib:2013tjd}. In the near future, LUX~expects to analyze 300~days of exposure within the next year, while the extension of Xenon100 to the one ton level will follow soon thereafter.
We can therefore discuss the discovery prospects for dark matter in two stages. First we determine what, if any, parameter space is already in conflict with existing results from Xenon100 and LUX. Then we ask how future enlargements of the data-taking on liquid xenon detectors will affect the remaining parameter space of the flux-compactified Type~IIB model.

\subsection{Bino-like LSPs}

\begin{figure}[t]
\begin{center}
\includegraphics[width=0.45\textwidth]{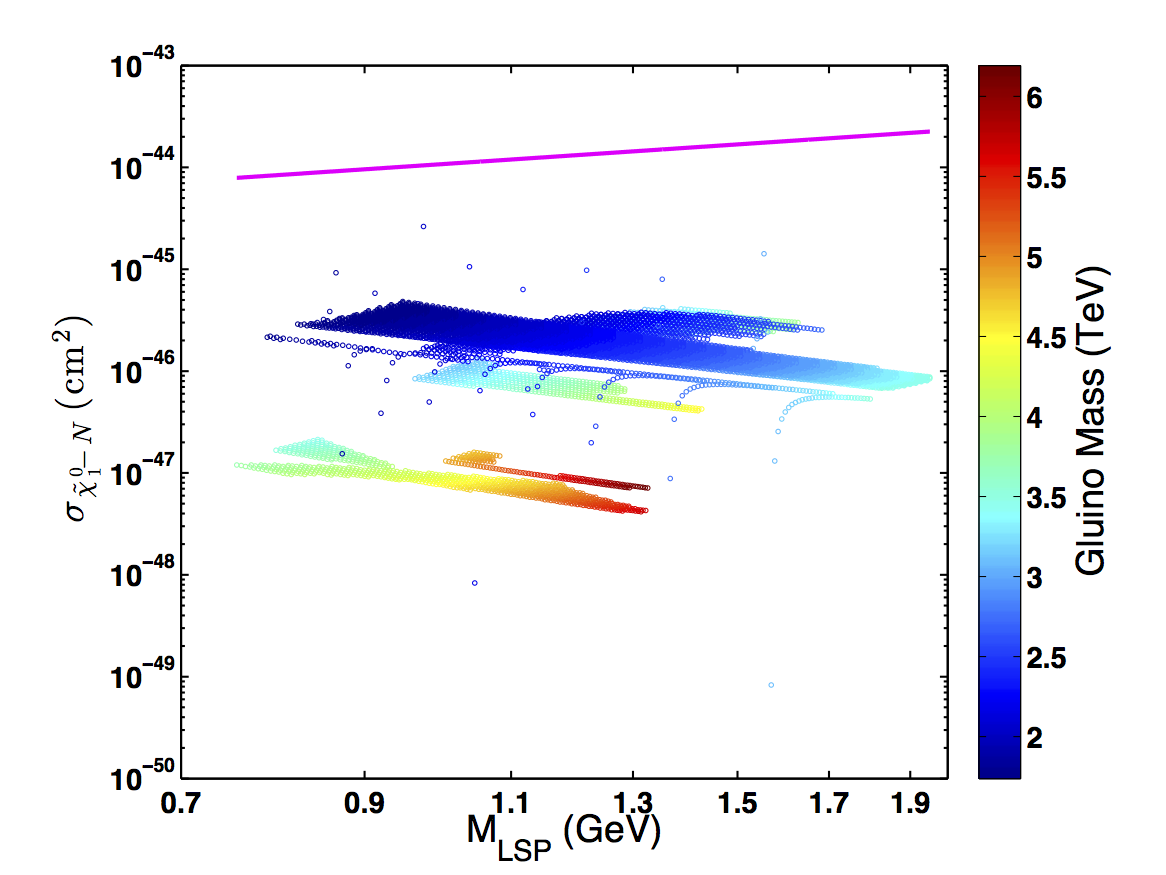}
\includegraphics[width=0.45\textwidth]{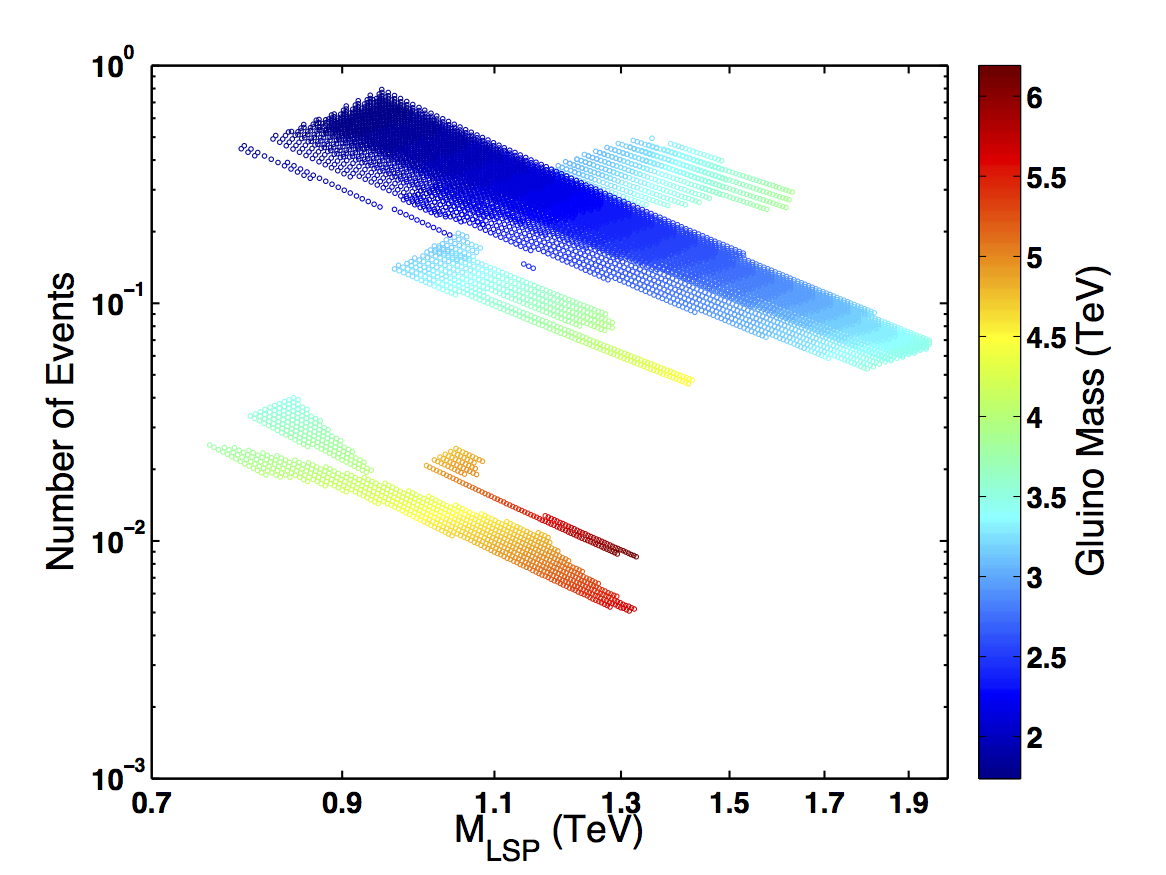}
\caption{\textbf{Dark Matter Detection Prospects for Bino-like LSP Points.} Left panel shows the distribution in neutralino-nucleon scattering cross-sections versus neutralino mass for the bino-like segment of the Type~IIB flux compactification scenario. The solid magenta line represents the limit set by the recent results from LUX. The right panel gives the rate of nuclear recoils, integrated over the recoil energy range of 5-25~keV, after one ton-year of exposure. Both panels aggregate all the cases with a bino-like LSP for all modular weight combinations.}
\label{fig:BinoXenonCompare}
\end{center}
\end{figure}

A nearly bino-like LSP is found in six of the nine modular weight combinations, summarized in Table~\ref{targetscan}. For the purposes of discussing dark matter phenomenology, it is convenient to aggregate these modular weight combinations and consider the bulk properties of all bino-like neutralino cases as one phenomenologically similar region. For this combined region, the LSP is heavy, ranging from 750-1950~GeV. The left panel in Figure~\ref{fig:BinoXenonCompare} shows the familiar neutralino-nucleon cross-section versus LSP mass for all of the targeted scan regions with bino-like LSPs. The top magenta line represents the results from the preliminary LUX data for LSPs in the appropriate mass range, corresponding to a fiducial volume of 118~kg and an exposure of 85~days. Because the fiducial volume was over triple the size of that used by Xenon100, LUX was able to surpass 224~days of Xenon100 exposure within three months.

While there are a handful of points with very large cross sections, the bulk of the bino-like parameter space in the KKLT Type~IIB flux compactification scenario is currently outside the reach of these experiments. Xenon1T and LUX expect to improve the limiting cross section on neutralino-nucleon scattering by an order of magnitude or more; for bino-like LSPs in the KKLT paradigm, that improvement will be insufficient. The right panel of Figure~\ref{fig:BinoXenonCompare} shows the number of expected events for an exposure of 300~days for 1000~kg of liquid Xenon within the recoil energy range of 5-25~keV. Even in the most favorable scenario, this much integrated exposure would yield only 0.3~events, which would still be below the estimated backgrounds for these experiments. In fact, the projected number of events in one ton-year of accumulated data on liquid xenon would be given by the right panel of Figure~\ref{fig:BinoXenonCompare}, suggesting that if these points are to have any hope of being discovered in the near future, it will have to be at the~LHC.

\subsection{Higgsino-like LSPs}

\begin{figure}[t]
\begin{center}
\includegraphics[width=0.45\textwidth]{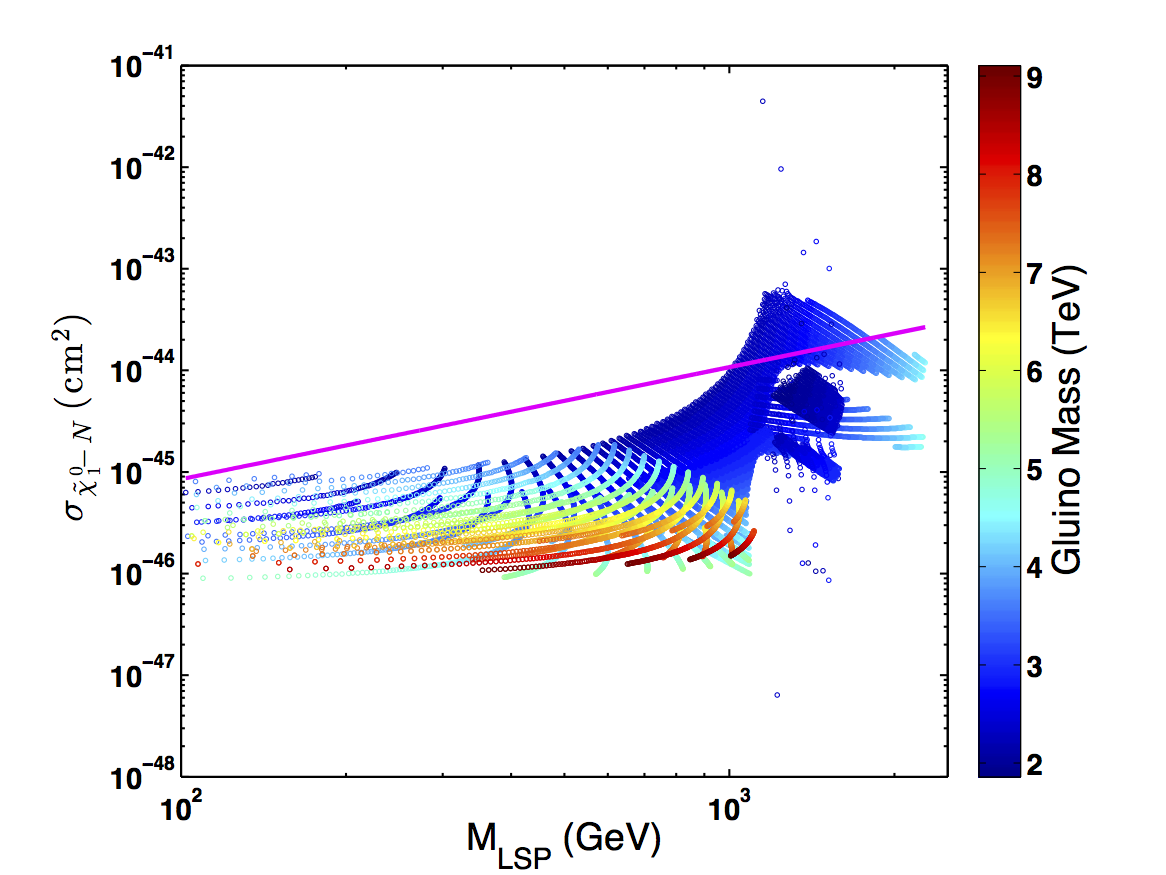}
\includegraphics[width=0.45\textwidth]{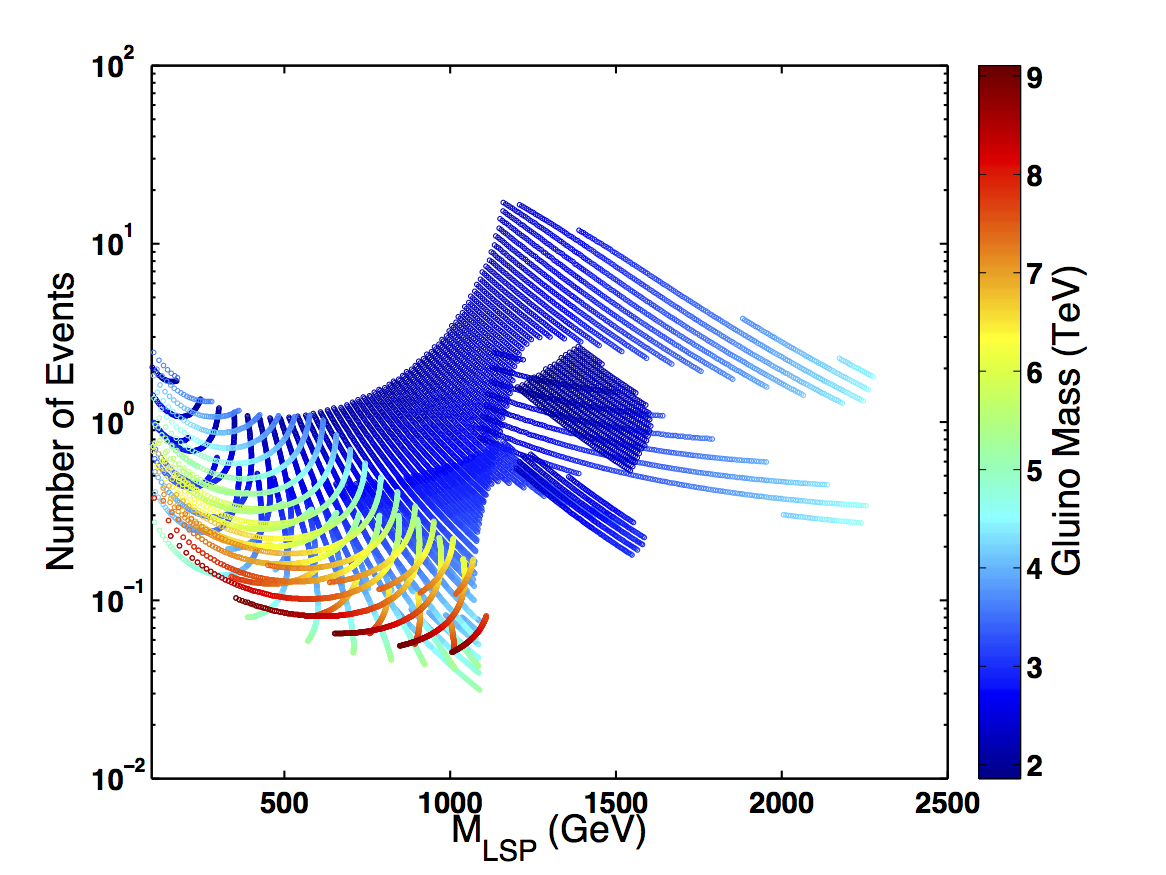}
\caption{\textbf{Dark Matter Detection Prospects for Higgsino-like LSP Points.} Left panel shows the distribution in neutralino-nucleon scattering cross-sections versus neutralino mass for the Higgsino-like segment of the Type~IIB flux compactification scenario. The solid magenta line represents the limit set by the recent results from LUX. The right panel gives the rate of nuclear recoils, integrated over the recoil energy range of 5-25~keV, after one ton-year of exposure. Both panels aggregate all the cases with a Higgsino-like LSP for all modular weight combinations.}
\label{fig:HinoXenonCompare}
\end{center}
\end{figure}

In the alternative case for which the neutralino is almost purely Higgsino-like, we can again aggregate all eight combinations of modular weights that admit a Higgsino-like LSP into a single region. For this combination, the LSP can be as heavy as 2.3~TeV, which is slightly heavier than the maximum value in the bino-like case. However, LSPs as light as 100~GeV are also present. The neutralino-nucleon scattering cross section is similarly spread over a wide range of values. The smallest cross section achieved is $6\times 10^{-48}\,\rm{cm}^2$, though the majority of the parameter space has a cross section of between $10^{-46}$ and $10^{-43}\,\rm{cm}^2$. 

The left panel of Figure~\ref{fig:HinoXenonCompare} shows the neutralino-nucleon cross-section versus LSP mass for the Higgsino-like LSP cases, where we again overlay the LUX~bound for 118~kg $\times$ 85.3~days of exposure (the solid magenta line in the figure). In contrast to the bino-like LSP case, LUX~has already begun to rule out some areas of the parameter space. The majority of the points lying above the magenta line correspond to the $(n_M,n_H)$ =  $(\frac{1}{2},0)$ case; in addition six additional points corresponding to $(n_M,n_H)$ =  $(0,0)$ and $(0,\frac{1}{2})$ are also ruled out by this experiment. These points represent cases with $1000\,{\rm GeV} \lappeq m_{\chi_1^0} \lappeq 1100\,{\rm GeV}$ and relatively light gluinos, though the points with the absolute lowest gluino masses in the data sample continue to exist just below the current LUX~limit. We thus expect that any improvement upon this bound will reduce the viable parameter space considerably, particularly for the low mass neutralinos in the 100~GeV range.

To gain a sense of how quickly each of the combinations of modular weights will be within experimental limits, we can consider the number of events that would be observed for a given fiducial volume and exposure time. The right panel of Figure~\ref{fig:HinoXenonCompare} again shows the expected number of events within the recoil energy range of 5-25~keV
for a baseline of 300~days of exposure and 1000~kg of volume, which we will consider one `ton-year'. LUX~claims a future background expectation of approximately 1~event per ton-year at these recoil energies~\cite{Akerib:2012ys}. We can therefore expect a very large fraction of the Higgsino-like outcomes in the Type~IIB flux compactification scenario to be within reach in the near future.

\begin{table}[t]
\begin{center}
\begin{tabular}{|cc||c|c|}\hline
\multicolumn{2}{|c||}{Higgsino LSP}  & \multicolumn{2}{|c|}{Recoil Events, 1 Ton-Year } \\
\parbox{1.0cm}{$n_M$} & \parbox{1.0cm}{$n_H$} &	\parbox{2.0cm}{Minimum}	&	\parbox{2.0cm}{Maximum} \\ \hline \hline
0 & 0 & 0.336	& 1.236	\\
0 & $\frac{1}{2}$ & 0.9858	& 4.899	\\
$\frac{1}{2}$ & 0	& 0.2172	& 27.81	\\
$\frac{1}{2}$ & $\frac{1}{2}$	& 0.0816	& 12.42	\\
$\frac{1}{2}$ & 1	& 0.0599	& 0.18	\\
1 & 0	& 0.2085	& 26.61	\\
1 & $\frac{1}{2}$	& 0.0978	& 9.78	\\
1 & 1	&0.0981	& 3.78	\\\hline
\multicolumn{2}{|c||}{Bino LSP}	& 0.0020	& 0.31	\\\hline
\end{tabular}
\caption{\footnotesize \textbf{Minimum and Maximum Event Rate for One Ton-Year Exposure on Liquid Xenon}. The Higgsino-like cases from Table~\ref{targetscan} are listed individually, with the minimum and maximum of nuclear recoil events in one ton-year of exposure listed for the entire parameter space. For comparison purposes, the minimum and maximum number of nuclear recoils for all bino-like cases is also listed. Recoil rates are integrated over the recoil energy range of 5-25~keV.}
\label{tonyearrate}
\end{center}
\end{table}

In Table~\ref{tonyearrate} we dis-aggregate the points in Figure~\ref{fig:HinoXenonCompare} and give the largest and smallest event rate in one ton-year for the eight modular weight combinations with a Higgsino-like LSP from Table~\ref{targetscan}. For comparison purposes, we also give the largest and smallest event rate for the aggregated Bino-like cases from the previous subsection. All entries in Table~\ref{tonyearrate} are integrated over the recoil energy range of 5-25~keV for liquid xenon targets. The most interesting cases, from a theoretical point of view, are the first four modular weight combinations, for which it is reasonable to expect a signal to emerge on the time scale of one ton-year of exposure. The one outlier in the Higgsino-like region is the case of  $(n_M,n_H)$ =  $(\frac{1}{2},1)$, which is nearly ruled out in any case simply by electroweak symmetry breaking constraints.
Thus we expect future results from LUX and Xenon1T to probe deep into this very interesting region of the parameter space of flux-compactified Type~IIB string theory. In addition, some of this parameter space also predicts a gluino mass which should be well within reach at the post-shutdown~LHC. We turn to a consideration of LHC~searches in the next section.

\section{LHC Implications}
\label{LHC}

As was discussed earlier, the spectrum of flux-stabilized Type~IIB models of the KKLT~type generally involves a large hierarchy between the electroweak gauginos and the much heavier squarks and sleptons. For that reason, we have emphasized thus far only the masses of the LSP neutralino and the gluino, as (for example) in Table~\ref{targetscan}. The scale of superpartner masses is generally pushed to large values by the requirement that the Higgs mass satisfy the recent~LHC measurements. We anticipate, therefore, that no point in the parameter space of this class of theories faces elimination from the data taken at $\sqrt{s} = 8\,{\rm TeV}$, once the Higgs mass constraint is satisfied. We will confirm this point below. Nevertheless, much of the parameter space will be accessible at the post-upgrade center of mass energy of $\sqrt{s}=14\,{\rm TeV}$. This is particularly welcome for the regions with a bino-like LSP, which have limited prospects for discovery in dark matter detection experiments.

\subsection{Benchmark Phenomenology}

To examine the extent to which this region has been probed by the $\sqrt{s}=8\,{\rm TeV}$ data, and to determine what will be of interest at $\sqrt{s}=14\,{\rm TeV}$, for every point in the flux-compactified Type~IIB parameter space would be computationally expensive. It would also be largely unnecessary, as much of the space has a similar phenomenology that can be described succinctly by a handful of examples.
For these reasons we will choose a small sample of benchmark points, representative of each of the two phenomenologically distinct regions we have established, and analyze this set of points with regard to LHC~searches for superpartners. This will also afford us the opportunity to take a closer look at the allowed physical mass spectra of the various modular weight combinations, in light of the PLANCK dark matter fits and LHC~Higgs mass measurements.

\begin{table}[t]
\begin{center}
\begin{tabular}{|c||c|c|c|c|c||c|c|c|c|c|c|c||c|c|c|}\hline
 & \multicolumn{5}{|c||}{Benchmark Inputs}  &  \multicolumn{7}{|c||}{Key Physical Masses (GeV)} & \multicolumn{3}{|c|}{Key Properties} \\ \hline
Name	& $n_M$ 		& $n_H$ 		& $M_0$ & $\alpha$ & $\tan\beta$ & $m_h$ & $m_{\tilde{\chi}^0_1}$ & $m_{\tilde{\chi}_1^\pm}$ &  $m_{A}$ & $m_{\tilde{\tau}}$ & $m_{\tilde{g}}$ & $m_{\tilde{t}_1}$ & $\Omega_{\chi}h^2$ & \parbox{2.0cm}{$\sigma_{\rm SUSY}^{\rm 8\,TeV}$ (fb)} &  \parbox{2.0cm}{$\sigma_{\rm SUSY}^{\rm 14\,TeV}$ (fb)} \\ \hline
Higgsino-A 	& 0 & 0	& 2.57 & 2.00 & 48 & 
124.1 & 1201 & 1204 & 1977 & 1334 & 2287 & 1340 & 0.072 & 0.04	& 4.7 \\
Higgsino-B 	& 0 & $\frac12$	& 2.07 & 2.00 & 48 & 
124.3 & 1203 & 1206 & 1505 & 1472 & 1857 & 1385 & 0.076	& 0.31 & 23.8 \\
Higgsino-C 	& $\frac12$	& 0	& 2.49 & 1.73 & 16 & 
124.2 & 145.6 & 147.8 & 3111 & 1661 & 2535 & 1369 & 0.003 & 1525 & 3542 \\
Higgsino-D 	& $\frac12$	& $\frac12$	& 4.35 & 1.76 & 30 & 
125.2 & 353.5 & 354.9 & 3731 & 2871 & 4280 & 2801 & 0.014 & 44.3 & 142.4 \\
Higgsino-E 	& $\frac12$	& 1	& 4.60 & 2.00 & 41 & 
126.4 & 826.1 & 827.3 & 1547 & 2992 & 3873 & 2530 & 0.076 & 0.42 & 4.8 \\
Higgsino-F 	& 1 & 0	& 4.07 & 0.69 & 15 & 
124.2 & 105.7 & 107.1 & 4666 & 1131	& 6490 & 3704 & 0.002 & $5.1\times 10^3$ & $11.0 \times 10^3$ \\
Higgsino-G 	& 1 & 1	& 5.97 & 1.13 & 24 & 
124.9 & 353.6 & 354.7 & 1070 & 1161	& 7822 & 4816 & 0.014 & 44.3 & 142.8 \\ \hline
Bino-A & 0 & 0 & 1.94 & 1.03 & 31 & 
126.4 & 1432 & 1758 & 3002 & 1460 & 2900 & 1607 & 0.127 & $0.95\times 10^{-3}$ & 0.41 \\
Bino-B & 0 & 0 & 1.24 & 1.17 & 24 & 
124.2 & 957 & 1125 & 2097 & 1098 & 1787 & 979 & 0.044 & 0.71 & 36.6 \\ 
Bino-C & 0 & $\frac12$ & 1.93 & 1.26 & 14 & 
125.2 & 1549 & 1775 & 3106 & 1947 & 2619 & 1782 & 0.128 & $0.79\times 10^{-3}$ & 0.95 \\ 
Bino-D & 0 & 1 & 1.75 & 0 & 52 & 
124.2 & 760 & 1451 & 1824 & 761 & 3773 & 2619 & 0.078 & $0.19\times 10^{-3}$ & $0.86\times 10^{-2}$ \\
Bino-E & $\frac12$ & $\frac12$ & 2.49 & 0.44 & 35 & 
125.7 & 1415 & 2138 & 2890 & 1423 & 4518 & 2999 & 0.115	& $0.18\times 10^{-3}$ & $0.57\times 10^{-2}$ \\
Bino-F & $\frac12$ & 1 & 2.37 & 0 & 54 & 
124.2 & 1032 & 1960 & 2164 & 1034 & 4978 & 3575 & 0.090 & $0.19\times 10^{-3}$ & $0.81\times 10^{-3}$ \\ 
\hline
\end{tabular}
\end{center}
\caption{\footnotesize \textbf{Spectra and Key Properties for Benchmark Cases from Allowed Parameter Space of the Type~IIB Flux Compactification Scenario}. The input value of the bulk mass parameter $M_0$ is listed in units of~TeV, while output physical masses for certain key superpartners are given in units of~GeV. For reference, we also give the thermal relic density for the lightest neutralino, as well as the total SUSY production cross-section at both $\sqrt{s}=8\,{\rm TeV}$ and $\sqrt{s} = 14\,{\rm TeV}$.}
\label{tbl:benchmarks}
\end{table}

We collect our representative benchmarks in Table~\ref{tbl:benchmarks}.
Of the regions with a Higgsino-like LSP, 7~points were chosen; of the regions with a bino-like LSP, 6~points were chosen. These sample a variety of Higgs masses, gluino masses, and dark matter relic densities, as well as different combinations of modular weights. Note that we have chosen two examples from the bino-like region of the $(n_M,n_H)$ =  $(0,0)$ case, to exemplify the variety of outcomes possible in this case. Note also that for Table~\ref{tbl:benchmarks} we have chosen not to create a specific benchmark for the $(n_M,n_H)$ =  $(1,\frac{1}{2})$ case due to its similarities with the $(n_M,n_H)$ =  $(1,0)$ and $(1,1)$ cases. The input value of the bulk mass parameter $M_0$ is listed in units of~TeV, while output physical masses for certain key superpartners are given in units of~GeV. For reference, we also give the thermal relic density for the lightest neutralino, as well as the total SUSY production cross-section at both $\sqrt{s}=8\,{\rm TeV}$ and $\sqrt{s} = 14\,{\rm TeV}$, as computed via PYTHIA~6.4~\cite{Sjostrand:2006za}.

The Higgsino-like cases tend to occur at the high end of the $\alpha$ range we considered, which is consistent with the scan results presented in Section~\ref{scan}. As expected, the cases with very light ($\order(100\,{\rm GeV})$) LSP neutralinos are associated with very low relic densities. Values approaching the PLANCK-preferred range are possible if the mass of the neutralino is increased. As we take an agnostic point of view as to the degree to which non-thermal production mechanisms may be operative in the early universe, we have not chosen to enforce a lower bound on $\Omega_{\chi}h^2$ on our choice of benchmark points. Very large total production cross-sections, of more than one picobarn, are possible for the cases with very light neutralinos. We therefore expect the Higgsino-like points in the KKLT parameter space to be producing substantial numbers of superpartners at the LHC, even at $\sqrt{s} = 8\,{\rm TeV}$. However, as we will discuss below, these points will nevertheless evade detection.

The cases where the lightest neutralino is bino-like generally fall at lower values of the parameter $\alpha$ and allow for slightly smaller values of the mass scale $M_0$. We note that cases Bino-D and Bino-F have $\alpha=0$, and thus have soft supersymmetry breaking masses for the gauginos which obey the relations familiar from minimal supergravity. The lightest neutralino in these cases is always much heavier than 100~GeV, and is often highly degenerate with the stau and/or lightest stop. Note that, with the exception of case Bino-E, all points obey the relation $m_A \simeq 2 m_{\chi_1^0}$. Thus the thermal relic abundance for these cases can be quite consistent with the PLANCK observations using co-annihilation and resonant annihilation channels. The heavy electroweak gauginos result in a much lower production cross-section at the LHC, particularly at $\sqrt{s} = 8\,{\rm TeV}$, though Bino-B with $m_{\tilde{g}} \simeq 1800\,{\rm GeV}$ has a reasonable cross-section of just under 1~fb at this center-of-mass energy.

The mass spectra represented by the benchmark points in Table~\ref{tbl:benchmarks} are representative of the entire Type~IIB flux-compactification parameter space, once the constraints arising from the Higgs mass measurement are imposed. In particular, it is always the case that the gluino is comparable in mass, but heavier than, the lightest stop. The mass difference ranges from 500~GeV to 1~TeV. This is easy to understand from the perspective of the boundary conditions in~(\ref{gaugmass}) and~(\ref{scalarmass}). When $n_M=1$, clearly the gauginos will begin heavier than the matter fields. But even for $n_M=0$ we expect a negative contribution to $M_3$ which is small relative to $M_0$. Using the definition in~(\ref{Mg}) we find that $M_g/M_0 = 0.12$ for $M_0 = 1800\,{\rm GeV}$ and $\alpha =0.6$, rising to $M_g/M_0 = 0.39$ for the same $M_0$ and $\alpha =2$. The renormalization group evolution only increases the gluino mass and decreases the lightest stop mass. 

\begin{table}[t]
\begin{center}
\begin{tabular}{|c||c|c|c|c|c|c|c|}
 \multicolumn{1}{c}{ }  & \multicolumn{7}{c}{Production Channel(s)} \\ \hline
 \multicolumn{1}{|c||}{ }  & \parbox{1.6cm}{$\tilde{g}\, \tilde{g}$} & \parbox{1.6cm}{$\tilde{q}\, \tilde{q}$} & \parbox{1.6cm}{$\tilde{q}\,\tilde{g}$} & \parbox{1.6cm}{$\tilde{t}\, \tilde{t}$} & \parbox{1.6cm}{$\tilde{q}\, \tilde{\chi}$} & \parbox{1.6cm}{$\tilde{\chi}\, \tilde{\chi}$} & \parbox{1.6cm}{Other} \\ \hline
Higgsino-A 	& 576 & 27834 & 10775 & 6257 & 164 & 4337 & 57 \\
Higgsino-B 	& 1233 & 28882 & 17602 & 1013 & 306 & 886 & 78 \\
Higgsino-C 	& 0	& 91 & 17 & 4 & 0 & 49888 & 0 \\
Higgsino-D 	& 0	& 14 & 0 & 0 & 0 & 49986 & 0 \\
Higgsino-E 	& 1	& 13430	& 178 & 8 & 3 & 36353 & 27 \\
Higgsino-F 	& 0 & 0	& 0 & 0 & 0 & 50000 & 0 \\
Higgsino-G 	& 0 & 0	& 0 & 0 & 0 & 49987 & 13 \\ \hline
Bino-A & 227 & 25312 & 6718 & 13938 & 1643 & 1406 & 711 \\
Bino-B & 1267 & 23378 & 15321 & 7913 & 1202 & 402 & 517 \\ 
Bino-C & 438 & 33524 & 11427 & 2306 & 1340 & 513 & 452 \\ 
Bino-D & 14 & 19988 & 1607 & 2140 & 6997 & 13224 & 6030 \\
Bino-E & 1 & 6472 & 255 & 701 & 2783 & 29111 & 10677 \\
Bino-F & 0 & 2509 & 20 & 68 & 5048 & 27120 & 15235 \\ 
\hline
\end{tabular}
\end{center}
\caption{\footnotesize \textbf{Relative Superpartner Production at $\sqrt{s} = 14\,{\rm TeV}$ for Benchmark Cases from Allowed Parameter Space of the KKLT Flux Compactification Scenario}. The distribution of 50,000~generated events at $\sqrt{s}=14\,{\rm TeV}$ is given for five different production channels. Columns two and three involve the aggregate across all light-flavor squarks and the scalar bottom quark, while the fourth column explicitly picks out stop pair production. The fifth column sums over all light-flavor squarks and all charginos and neutralinos, while column six sums over all charginos and neutralino combinations. The last column generally involves occasional associated production of a glunio with an electroweak gaugino, and slepton pair production where kinematically favorable.}
\label{tbl:prodXSec}
\end{table}

We therefore expect squark pair production to dominate over gluino pair production at the LHC, with the exception of cases with relatively light gluinos, for which gluino production in association with a squark can be sizeable. In general, however, much of the supersymmetric production cross-section will come in the form of pair production of electroweak gauginos, particularly for the Higgsino-like cases in Table~\ref{tbl:benchmarks}. In Table~\ref{tbl:prodXSec} we show the distribution of 50,000~generated events at $\sqrt{s}=14\,{\rm TeV}$, for each benchmark case, across five different production channels. Columns two and three involve the aggregate across all light-flavor squarks and the scalar bottom quark, while the fourth column explicitly picks out stop pair production. The fifth column sums over all light-flavor squarks and all charginos and neutralinos, while column six sums over all charginos and neutralino combinations. The last column generally involves occasional associated production of a glunio with an electroweak gaugino, and slepton pair production where kinematically favorable. Table~\ref{tbl:prodXSec} will inform the choices we make when we select certain LHC~search results for comparison to the parameter space of Type~IIB flux compactifications, in the next subsection.

\subsection{Summary of Relevant ATLAS Supersymmetry Searches}
\label{ATLAS}

The two general purpose detectors at the~LHC have each published multiple search results looking for superpartners in a variety of event topologies. To date, no signal above background has been detected, and the two experiments generally place very similar bounds on the effective supersymmetric production cross-section into various final states. For simplicity, therefore, we will consider only the ATLAS search results, as the published searches for this collaboration tend to involve simple geometric cuts and signal region definitions which are better suited to reproduction with a simplified detector simulator such as PGS4~\cite{PGS4}.


Since the~LHC began collecting data at $\sqrt{s}=8\,{\rm TeV}$, over three~dozen different searches have been performed by~ATLAS to attempt to discover evidence of supersymmetry. We can reduce this list by using the properties of the signals, as represented by the benchmarks in Tables~\ref{tbl:benchmarks} and~\ref{tbl:prodXSec}, to find those searches whose signal regions match well with the phenomenology of the Type~IIB flux compactification model. 

We can begin by eliminating searches which rely heavily on high-p$_T$ leptons in the final state. Here we are considering strictly electrons and muons with at least 10~GeV of transverse momenta. Benchmarks with bino-like LSPs, and a sizeable mass gap between the LSP and the next-to-lightest superpartner, can yield as many as two high-p$_T$ leptons in 5-10\% of produced events, but other benchmarks yield negligible numbers of leptons in the final state. We computed the expected signal at $\sqrt{s} = 8\,{\rm TeV}$ for the ATLAS single lepton search~\cite{ATLAS:2012tna} and same-sign dilepton search~\cite{ATLAS:2012sna} and found that none of the benchmark models in Table~\ref{tbl:benchmarks} would have produced a single event in the described signal regions when scaled to the appropriate integrated luminosity. Even at $\sqrt{s} = 14\,{\rm TeV}$, we expect only a handful of events in the single lepton and dilepton channels.
Though a large fraction of the total cross-section for benchmarks Bino-D, E~and~F involve stau production, the overall cross-section for these points is rather small. Comparing with the ATLAS search for pairs of hadronically-decaying taus~\cite{ATLAS:2013yla}, we again find that no pairs of opposite-sign taus would pass the trigger requirements at $\sqrt{s} = 8\,{\rm TeV}$ when scaled to 20 fb$^{-1}$. We therefore focus our attention on searches which invoke a lepton veto for the remainder of this section.

A striking feature of many of the benchmarks with Higgsino-like LSPs is the overwhelming tendency to produce pairs of electroweak gauginos. This is true of the benchmarks with the highest overall cross-sections. These gauginos tend to be from the degenerate system of low-lying neutralinos and charginos, so the decay products are soft and do not generally reconstruct as jets. Lepton multiplicities for these benchmarks are also exceedingly low. Such events are best sought after via mono-jet topologies, though discovery in these cases will likely be quite difficult~\cite{Cheung:2005pv,Baer:2011ec,Han:2013kza}. We expect this channel to be especially important for benchmarks Higgsino-C through Higgsino-F, whose jet multiplicities peak at a single high-p$_T$ jet, and drop rapidly thereafter.

For the remainder of the cases we find that typical jet multiplicities tend to be low. Benchmarks Higgsino-A and Higgsino-B, and all of the bino-like benchmarks, have significant production of $SU(3)$-charged superpartners. Here jet multiplicities peak in the range $3 \leq N_{\rm jet} \leq 5$. We will therefore consider the low-multiplicity multijet search with missing transverse energy and a leptonic veto. In addition, stop pair production can be significant for many of these points. Indeed, often the `light' flavored squark in Table~\ref{tbl:prodXSec} is, in fact a scalar bottom quark. Thus we will also consider the two principal searches that utilize b-tagged jets with a leptonic veto.


For each of the searches, the search strategy is divided between object reconstruction, which sets criteria used to define each object within an event, and signal region definitions that make selections based on the properties of these objects. For the ATLAS searches, jets are reconstructed using the anti-$k_T$ algorithm with a radius parameter of 0.4. Jets must be isolated from leptons using the following prescription: jets within $\Delta R\equiv\sqrt{\Delta\phi^2+\Delta\eta^2}=2$ of an electron are discarded. If any lepton is within $\Delta R=0.4$ of a jet, the lepton is discarded. The missing energy, denoted $\slashed E_T$, is the vector sum of the $p_T$ of any reconstructed objects, and any other calorimeter clusters with $|\eta|<4.9$ not belonging to other reconstructed objects. For jet and lepton candidates, a requirement is placed on both $p_T$ and $|\eta|$ that varies between each of the searches. 
Isolation requirements are placed on electrons and muons that are equivalent to, or looser than, those required by the PGS4 reconstruction.
In addition, further requirements may be placed on shower shape and track-selection criteria. Because we will be performing our simulation using PGS4, we will use the default PGS4 reconstruction, supplemented by these requirements. The missing energy and effective mass is then recalculated using these new definitions. Once these objects are reconstructed, we can now define the signal regions used by the relevant searches. The signal regions used by each of these searches, as well as the observed results, are listed below.

\begin{description}
\item[Low Multiplicity Multijets~\cite{TheATLAScollaboration:2013fha}] This search was conducted with $20.3\,\rm{fb}^{-1}$ of integrated luminosity. Electrons were required to have $p_T>10\GeV$ and $|\eta|<2.47$. Muon candidates must have $p_T>10\GeV$ and $|\eta|<2.4$. Jet candidates are required to have $p_T>20\GeV$ and $|\eta|<4.5$. With this reconstruction, signal region jets are required to have $p_T>40\GeV$ and $|\eta|<2.5$. 
For the 2-jet and 3-jet signal regions, $\Delta\phi$ between any jet and the direction of $\slashed E_T$ must be greater than 0.4 for the first two jets, and the third jet should have $p_T>40\GeV$. For the 4-jet, 5-jet, and 6-jet signal regions, $\Delta\phi>0.4$ for the first three jets, and $\Delta\phi>0.2$ for any additional jets with $p_T>40\GeV$.

\begin{table}[t]
\begin{center}
\begin{tabular}{|c||c|c||c|c||c|c||c||c|c|c|}\hline
\multirow{2}{*}{Requirement} & \multicolumn{9}{c}{Channel} &\\
 & 2JL & 2JM & 3JM & 3JT & 4JM & 4JT & 5J & 6JL & 6JM & 6JT \\\hline
$\slashed E_T$ & \multicolumn{9}{c}{$\geq 160\,{\rm GeV}$} & \\\hline
$p_T^{j_1}$ & \multicolumn{9}{c}{$\geq 130\,{\rm GeV}$} & \\\hline
$p_T^{j_n}$ & \multicolumn{9}{c}{$\geq 60\,{\rm GeV}$} & \\\hline
$\slashed E_T/M_{\rm eff}$ & 0.2 & -- & 0.3 & 0.4 & 0.25 & 0.25 & 0.2 & 0.15 & 0.2 & 0.25 \\\hline
$M_{\rm eff}^{\rm incl}$ (GeV) & 1000 & 1600 & 1800 & 2200 & 1200 & 2200 & 1600 & 1000 & 1200 & 1500 \\\hline\hline
Observed & 5333 & 135 & 29 & 4 & 228 & 0 & 18 & 166 & 41 & 5 \\\hline
$N_{95}$ & 1341.2 & 51.3 & 14.9  & 6.7 & 81.2 & 2.4 & 15.5 & 92.4 & 28.6 & 8.3 \\\hline
\end{tabular}
\end{center}
\caption{\footnotesize \textbf{Signal Region Definitions for the Multijet Search of~\cite{TheATLAScollaboration:2013fha}}. Requirements on the amount of missing transverse energy ($\slashed E_T$), transverse momentum of the leading jet ($p_T^{j_1}$), and transverse momenta of all additional required jet(s) ($p_T^{j_n}$), are given in units of~GeV. These requirements are universal across all signal regions. Also given is the minimum required inclusive effective mass ($M_{\rm eff}^{\rm incl}$), defined as the sum of the missing transverse momentum and all reconstructed jets with $p_T^{j} > 40\,{\rm GeV}$, and the ratio $\slashed E_T/M_{\rm eff}$, where the $M_{\rm eff}$ in the denominator sums only over the leading N~jets. Also listed is the number of observed events in each channel, and the corresponding value of $N_{95}$.}
\label{tbl:multijets}
\end{table}

Signal regions are then defined in terms of the number of jets, all of which must meet the minimum requirements listed above. When selecting events with at least N~jets, the quantity $M_{\rm eff}$ is defined to be the scalar sum of the transverse momenta of the leading N~jets and $\slashed E_T$. Signal regions are then defined by a minimum value of missing transverse energy ($\slashed E_T$), transverse momentum of the leading jet ($p_T^{j_1}$), and transverse momenta of all additional required jet(s) ($p_T^{j_n}$), in addition to a minimum for the ratio $\slashed E_T/M_{\rm eff}$. Signal regions of a given jet multiplicity are further subdivided into `loose', `medium', and `tight' subcategories based on the minimum value required for the inclusive effective mass ($M_{\rm eff}^{\rm incl}$), defined as the sum of the missing transverse momentum and all reconstructed jets with $p_T^{j} > 40\,{\rm GeV}$. These minimum requirements are collected in Table~\ref{tbl:multijets}, which also gives the number of observed events in each channel, and the corresponding upper value on the number of events arising from non-Standard Model processes in this channel, at the 95\% confidence level.

%
\item[Multijets with Two B-Tagged Jets~\cite{ATLAS:2013cma}] This search was conducted with 20.5 $\rm{fb}^{-1}$ integrated luminosity. Electrons were subject to "loose" shower shape requirements, and were required to have $p_T>10\GeV$ and $|\eta|<2.47$. Muon candidates must have $p_T>10\GeV$ and $|\eta|<2.4$. Jet candidates are required to have $p_T>20\GeV$ and $|\eta|<4.5$. With this reconstruction, signal region jets are required to have $p_T>35\GeV$ and $|\eta|<2.5$. B-tagged jets increase this requirement to $p_T>40\GeV$ and $|\eta|<2.5$.

Three signal regions are defined. For each of these, there is a requirement that there be zero leptons, at least two b-tagged jets, and an invariant jet mass for the first three jets of between 80 and 270~GeV. The two hardest jets must satisfy $p_T \geq 80\GeV$, while additional jets are required to have $p_T \geq 35\GeV$. The jets must be separated from the direction of the $\slashed E_T$ by $\Delta\phi>\pi/5$. The transverse mass $m_T$ constructed from the hardest b-jet and $\slashed E_T$ must be at least 175~GeV. The three signal regions are defined by requiring $\slashed E_T$ to be 200, 300, and 350~GeV, respectively. Labeling these signal regions~SR1, SR2~and~SR3, the ATLAS collaboration reports observing~15, 2~and~1 event, respectively in these channels. From this data it was possible to establish an upper bound to the number of events for contributions beyond that of the Standard Model at the 95\% confidence level, denoted $N_{95}$. For the three signal regions of this search, that number was 10, 3.6 and 3.9 events, respectively.

%
\item[Monojets~\cite{TheATLAScollaboration:2013aia}] This search was conducted with $20.3\rm{ fb}^{-1}$ integrated luminosity. 
A pre-selection is defined by requiring $\slashed E_T>120\GeV$, zero reconstructed leptons, and at least one jet with $p_T>120\GeV$ and $|\eta|<2.8$. A monojet-like signal region is defined by requiring at most three jets with $p_T>30\GeV$ and $|\eta|<2.8$, and $\Delta\phi>0.4$ between each jet and the missing transverse energy. The leading jet must have $p_T>280\GeV$, and the missing transverse energy must satisfy $\slashed E_T>220\GeV$. 
In this signal region, a total of 30793~events were observed, corresponding to $N_{95}=2770$ events. A second signal region involving charm-tagged jets will be ignored, as~PGS4 does not implement a charm-tagging algorithm.

%
\item[Stop Pair Production~\cite{Aad:2013ija}]

This search was conducted with $20.1\,{\rm fb}^{-1}$ of integrated luminosity. Jet candidates are required to have $p_T>20\GeV$ and $|\eta|<4.5$. Signal region jets are required to have $|\eta|<2.5$.  Electrons were required to have $p_T>7\GeV$ and $|\eta|<2.47$, while muon candidates must have $p_T>6\GeV$ and $|\eta|<2.4$. Events were rejected if any such electrons or muons were present in the final state. Nevertheless, for this search $\slashed{E}_T$ is constructed from all electrons and muons satisfying $p_T>10\GeV$ and all jets with $p_T>20\GeV$.

\begin{table}[t]
\begin{center}
\begin{tabular}{|c||c|c|c|c|c||c|}\hline
Requirement		& \multicolumn{5}{c||}{SRA}	& SRB \\\hline\hline
$\slashed{E}_T$	& \multicolumn{5}{c||}{$\geq 150\,{\rm GeV}$} & $\geq 250\,{\rm GeV}$  \\
$p_T^{\rm j_1}$	& \multicolumn{5}{c||}{$\geq 130\,{\rm GeV}$}	& $\geq 150\,{\rm GeV}$ \\
$p_T^{\rm j_2}$		& \multicolumn{5}{c||}{$\geq 50\,{\rm GeV}$}	& $\geq 30\,{\rm GeV}$ \\
$p_T^{\rm j_3}$		& \multicolumn{5}{c||}{--} & $\geq 30\,{\rm GeV}$ \\ \hline
$\Delta\phi(\slashed{E}_T,j_1)$	& \multicolumn{5}{c||}{--}  & $>2.5$\\
$\Delta\phi_{\rm min}$	& \multicolumn{5}{c||}{$>0.4$} & $>0.4$\\
$\slashed{E}_T/M_{\rm eff}$		& \multicolumn{5}{c||}{$>0.35$}	& $>0.25$\\\hline
$H_{T,3}$ &  \multicolumn{5}{c||}{--} & $\leq 50\,{\rm GeV}$ \\
$m_{\rm inv}^{bb}$ & \multicolumn{5}{c||}{$\geq 200\,{\rm GeV}$}	& -- \\
$m_{CT}$	& $\geq 150\,{\rm GeV}$ & $\geq 200\,{\rm GeV}$ & $\geq 250\,{\rm GeV}$ & $\geq 300\,{\rm GeV}$ & $\geq 350\,{\rm GeV}$ & -- \\  \hline\hline
Observed & 102 & 48 & 14 & 7 & 3 & 65\\
$N_{95}$ & 38 & 26 & 9 & 7.5 & 5.2 & 27 \\\hline
\end{tabular}
\end{center}
\caption{\footnotesize \textbf{Signal Region Definitions for the Stop Search of~\cite{Aad:2013ija}}. Requirements on the amount of missing transverse energy ($\slashed E_T$) and the transverse momenta of the leading three jets (where applicable) are given in~GeV. Also given are the values of separation requirements between various jet objects and the missing transverse energy, as well as the ratio of $\slashed E_T$ to the appropriate effective mass variable. Signal region specific cuts are described in the text. Also listed is the number of observed events in each channel, and the corresponding value of $N_{95}$.}
\label{tbl:bjets}
\end{table}

This search targets two independent types of events, and thus two signal regions are defined. The first, SRA, requires only two jets, both of which must be b-tagged. This particular signal region vetoes events with a third jet satisfying $p_T^{\rm j_3} > 50\,{\rm GeV}$. The second, SRB, allows three jets, with the second and third hardest jet b-tagged. Requirements on the minimum value of $\slashed{E}_{T}$ and the hardest N~jets are collected in Table~\ref{tbl:bjets}. Both signal regions introduce the kinematic variable $\Delta\phi_{min}$, defined as the minimum azimuthal distance between any of the three hardest jets and the direction of $\slashed{E}_{T}$, and require $\Delta\phi_{min} > 0.4$. Both also place a minimum value on the ratio $\slashed E_T/M_{\rm eff}$, where $M_{\rm eff}$ is defined as the scalar sum of the $\slashed{E}_{T}$ and the two (three) hardest jets for signal region SRA~(SRB).

Additional kinematic requirements are signal region specific. Signal region~SRB requires $H_{T,3} \leq 50\,{\rm GeV}$, where $H_{T,3}$ is the scalar sum of $\slashed{E}_T$ and the $p_T$ of all but the three hardest jets. It also requires $\Delta\phi$ between the leading (non b-tagged) jet and the direction of $\slashed E_T$ to be greater than 2.5. For signal region~SRA, the invariant mass of the tagged b-jets must satisfy $m_{\rm inv}^{bb} \geq 200\,{\rm GeV}$. The signal region is subdivided according to the value of the con-transverse mass, defined as
\begin{equation}
m_{CT}(v_1,v_2)=\sqrt{[E_T(v1)+E_T(v_2)]^2-[p_T(v1)-p_T(v_2)]^2}\, ,
\end{equation}
where $v_1$ and $v_2$ represent the two b-tagged jets. The requirements on these quantities for the various signal regions are collected in Table~\ref{tbl:bjets}, as are the corresponding upper value on the number of events arising from non-Standard Model processes in this channel, at the 95\% confidence level.
\end{description}

\subsection{Discoverability Prospects at $\sqrt{s}=8\,{\rm TeV}$ and $\sqrt{s}=14\,{\rm TeV}$}

To compare the Type~IIB flux compactification scenario with LHC data, we use our benchmark cases in Table~\ref{tbl:benchmarks} as proxies for the various pockets of parameter space that meet all phenomenological criteria, determined by the targeted scan in Section~\ref{scan}. All supersymmetric signals are generated by first calculating proper decay widths and branching ratios using SUSY-HIT. The output is passed to PYTHIA~6.4 for event generation and PGS4 to simulate the detector response. In the analysis that follows, b-tagging will prove to be important. For that reason, we use a modified version of PGS4 with an improved b-tagging algorithm~\cite{Altunkaynak:2010we} designed to more accurately mimic the b-tagging efficiency as a function of pseudorapidity $|\eta|$ and jet-p$_T$ reported in the ATLAS and CMS Technical Design Reports. Signals are computed for a fixed 50,000~events, generated with Level-0 triggers, at both $\sqrt{s}=8\,{\rm TeV}$ and $\sqrt{s}=14\,{\rm TeV}$ center-of-mass energies, and the results are scaled to the appropriate integrated luminosity to compare with LHC~measurements. 

Our analysis suggests that none of the models in Table~\ref{tbl:benchmarks} would present a signal in the data taken thus far. The strongest signals come from the monojet searches in the Higgsino-like cases with large cross-sections. This would be Higgsino-C (23 signal events in 20 fb$^{-1}$) and Higgsino-F (63 signal events in 20 fb$^{-1}$). This is to be compared with an $N_{95}$ value of~2770 in 20.3~fb$^{-1}$. The low-multiplicity multijet searches produce less than ten events in the two- and three-jet categories for those benchmarks that produce any signal at all in 20~fb$^{-1}$. These are all well below the $N_{95}$ values reported by the ATLAS experiment. Finally, despite very low background estimates (and thus, very low $N_{95}$ values) for the various stop searches involving lepton vetoes and b-tagged jets, the scaled signal expectation for our benchmark cases is always less than one event in 20~fb$^{-1}$ of integrated luminosity. 

One could ask if {\em any} of the allowed parameter space, identified in Section~\ref{scan}, could have detectable superpartners in the data collected thus far at $\sqrt{s} = 8\,{\rm TeV}$. To check this, we also generated 50,000 signal events for the model point from each region described in Table~\ref{targetscan} with the lightest gluino. The values of $\lbrace \alpha,M_0 \rbrace$ corresponding to the lightest gluino can be estimated from the figures found in Section~\ref{sec:fine}. In all but one case -- the bino-like region with vanishing modular weights -- this was also the point with the lightest stop mass. None of these lightest-gluino cases would give a signal above background in any of the ATLAS~searches described above. The best prospects would be for the Higgsino-like LSP point with $(n_M,n_H)$ =  $(\frac{1}{2},0)$ and $m_{\tilde{g}} = 2038\,{\rm GeV}$. This point would yield six~events in the three-jet `medium' bin of Table~\ref{tbl:multijets}, versus $N_{95} = 14.9$ events. We also estimate that this point would produce 20~events in~SRA with $m_{CT} \geq 150\,{\rm GeV}$ of Table~\ref{tbl:bjets}, versus $N_{95} = 38$ events.

We therefore turn our attention to $\sqrt{s}=14\,{\rm TeV}$, where the LHC will begin taking data in 2015. For future supersymmetry searches we cannot rely on published numbers such as the $N_{95}$ value, but must attempt to estimate the signal significance by calculating the contribution from Standard Model backgrounds to the signal regions described in Section~\ref{ATLAS}. For the purpose of this paper we will content ourselves with a rather crude estimate of these backgrounds, generated at the level of PYTHIA with level-one triggers within PGS4. An appropriately weighted sample representing 5~fb$^{-1}$ each of  $b$/$\bar{b}$ pair production, high-$p_T$ QCD dijet production, single $W^{\pm}$ and $Z$-boson production, pair production of electroweak gauge bosons ($W^+\,W^-$, $W^{\pm}\,Z$ and $Z\,Z$), and Drell-Yan processes, was generated at $\sqrt{s}=14\,{\rm TeV}$, as well as 20~fb$^{-1}$ of $t$/$\bar{t}$ pair production. Both the signal and the background was then scaled to the desired integrated luminosity, where the ratio of the signal events to the square root of the background events ($S/\sqrt{B}$) could be computed.

We note that the cut on the contransverse mass $m_{CT}$, employed in~SRA of the dedicated stop search of Reference~\cite{Aad:2013ija}, is extremely effective at reducing the backgrounds from pair-production of heavy-flavored quarks. This is reflected in the very low numbers of observed events, and $N_{95}$ values in Table~\ref{tbl:bjets}. In fact, the ATLAS collaboration estimates that their dominant background in these channels is production of a $Z$-boson in association with a single  heavy-flavor jet, with the $Z$-boson then decaying to two neutrinos. This is particularly true as the value of the minimum contransverse mass is increased. This particular background is poorly reproduced with PYTHIA, and thus we will not consider this search at $\sqrt{s} = 14\,{\rm TeV}$.


\begin{table}[t]
\begin{center}
\begin{tabular}{|c|c|c||c|c|c|c|c|c|c|c|c|c||c||c|c|c|}\hline
\multicolumn{3}{|c||}{Benchmark} 
& \multicolumn{10}{|c||}{Multijets}  &  & \multicolumn{3}{|c|}{Two B-Tagged Jets} \\ \hline
Name & $n_M$ & $n_H$ & \parbox{0.6cm}{2JL} & \parbox{0.6cm}{2JM} & \parbox{0.6cm}{3JM} & \parbox{0.6cm}{3JT} & \parbox{0.6cm}{4JM} & \parbox{0.6cm}{4JT} & \parbox{0.6cm}{5J} & \parbox{0.6cm}{6JL} & \parbox{0.6cm}{6JM} & \parbox{0.6cm}{6JT} & Monojet 
& \parbox{1.0cm}{SR1} & \parbox{1.0cm}{SR2} & \parbox{1.0cm}{SR3} \\ \hline
Higgsino-A 	& 0 & 0	& 33 & 31 & 17 & 8 & 8 & 3 & 4 & 2 & 2 & 1 
& 2 & 0	& 0 & 0 \\
Higgsino-B 	& 0 & $\frac12$	& 186 & 107 & 42 & 7 & 20 & 4 & 15 & 8 & 7 & 5 
& 11 & 3 & 3 & 3 \\
Higgsino-C 	& $\frac12$	& 0	& 193 & 123 & 47 & 23 & 20 & 10 & 9 & 3 & 3 & 3 
& 75 & 1 & 1 & 1 \\
Higgsino-D 	& $\frac12$	& $\frac12$	& 9 & 2 & 1 & 0 & 0 & 0 & 0 & 0 & 0 & 0 
& 12 & 0 & 0 & 0 \\
Higgsino-E 	& $\frac12$	& 1	& 13 & 14 & 7 & 4 & 2 & 1 & 1 & 0 & 0 & 0 
& 2 & 0 & 0 & 0 \\
Higgsino-F 	& 1 & 0	& 92 & 13 & 4 & 0 & 4 & 0 & 0 & 0 & 0 & 0 
& 163 & 0 & 0 & 0 \\
Higgsino-G 	& 1 & 1	& 8 & 1 & 1 & 0 & 0 & 0 & 0 & 0 & 0 & 0 
& 12 & 0 & 0 & 0 \\ \hline
Bino-A & 0 & 0 & 2 & 2 & 1 & 1 & 1 & 0 & 0 & 0 & 0 & 0 
& 0 & 0 & 0 & 0 \\
Bino-B & 0 & 0 & 274 & 216 & 98 & 28 & 47 & 13 & 30 & 12 & 10 & 8 
& 9 & 4 & 3 & 3 \\ 
Bino-C & 0 & $\frac12$ & 8 & 8 & 4 & 2 & 2 & 1 & 1 & 0 & 0 & 0 
& 0 & 0 & 0 &0  \\ 
Bino-D & 0 & 1 & 0 & 0 & 0 & 0 & 0 & 0 & 0 & 0 & 0 & 0 
& 0 & 0 & 0 & 0 \\
Bino-E & $\frac12$ & $\frac12$ & 0 & 0 & 0 & 0 & 0 & 0 & 0 & 0 & 0 & 0 
& 0	& 0 & 0 & 0 \\
Bino-F & $\frac12$ & 1 & 1 & 1 & 1 & 1 & 0 & 0 & 0 & 0 & 0 & 0 
& 0 & 0 & 0 & 0 \\ \hline \hline
\multicolumn{3}{|c||}{SM Background} & 63231 & 31157 & 307 & 24 & 69 & 8 & 155 & 585 & 227 & 39 & 20008 & 113 & 19 & 15 \\
\hline
\end{tabular}
\end{center}
\caption{\footnotesize \textbf{Estimated Signal Counts Produced by Benchmark Points of Table~\ref{tbl:benchmarks} in 20~fb$^{-1}$ at $\sqrt{s}=14\,{\rm TeV}$, for Relevant LHC8 Searches}. Signal counts are computed using the kinematic cuts defined by references~\cite{TheATLAScollaboration:2013fha}, \cite{ATLAS:2013cma}~and~\cite{TheATLAScollaboration:2013aia} for the $\sqrt{s}=8\,{\rm TeV}$ data set. We do not consider the dedicated stop search of Reference~\cite{Aad:2013ija} at $\sqrt{s}=14\,{\rm TeV}$. Also given is our estimate of the background contribution to each channel.}
\label{tbl:signal14TeV}
\end{table}

We begin by simply applying the signal region definitions described above to the signal and background samples at $\sqrt{s} = 14\,{\rm TeV}$. For convenience, we will refer to the set of cuts and object requirements in references~\cite{TheATLAScollaboration:2013fha}, \cite{ATLAS:2013cma}~and~\cite{TheATLAScollaboration:2013aia}, collectively, as `LHC8 searches.' Our estimation of the event counts in the~14 signal regions, with reliable background estimates, is given in Table~\ref{tbl:signal14TeV} for our benchmark points, and for the sum total of all background samples generated. The data in Table~\ref{tbl:signal14TeV} is normalized to 20~fb$^{-1}$ of integrated luminosity. We consider this to be a conservative estimate for the first year of data collection after the LHC~resumes operation in~2015.

Benchmarks Higgsino~C and Bino~B produce a signal of comparable size to our estimate of the backgrounds in both the three-jet and four-jet bins for the multijet analysis. In both cases the signal significance is greatly boosted by the application of a large lower bound on the inclusive effective mass in the event, set to 2.2~TeV for the `tight' sub-channels. Case Bino-B will provide a 3$\sigma$~excess in events over background in all the three-jet and four-jet channels within the first 10~fb$^{-1}$, and a 5$\sigma$ discovery in three of the four channels within the first 16~fb$^{-1}$. Higgsino-C will require slightly more than 20~fb$^{-1}$ to achieve a five-sigma discovery in any given channel, though it is likely that a discovery could be made here too by combining channels in the multijet + $\slashed{E}_{T}$ analysis. It is important to note that despite being a good candidate for a mono-jet search strategy, the signal in this channel produces less than a standard deviation of excess events over a quite substantial background in 20~fb$^{-1}$.

\begin{table}[t]
\begin{center}
\begin{tabular}{|c|c|c||c|c|c|c|c|c|c|c|c|c||c||c|c|c|}\hline
\multicolumn{3}{|c||}{Benchmark} 
& \multicolumn{10}{|c||}{Multijets}  &  & \multicolumn{3}{|c|}{Two B-Tagged Jets} \\ \hline
Name & $n_M$ & $n_H$ & \parbox{0.6cm}{2JL} & \parbox{0.6cm}{2JM} & \parbox{0.6cm}{3JM} & \parbox{0.6cm}{3JT} & \parbox{0.6cm}{4JM} & \parbox{0.6cm}{4JT} & \parbox{0.6cm}{5J} & \parbox{0.6cm}{6JL} & \parbox{0.6cm}{6JM} & \parbox{0.6cm}{6JT} & Monojet 
& \parbox{1.0cm}{SR1} & \parbox{1.0cm}{SR2} & \parbox{1.0cm}{SR3} \\ \hline
 	& 0 & 0	& 37 & 34 & 18 & 8 & 8 & 4 & 5 & 2 & 2 & 1 
& 2 & 1	& 1 & 1 \\
& $\frac12$	& 0	& 445 & 425 & 202 & 100 & 76 & 40 & 36 & 13 & 12 & 11 
& 69 & 2 & 2 & 2 \\
Higgsino & $\frac12$ & $\frac12$ & 123 & 109 & 58 & 31 & 22 & 12 & 10 & 4 & 4 & 4 
& 43 & 0 & 0 & 0 \\
& 1 & 0	& 27 & 4 & 1 & 0 & 0 & 0 & 0 & 0 & 0 & 0 
& 45 & 0 & 0 & 0 \\
& 1 & $\frac12$	& 26 & 4 & 2 & 1 & 1 & 0 & 0 & 0 & 0 & 0 
& 32 & 0 & 0 & 0 \\
& 1 & 1	& 19 & 4 & 2 & 1 & 1 & 0 & 0 & 0 & 0 & 0 
& 33 & 0 & 0 & 0 \\ \hline
& 0 & 0 & 403 & 302 & 109 & 28 & 52 & 14 & 33 & 13 & 12 & 9 
& 12 & 3 & 3 & 3 \\
& 0 & $\frac12$ & 79 & 71 & 35 & 12 & 18 & 6 & 13 & 6 & 6 & 5 
& 3 & 1 & 1 & 1 \\ 
Bino & 0 & 1 & 0 & 0 & 0 & 0 & 0 & 0 & 0 & 0 & 0 & 0 
& 0 & 0 & 0 & 0 \\ 
& $\frac12$ & $\frac12$ & 3 & 3 & 2 & 1 & 1 & 0 & 0 & 0 & 0 & 0 
& 0	& 0 & 0 & 0 \\
& $\frac12$ & 1 & 0 & 0 & 0 & 0 & 0 & 0 & 0 & 0 & 0 & 0 
& 0 & 0 & 0 & 0 \\ \hline \hline
\multicolumn{3}{|c||}{SM Background} & 63231 & 31157 & 307 & 24 & 69 & 8 & 155 & 585 & 227 & 39 & 20008 & 113 & 19 & 15 \\
\hline
\end{tabular}
\end{center}
\caption{\footnotesize \textbf{Estimated Signal Counts Produced by Lightest-Gluino Points in 20~fb$^{-1}$ at $\sqrt{s}=14\,{\rm TeV}$ for Relevant LHC8 Searches}. Simulated signals for the points in parameter space with the lightest gluino, for each distinct region identified in Table~\ref{targetscan}. The signal regions are the same as those for Table~\ref{tbl:signal14TeV}. Note that the points with the lightest gluino for the Higgsino-like LSP cases with $(n_M,n_H)$ =  $(0,\frac{1}{2})$ and $(\frac{1}{2},1)$ are our benchmarks Higgsino-B and Higgsino-E, respectively, which are already listed in Table~\ref{tbl:signal14TeV}.}
\label{tbl:signal14TeVlow}
\end{table}

The outcomes for these particular benchmarks are generally representative of the allowed parameter space identified in Table~\ref{targetscan}. In Table~\ref{tbl:signal14TeVlow} we repeat the exercise for the points in each region of Table~\ref{targetscan} with the lightest gluino. Here again we see that the representative with vanishing modular weights and Bino-like LSP yields a signal comparable to the background in the three- and four-jet channels, proving a three-sigma excess in 5~fb$^{-1}$ at $\sqrt{s}=14\,{\rm TeV}$, and a five-sigma discovery in three of the four sub-channels within 15~fb$^{-1}$ of data-taking. The point with the lightest gluino in the $(n_M,n_H)$ =  $(\frac{1}{2},\frac{1}{2})$ region is similarly detectable. Most striking is the point with Higgsino-like LSP and $(n_M,n_H)$ =  $(\frac{1}{2},0)$. This was the point that was closest to the discovery threshold in the $\sqrt{s}=8\,{\rm TeV}$ data. We estimate a 3$\sigma$~excess in all three- and four-jet multijet channels almost immediately after data-taking resumes, with a 5$\sigma$~discovery in all four channels within the first 6~fb$^{-1}$ of integrated luminosity.

\begin{table}[t]
\begin{center}
\begin{tabular}{|c|c|c||c|c|c|c|c|c|c|c||c|c|c|c|c|c|c|c|}\hline
\multicolumn{3}{|c||}{Benchmark} 
& \multicolumn{8}{|c||}{20~fb$^{-1}$}  & \multicolumn{8}{|c|}{300~fb$^{-1}$} \\ \hline
Name & $n_M$ & $n_H$ & Monojet & \parbox{0.65cm}{2JM} & \parbox{0.65cm}{3JM} & \parbox{0.65cm}{3JT} & \parbox{0.65cm}{4JM} & \parbox{0.65cm}{4JT} & \parbox{0.65cm}{5J} & \parbox{0.65cm}{6JT} &  Monojet & \parbox{0.65cm}{2JM} & \parbox{0.65cm}{3JM} & \parbox{0.65cm}{3JT} & \parbox{0.65cm}{4JM} & \parbox{0.65cm}{4JT} & \parbox{0.65cm}{5J} & \parbox{0.65cm}{6JT} \\ \hline
Higgsino-A 	& 0 & 0	& 0.01 & 0.18 & 0.95 & 1.53 & 0.91 & 1.14 & 0.34 & 0.23 
& 0.06 & 0.69 & 3.67 & 5.93 & 3.51 & 4.42 & 1.32 & 0.89 \\
Higgsino-B 	& 0 & $\frac12$	& 0.08 & 0.61 & 2.39 & 1.50 & 2.44 & 1.32 & 1.23 & 0.83 
& 0.30 & 2.36 & 9.25 & 5.82 & 9.45 & 5.12 & 4.77 & 3.22 \\
Higgsino-C 	& $\frac12$	& 0 & 0.53 & 0.70 & 2.67 & 4.59 & 2.38 & 3.46 & 0.68 & 0.45  
& 2.06 & 2.70 & 10.3 & 17.8 & 9.23 & 13.4 & 2.64 & 1.76 \\
Higgsino-E 	& $\frac12$	& 1 & 0.01 & 0.08 & 0.42 & 0.85 & 0.29 & 0.47 & 0.07 & 0.02
& 0.05 & 0.31 & 1.63 & 3.28 & 1.13 & 1.80 & 0.26 & 0.08 \\
Higgsino-F 	& 1 & 0	& 1.15 & 0.07 & 0.25 & -- & 0.53 & -- & -- & --
& 4.45 & 0.29 & 0.97 & 0.02 & 2.05 & 0.02 & -- & -- \\
Bino-B & 0 & 0 & 0.06 & 1.22 & 5.57 & 5.67 & 5.63 & 4.53 & 2.41 & 1.36 
& 0.24 & 4.74 & 21.6 & 21.9 & 21.8 & 17.5 & 9.24 & 5.27 \\ \hline \hline
Higgsino 	& 0 & 0	& 0.02 & 0.19 & 1.02 & 1.64 & 0.97 & 1.24 & 0.37 & 0.23 
& 0.07 & 0.76 & 3.95 & 6.34 & 3.76 & 4.81 & 1.42 & 0.91 \\
Higgsino 	& $\frac12$	& 0 & 0.49 & 2.41 & 11.5 & 20.3 & 9.11 & 14.0 & 2.93 & 1.79  
& 1.89 & 9.32 & 44.7 & 78.7 & 35.3 & 54.3 & 11.3 & 6.92 \\
Higgsino 	& $\frac12$	& $\frac12$ & 0.31 & 0.61 & 3.30 & 6.32 & 2.63 & 4.22 & 0.81 & 0.63
& 1.19 & 2.38 & 12.8 & 24.5 & 10.2 & 16.4 & 3.16 & 2.42 \\
Bino & 0 & 0 & 0.08 & 1.71 & 6.23 & 5.73 & 6.25 & 4.87 & 2.63 & 1.49 
& 0.32 & 6.64 & 24.1 & 22.2 & 24.2 & 18.9 & 10.2 & 5.77 \\
Bino & 0 & $\frac12$ & 0.02 & 0.40 & 1.98 & 2.49 & 2.16 & 2.14 & 1.02 & 0.76 
& 0.08 & 1.56 & 7.67 & 9.65 & 8.37 & 8.29 & 3.96 & 2.93 \\ \hline
\end{tabular}
\end{center}
\caption{\footnotesize \textbf{Estimated Signal Significance for Selected Benchmark Points of Tables~\ref{tbl:signal14TeV} and~\ref{tbl:signal14TeVlow} in 20~fb$^{-1}$ and 300~fb$^{-1}$}. Signal significance, defined as the number of signal events divided by the square root of the number of background events ($S/\sqrt{B}$), is given for all benchmarks for which $S/\sqrt{B} \geq 3$ for at least one LHC8~channel with 300~fb$^{-1}$ of data. Only those channels that give such a signal are included in the table. The dashes imply zero signal events for that channel.}
\label{tbl:signalsig}
\end{table}

Clearly, some LHC8~channels, and some benchmark points, do not look promising, even if the signal and background is extrapolated to very large integrated luminosities. Table~\ref{tbl:signalsig} gives the signal significance for all benchmarks for which $S/\sqrt{B} \geq 3$ in at least one LHC8~channel with 300~fb$^{-1}$ of data. Only those channels that give such a signal are included in the table. Taking 300~fb$^{-1}$ to be a reasonable guess as to the total data set accumulated before the next shut-down, we expect only very favorable cases with a bino-like LSP to be accessible in the near future at the~LHC. 

It is promising, however, that all cases with a Higgsino-like LSP, that involve embedding the Standard Model field content exclusively into a system of $D7$ branes, will yield some testable parameter space in the next run at $\sqrt{s}=14\,{\rm TeV}$. Only cases in which one or both of the modular weights are unity would fail to yield an excess in 300~fb$^{-1}$, and in some exceptional cases, such as Higgsino-F, the LHC8 monojet search would eventually yield a discovery at this level of integrated luminosity.

\section{Conclusions}

Type~IIB string theory compactified on Calabi-Yau orientifolds, in the manner first described by Kachru et al., has remained one of the best-studied string-motivated effective supergravity models for almost a decade. Such models naturally give rise to a mirage pattern of gaugino masses, and ultimately provided the very name for this paradigm of supersymmetry breaking. From a low-energy effective field theory point of view, the model class studied here can be considered a generalized modulus-dominated scenario, and therefore forms a natural complement to the generalized dilaton-domination scenario considered by the authors in Reference~\cite{Kaufman:2013pya}. 

As a model in which a K\"ahler modulus transmits the supersymmetry breaking to the observable sector, the weights of the various matter representations under $SL(2,Z)$ modular transformations are relevant for the scale and pattern of supersymmetry-breaking scalar masses. This motivates breaking the total parameter space of the theory into disjoint cases, which map onto different ways in which the Standard Model field content can be realized locally in terms of systems of $D3$ and/or $D7$ branes. This, in turn, has the very attractive feature that properties of the theory testable at the~LHC, or at various dark matter detection experiments, can be directly related to the nature of the compact space at the string scale.

A scan over the available free parameters of the Type~IIB flux compactification model reveals that the observation of a Standard Model-like Higgs boson with $m_h \simeq 125\,{\rm GeV}$ is already severely constraining on the model space, particularly if one is to insist on no more thermal neutralino relic abundance than that indicated by the PLANCK and WMAP satellite data. These measurements alone already suggest gluino masses at or above 2~TeV, with no expected signal above background in supersymmetry searches performed thus far at the~LHC. The original prediction of Kachru et al., that $\alpha = 1$ when vacuum uplift is achieved through anti-$D3$ branes, is allowed in only a handful of modular weight combinations -- intriguingly, those combinations associated with semi-realistic Type~IIB model building: $(n_M,n_H)$ =  $(0,0)$, $(0,\frac{1}{2})$ and $(\frac{1}{2},0)$. If this is the correct theory of Nature, then the LSP will be overwhelmingly bino-like with a mass on the order of 1~TeV. Such a dark matter candidate will likely remain inaccessible to direct search experiments for the foreseeable future. The strongly-interacting superpartners will likely be detectable at the~LHC, however. For example, the heaviest gluino for the $(n_M,n_H)$ =  $(0,\frac{1}{2})$ case with $0.95 \leq \alpha \leq 1.05$ has a mass of just over 2300~GeV. This point will produce a one-sigma excess over the background in events with three and four jets plus missing transverse energy with 20~fb$^{-1}$ at $\sqrt{s}=14\,{\rm TeV}$. This should be a three-sigma excess with 100~fb$^{-1}$, and a five-sigma discovery with 300~fb$^{-1}$ in at least one channel from Table~\ref{tbl:multijets}.

When one takes a more agnostic point of view with regards to the eventual uplift mechanism, more flexibility in the parameter $\alpha$ is allowed. In these cases we find that the bulk of the parameter space prefers a Higgsino-like LSP with a relatively good chance of being detected in 1-3~ton-years of exposure in liquid xenon-based dark matter detectors. Such model points continue to have a heavy gluino and heavy squarks, making observation at the~LHC at $\sqrt{s}=14\,{\rm TeV}$ challenging, but not hopeless. The first 300~fb$^{-1}$ will be enough data to begin to probe these most-promising regions of the flux-compactified Type~IIB model. Even using the signal region definitions employed at $\sqrt{s}=8\,{\rm TeV}$, we anticipate a significant reach for Higgsino-like LSP outcomes. The cuts can be adjusted to take advantage of improved signal-to-background ratios available at the higher center-of-mass energy, and multiple channels can be combined to extend the reach still further. Yet complete coverage of the Type~IIB model, compactified on an orientifold in the presence of fluxes, will likely require a next-generation proton collider with $\sqrt{s} \simeq 100\,{\rm TeV}$~\cite{Jung:2013zya}.

\section{Acknowledgements}
BK would like to thank Sujeet Akula and Baris Altunkaynak for technical assistance in the early stages of this work. BK and BDN are supported by the NSF under grant PHY-0757959. We would also like to thank the Boston Red Sox for winning the World Series during the preparation of this work.

\section{Appendix}

Our conventions for the coefficients in~(\ref{Aterm}) and~(\ref{scalarmass}) follow those of the Appendix to Ref.~\cite{Altunkaynak:2010xe}. In particular, we work in the approximation that generational mixing can be neglected, so that only third-generation Yukawa couplings are relevant. At one loop, the anomalous dimensions are given by 
\begin{eqnarray}
\gamma_i = 2 \sum_a g_a^2 c_a(\Phi_i) - \frac{1}{2}\sum_{lm} |y_{ilm}|^2,
\label{gammaexp}
\end{eqnarray}
in which $c_a$ is the quadratic Casimir, and $y_{ilm}$ are the normalized Yukawa couplings. For the MSSM fields $Q$, $U^c$, $D^c$, $L$, $E^c$, $H_u$ and $H_d$,
the anomalous dimensions are
\begin{eqnarray}
\gamma_{Q,i} &=& \frac{8}{3} g_3^2 + \frac{3}{2} g_2^2 + \frac{1}{30} g_1^2
- (y_t^2 + y_b^2) \delta_{i3}\nonumber \\
\gamma_{U,i} &=& \frac{8}{3} g_3^2 + \frac{8}{15} g_1^2
- 2 y_t^2 \delta_{i3},\;\; 
\gamma_{D,i} =
\frac{8}{3} g_3^2 + \frac{2}{15} g_1^2
- 2 y_b^2 \delta_{i3},\nonumber \\
\gamma_{L,i} &=& \frac{3}{2} g_2^2 + \frac{3}{10} g_1^2
-y_\tau^2 \delta_{i3}, \;\; 
\gamma_{E,i} =
\frac{6}{5} g_1^2
-2 y_\tau^2 \delta_{i3}, \nonumber \\
\gamma_{H_u} &=& \frac{3}{2} g_2^2 + \frac{3}{10} g_1^2
-3 y_t^2, \;\; 
\gamma_{H_d} =
\frac{3}{2} g_2^2 + \frac{3}{10} g_1^2
- 3 y_b^2 - y_{\tau}^2\, .
\end{eqnarray}
The $\dot{\gamma}_i$'s are given by the expression
\begin{eqnarray}
\dot{\gamma}_i=2\sum_a g_a^4b_a c_a(\Phi_i) - \sum_{lm} |y_{ilm}|^2b_{y_{ilm}},
\end{eqnarray}
in which $b_{y_{ilm}}$ is the beta function for the Yukawa coupling $y_{ilm}$.  The 
$\dot{\gamma}_i$'s are given by
\begin{eqnarray}
\dot\gamma_{Q,i}
&=& \frac{8}{3} b_3 g_3^4 + \frac{3}{2} b_2 g_2^4 + \frac{1}{30} b_1 g_1^4
- (y_t^2 b_t + y_b^2 b_b ) \delta_{i3} \nonumber \\
\dot\gamma_{U,i}
&=& \frac{8}{3} b_3 g_3^4 + \frac{8}{15} b_1 g_1^4
- 2 y^2_t b_t \delta_{i3}, \;\; 
\dot\gamma_{D,i}=
\frac{8}{3} b_3 g_3^4 + \frac{2}{15} b_1 g_1^4
- 2 y^2_b b_b \delta_{i3} \nonumber \\
\dot\gamma_{L,i}
&=& \frac{3}{2} b_2 g_2^4 + \frac{3}{10} b_1 g_1^4
 - y_\tau^2 b_\tau \delta_{i3},\;\; 
\dot\gamma_{E,i}=
\frac{6}{5} b_1 g_1^4
 - 2 y_\tau^2 b_\tau \delta_{i3} \nonumber \\
\dot\gamma_{H_u}
&=& \frac{3}{2} b_2 g_2^4 + \frac{3}{10} b_1 g_1^4
 - 3 y^2_t b_t,\;\; 
\dot\gamma_{H_d}
=\frac{3}{2} b_2 g_2^4 + \frac{3}{10} b_1 g_1^4
 - 3 y_b^2 b_b - y^2_\tau b_\tau,  \label{dotgammaexp}
\end{eqnarray}
where
$b_t = 6 y_t^2 + y_b^2 -\frac{16}{3} g_3^2 - 3 g_2^2 - \frac{13}{15} g_1^2$, 
$b_b =
y_t^2 + 6 y_b^2 + y_\tau^2 -\frac{16}{3} g_3^2 - 3 g_2^2
- \frac{7}{15} g_1^2$ 
and $b_\tau = 3 y_b^2 + 4 y_\tau^2 - 3 g_2^2 -\frac{9}{5} g_1^2$.
Finally, ${\theta_i}$, which appears in the mixed modulus-anomaly term in the soft scalar mass-squared parameters, is given by
\begin{eqnarray}
\theta_i = 4 \sum_a g_a^2 c_a(Q_i) - \sum_{i,j,k} |y_{ijk}|^2
( 3- n_i -n_j- n_k).
\end{eqnarray}
For the MSSM fields, they take the form
\begin{eqnarray}
\theta_{Q,i} &=& \frac{16}{3} g_3^2 + 3 g_2^2 + \frac{1}{15} g_1^2
-2 ( y_t^2 (3-n_{H_u}-n_Q -n_U) + y_b^2 (3-n_{H_d} -n_Q - n_D )) \delta_{i3},
\nonumber \\
\theta_{U,i} &=& \frac{16}{3} g_3^2 + \frac{16}{15} g_1^2
- 4 y_t^2 (3-n_{H_u} - n_Q - n_U ) \delta_{i3}\nonumber \\
\theta_{D,i}&=& 
\frac{16}{3} g_3^2 + \frac{4}{15} g_1^2
- 4 y_b^2 (3- n_{H_d} - n_Q - n_D ) \delta_{i3}, \nonumber \\
\theta_{L,i} &=& 3 g_2^2 + \frac{3}{5} g_1^2
-2 y_\tau^2 ( 3- n_{H_d} - n_L - n_E ) \delta_{i3}\nonumber \\
\theta_{E,i} &=& 
\frac{12}{5} g_1^2
- 4 y_\tau^2 (3-n_{H_d} -n_L -n_E ) \delta_{i3}, \nonumber \\
\theta_{H_u} &=& 3 g_2^2 + \frac{3}{5} g_1^2
- 6 y_t^2 ( 3- n_{H_u} - n_Q - n_U )\nonumber \\
\theta_{H_d} &=& 
3 g_2^2 + \frac{3}{5} g_1^2
-6 y_b^2 ( 3- n_{H_d} - n_Q - n_D )
-2 y_\tau^2 ( 3- n_{H_d} - n_L - n_E ).
\end{eqnarray}

\end{document}